\documentclass[9pt,twocolumn,twoside]{IEEEtran}
\usepackage{amssymb}

\usepackage{amsmath}
\usepackage{array}
\newcolumntype{C}[1]{>{\centering\arraybackslash}p{#1}}

\DeclareSymbolFont{letters}{OML}{ztmcm}{m}{it}
\DeclareSymbolFontAlphabet{\mathnormal}{letters}

\usepackage{balance}
\usepackage{boldline}			
\usepackage{color}
\usepackage[inline]{enumitem}
\usepackage{balance}

\usepackage{graphicx}
\usepackage{multibib}
\usepackage{multirow}
\usepackage{placeins}
\usepackage{rotating}
\usepackage{resizegather}
\usepackage[caption=false]{subfig}
\usepackage{textcomp}
\usepackage{url}
\usepackage[rawfloats=true]{floatrow}
\usepackage[flushleft]{threeparttable}
\usepackage{lipsum}

\usepackage{cite}

\usepackage{pifont}
\newcommand{\cmark}{\ding{51}}%
\newcommand{\xmark}{\ding{55}}%

\newcolumntype{P}[1]{>{\centering\arraybackslash}p{#1}}

\newcommand{\Csharp}{%
  {\settoheight{\dimen0}{C}C\kern-.05em \resizebox{!}{\dimen0}{\raisebox{\depth}{\#}}}}
  
\begin{document}

\title{Peer-to-Peer based Social Networks: A Comprehensive Survey}

\author{\IEEEauthorblockN{Newton Masinde and Kalman Graffi}
\IEEEauthorblockA{\\Technology of Social Networks\\Heinrich Heine University,\\ Universit{\"a}tsstrasse 1, 40225 D{\"u}sseldorf, Germany\\email: \{newton.masinde, graffi\}@hhu.de\\web: http://tsn.hhu.de/}}

\maketitle

\begin{abstract}

Online social networks, such as Facebook and twitter, are a growing phenomenon in today's world, with various platforms providing capabilities for individuals to collaborate through messaging and chatting as well as sharing of content such as videos and photos. Most, if not all, of these platforms are based on centralized computing systems, meaning that the control and management of the systems lies in the hand of one provider, which must be trusted to treat the data and communication traces securely. While users aim for privacy and data sovereignty, often the providers aim to monetize the data they store. Even, federated privately run social networks require a few enthusiasts that serve the community and have, through that, access to the data they manage. As a zero-trust alternative, peer-to-peer (P2P) technologies promise networks that are self organizing and secure-by-design, in which the final data sovereignty lies at the corresponding user. Such networks support end-to-end communication, uncompromising access control, anonymity and resilience against censorship and massive data leaks through misused trust. The goals of this survey are three-fold. Firstly, the survey elaborates the properties of P2P-based online social networks and defines the requirements for such (zero-trust) platforms. Secondly, it elaborates on the building blocks for P2P frameworks that allow the creation of such sophisticated and demanding applications, such as user/identity management, reliable data storage, secure communication, access control and general-purpose extensibility, features that are not addressed in other P2P surveys. As a third point, it gives an overview of proposed P2P-based online social network applications, frameworks and architectures. In specific, it explores the technical details, inter-dependencies and maturity of the available solutions.

\end{abstract}


\section{Introduction}
\label{sec:Intro}
Social networking as a means of online interaction, has over the past few years experienced unparalleled growth with a massive increase in the number of users over the last 10 years as shown in Table~\ref{tab:TopOSNs}.
This growth has been realized by the evergrowing number of users of the different online social network (OSN) users, in which virtually every age group is effectively represented.

\begin{table}[t]
 \scriptsize \centering
 \label{tab:}
 \caption{Structure of the Survey}
 \label{tab:Struct}
 \begin{tabular}{p{8.0cm}}
  \hline \makeatletter \@starttoc{toc} \makeatother\\
  \hline
 \end{tabular}
\end{table}

Many studies that have been carried out on the current popular OSNs such as~\cite{ARM+11,HKP12,CCM14,MKI+16,KIa17} have uncovered several challenges that must be carefully considered and addressed.
These issues include, but are not limited to, seamless scaling of the network without straining of the available resources (both monetary and physical) and ability of users to control their data and maintain their privacy while using the social networks.
While the fist issue, the technical feasibility, has been mastered by the providers, the privacy and trust issue could not yet be solved. 
Let us consider the \textit{centralized} service provision model currently used by the listed OSN platforms in Table \ref{tab:TopOSNs}. 
Here, a single operator hosts the platform, maintains its availability and uses its access to all data stored by the ``customers'' to provide social networking services. 
The data comprises profile data, communication traces, all content uploaded and downloaded and all interaction traces. 
While the user of the platform typically only aims to interact with his friends or followers, in private or as group and sometimes in public, he has no chance to use the service without the provider to see and know. 
Thus, trust in the provider is required as the users to do not have the sovereignty on their data. 
The provider on the other hand, besides aiming for appealing services for the users, has also a high interest in monetizing the data of the users. 
Having to accept (personalized) ads is an annoying cumbersomeness, but further misuse happens. 
Examples of such misuse include the Facebook and Cambridge-Analytica scandal\footnote{https://www.cnbc.com/2018/03/21/facebook-cambridge-analytica-scandal-everything-you-need-to-know.html}.
There are also other concerns due to the misuse such as government surveillance to infringe privacy such as the Chinese Social Credit System~\cite{CCh17,LDK+18}, location tracking\footnote{https://www.businessinsider.com/three-ways-social-media-is-tracking-you-2015-5?IR=T}~\cite{LZG+14}, social media data mining for terrorist sentiments~\cite{AAA+19,AXu19} which may infringes on free speech, among other effects.

The privacy issue stems from the choice of the data model that is used in the design of the system.
Besides the common \textit{centralized} data model observed in the design of OSNs, \textit{decentralized federated} and \textit{peer-to-peer}~\cite{PZD11,Jus14} data models exist, that promise to both be technically feasible to host billions of users in a social networking platform and at the same time shift the  data sovereignty to the user. 


\begin{table}
 \centering \scriptsize
 \caption{Top 10 online social networks as of July 2019 (in millions), from https://www.statista.com}
 \label{tab:TopOSNs}
 \begin{tabular}{|p{2.2cm}|P{0.8cm}|}
  \hline \rule{0pt}{2ex}\textbf{OSN Platforms}&%
  \textbf{Users}\\
  \hline\hline \rule{0pt}{2ex}\textbf{Facebook}&%
  2,375\\
  \rule{0pt}{2ex}\textbf{YouTube}&%
  2,000\\
  \rule{0pt}{2ex}\textbf{WhatsApp}&%
  1,600\\
  \rule{0pt}{2ex}\textbf{Facebook Messenger}&%
  1,300\\
  \rule{0pt}{2ex}\textbf{WeChat}&%
  1,112\\
  \rule{0pt}{2ex}\textbf{Instagram}&%
  1,000\\
  \rule{0pt}{2ex}\textbf{QQ}&%
  832\\
  \rule{0pt}{2ex}\textbf{QZone}&%
  572\\
  \rule{0pt}{2ex}\textbf{Douyin/Tik Tok}&%
  500\\
  \rule{0pt}{2ex}\textbf{Sina Weibo}&%
  465\\
  \hline
 \end{tabular}
\end{table}

The federated decentralized model is a break away from the centralized models, in which there is no single owner of the network but rather users host parts of the OSN and federate for a complete network. 
While single node providers manage the data of a few members, node lists exist, that present new users possible connection points to the OSN they can attach to. 
There over 30 federated social network projects listed as being active (https://the-federation.info/\footnote{https://the-federation.info/}) and in Table~\ref{tab:FederatedOSNs} is a list the of the top ten based on the number of users active in the network.
However, although this solution promises to handle scalability concerns, the security and privacy concerns are very similar to the centralized model, as shown in~\cite{KIa17}. 
It appears that the owners of the nodes are able to access the private information of the users that connect to that particular node as well as having control of the content stored on the servers. 
Thus, the owners might not have access to all data, they still can access and modify the data of tens to hundreds of users. 
Taking into account, that OSNs are used for sensitive communication, it is even more frightening if the user is known to the owner of the node. 
Also here, data sovereignty is not in the hand of the user. 



\begin{table}
 \centering \scriptsize
 \caption{Top 10 federated online social networks (Nov.
  2, 2019)}
 \label{tab:FederatedOSNs}
 \begin{tabular}{|p{1.5cm}|P{0.8cm}|P{1.0cm}|P{1.8cm}|}
  \hline \rule{0pt}{2ex}\textbf{Project}&%
  \textbf{Nodes}&%
  \textbf{Users} &%
  \textbf{Users per node}\\
  \hline\hline \rule{0pt}{2ex}\textbf{Mastodon}&%
  2771&%
  2,525,434&%
  912\\
  \rule{0pt}{2ex}\textbf{Diaspora}&%
  202&%
  705,662&%
  3494\\
  \rule{0pt}{2ex}\textbf{Pleroma}&%
  590&%
  27,770&%
  48\\
  \rule{0pt}{2ex}\textbf{PeerTube}&%
  331&%
  20,018&%
  61\\
  \rule{0pt}{2ex}\textbf{Juick}&%
  1&%
  18,053&%
  18,053\\
  \rule{0pt}{2ex}\textbf{Friendica}&%
  107&%
  14,632&%
  137\\
  \rule{0pt}{2ex}\textbf{Hubzilla}&%
  119&%
  7,045&%
  60\\
  \rule{0pt}{2ex}\textbf{Write Freely}&%
  183&%
  5,023&%
  28\\
  \rule{0pt}{2ex}\textbf{PixelFed}&%
  101&%
  4,644&%
  46\\
  \rule{0pt}{2ex}\textbf{FunkWhale}&%
  30&%
  2,659&%
  89\\
  \hline
 \end{tabular}
\end{table}

Based on these concerns identified in the centralized and federated data models for the OSNs, our take home is that any suitable solutions must strive at:
\begin
 {enumerate*} [label=\itshape\alph*\upshape)]
\item
 being financially viable, \item alleviating the security and privacy concerns of users, and \item supporting dynamic system growth seamlessly.
\end
{enumerate*} It is evident that such a solution has thus far not been viable on the centralized OSNs as well as on the decentralized federated solutions.

We therefore aim at the third data model, peer-to-peer (P2P), to show that it can provide the required properties of scale, security/privacy and organic growth and in addition alleviate the need for monetizing. 
The P2P model has the advantages of being easily scalable depending on the type of overlay chosen.
Peers that join the network bring their own resources, such as a part of their available storage space, flat rate bandwidth and unused computing cycles, leading to an accumulation of ``free'' resources, that is by far sufficient to host a fully-decentralized OSN.
In addition, P2P mechanisms are self-organizing and also support zero-trust requirements, with the aim to remain functional under churn (node online dynamics), strong heterogeneity of resources and workload, the presence of malicious nodes in the network and tailored attacks. 
As all peers are equal, no role bearing higher rights, such as an administrator or operator, exist. 
In the code, the data sovereignty of the user, security and privacy can be enforced. 
To build a purely P2P-based OSN seems highly preferable: no operational costs, harnessing of ``free'' resources and thus more powerful functions, data sovereignty of the user and inability to turn the system down or to censor the content. 
However, building a purely P2P-based OSN is also highly challenging as key functionalities need to be considered such as routing methods, data storage, distribution and replication mechanisms, messaging-handling techniques, communication schemes, appealing apps as well as through all functions efficiency, security and privacy. 
These requirements altogether are highly challenging to solve at once in a system and have not yet been discussed in the detail and completeness as in this paper. 

\subsection{Identifying the gaps} There have been several recent studies that cover distributed (or decentralized) online social networks (DOSNs) under which P2P-based solutions are also discussed.
Different researchers focus on different aspects of the DOSNs such as P2P architectures for DOSNs~\cite{GCR13,KGu14}, DOSN design decisions (storage, access control, and interaction and signaling mechanisms)~\cite{PFS14}, security and privacy in DOSNs~\cite{GKB12,TKO15,DMR18,ORP19}, as well as general surveys on DOSNs~\cite{PFS14,CRS+15}.
Most of these surveys give insightful information of various aspects of DOSN, while at the same time including P2P-based OSNs.
However, they do not analyze the strong interdependencies stated by the requirements for a secure OSN with the (often limited) P2P technologies. 
Analyzing parts of the system, unfortunately, does not describe its property in interaction with further required OSN elements. 
In this sense, there are surveys that cover different component aspects of the P2P networks such as overlays~\cite{KGu14,Mal15}, replication~\cite{SBX14,SBX+15}, searching and indexing~\cite{RiM06,Kan11,ZXT+11}, security~\cite{TrK12,TKh12,GRS+14} and so on, as well as several good reference books that cover P2P networks in general such as~\cite{SYB10} and \cite{Kwo12}. 
In our experience, the challenges in building a P2P-based OSN appear when these widely discussed mechanism need to be combined and need to be secured. 
Thus pure P2P surveys only give a vary limited view on applicability of the P2P technologies for the purpose of building a P2P-based OSNs.

The study of this literature has enable us to identify several not so visible gaps, which it is our aim in this work to try to address.
\begin {itemize}
\item
 Centralized systems have matured and have gone through several steps into standardization in many areas.
 In order for P2P systems to, at a minimum, compete at the same level, there is need for clear discussion on steps into standardization within P2P systems.
 In this respect, we have observed that most descriptions of the requirements (functional, non-functional and technical) for the P2P-based OSNs are more or less focused on the system being proposed.
 These descriptions are useful in discussions that lead into a standard format for designing P2P-based applications.
\item
 Most of the discussions on P2P networks focus on the different mechanisms offered.
 It would be interesting to find a discussion that gives a clear comparison on different options available for a given P2P mechanism and the advantages and disadvantages of using one over the other.
 This would be beneficial to a designer by giving a general direction on how these components can be adapted and integrated into a fully functional application.
\item
 The general trend has been towards looking at P2P-based OSNs as a subset of the larger DOSNs, as opposed to having P2P-based OSNs as the singular point of focus.
 Even then, most comparisons of DOSN proposals give a one-dimensional approach in comparing them, that is, how they differ based on a particular focus such as, storage management and security.
 There are very few, if any, discussions that give a full comparison based on the overall composition of the proposed solutions.
\end
{itemize}

\subsection{Our Contributions} In line with the gaps identified in our study, we make the following contributions:
\begin
 {enumerate*} [label=\textbf{\itshape\arabic*\upshape})]
\item
 We identify key features common in all OSNs, as well the functional and non-functional requirements needed for a working OSN.
\item
 We give an overview of the concerns raised that led into research on decentralized solutions, and in particular P2P-based options, for the OSNs.
\item
 We give a general layout for the technical requirements needed to achieve the functional and non-functional requirements, and use these technical aspects to further define a P2P framework for an OSN.
\item
 Based on the technical requirements, we give a study on the core P2P network mechanisms that are integral in the design of any P2P application, with considerations into the security aspects for each mechanism.
\item
 We present a comprehensive survey of a number of the proposed P2P-based OSNs.
\item
 We provide a roadmap for possible future research into P2P-based OSNs.
\end
{enumerate*} The structure of the survey is shown in Table \ref{tab:Struct}.


\section{Social Networks}
\label{sec:SNs}
A social network in the classical sense involves real people interacting in the real world~\cite{GHW97}.
A more technical view is a social network as a directed graph structure~\cite{MWe11,Agg11}.
However, the term ``social network'' as it is used today tends to refer to a combination of the real and virtual world via web-based services~\cite{BoE07}.
As it stands today, SNs have distinguished themselves as the easiest and fastest means for connecting with other users for various reasons such as for friendship, business, entertainment and even knowledge sharing.
An emerging trend is the use of SNs for purposes of commercialization, which is highly dependent on the number of users that the site attracts~\cite{SSW10}.

There are different types of social network applications that have been studied.
However, for purposes of clarity that will become apparent in this work, we would like to focus on general social networks and microblogs.
Although microblogs are essentially also social networks, there is a clear line of distinction separating them from the common SNs.
Social networks allow participating members to share unlimited and unrestricted amounts of information through different ways such as microblogging, subscriptions, status, mobile text alerts, blogs, instant messaging, and forums and so on~\cite{DeW10}.
Blogs (short for ``weblog''), possible due to Web.2.0, allow users to maintain and publish a personal log or diary on the web and are quite similar to personal websites.
 They differ from websites on several respects such as ease in the ease of blog creation against the website.
 Microblogs are fundamentally blogs that impose a limitation on the amount of content that a user can upload to their personal space or share with others.
 The posts in microblogs are usually in the form of text, pictures, links, short videos or other forms of web media~\cite{JSF+07}.
 In this respect, that is, the size of content, microblogs such as twitter differ from other common social networks such as Facebook.

 In the rest of this section, we present the various ways of classifying OSNs.
 Thereafter, we consider the
 desired functions in the OSNs that guarantee meet the user's online needs.
 Further, the functional and non-functional SN requirements will be presented.
 Lastly, we look at the main concerns raised that have necessitated research into decentralized solutions for social networks, and in particular P2P-based solutions.

 \subsection{Social Network Classifications}
 \label{subsec:SNClass}
\begin{figure*}
  [htbp!] \centering \includegraphics[scale=0.6]{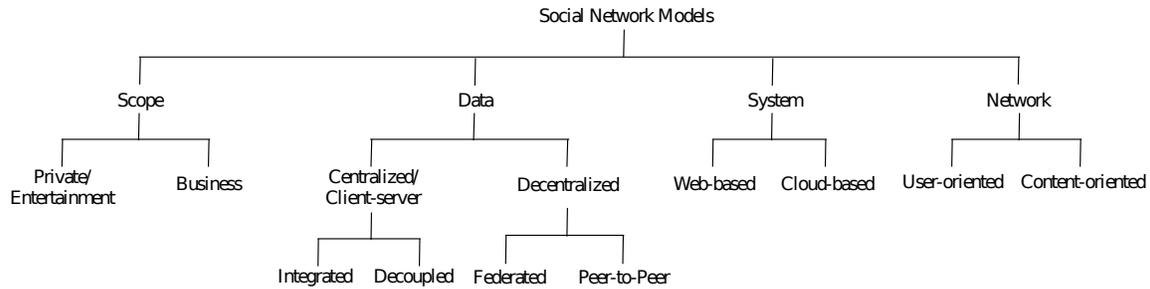}
  \caption{Social network classifications}
  \label{fig:SNS_Models}
 \end{figure*}

 Various proposals have been put forward for the classification of SNs as shown in Fig.~\ref{fig:SNS_Models} and discussed herein.

 \subsubsection{The scope model} The scope model~\cite{PZD11} considers the core activities of the SN which are grouped into two categories.
 The first category, \textit{entertainment} (or \textit{private}~\cite{HKP12}), focuses on the delivery of fun and interactive social experience to registered users, for example, Facebook, Flickr, MySpace and Hi5.
 The second category is \textit{business}, which focuses on connecting professionals for purposes of productivity and success, for example, LinkedIn and Xing.

 \subsubsection{The data model} The data model~\cite{PZD11} (or \textit{programming paradigm model}~\cite{Jus14}), classifies the SNs as centralized or decentralized.
 In the \textit{centralized model}, the storage of data is in a single administrative domain.
 It is further divided into \textit{integrated client-server} and \textit{decoupled client-server}.
 In the integrated client-server model, application developers utilize their own servers to manage and store social relationships and also provide required resources supporting content sharing.
 On the other hand, for the decoupled client-server model, users have the capability to manage their own social relationships but only core social services and social relationships which are linked to their accounts are updated centrally.
 For the \textit{decentralized model}, data is distributed across multiple administrative domains, and it has two sub-categories: \textit{decentralized federated} and \textit{peer-to-peer}.
 The decentralized federated model offers no centralized infrastructure from the application developers but there is reliance on existing decentralized and federated messaging system, such as Extensible Messaging and Presence Protocol (XMPP).
 Thus users choose the service provider of their choice so long as they are part of the same federation.
 The peer-to-peer model is fully decentralized and users directly connect to their trusted friends and share content.
 Developers either write their own P2P protocol or rely on existing P2P technologies.

 \subsubsection{The system model} This view of SNs looks at the application servers in terms of the hosting and content distribution.
 Pallis et al.~\cite{PZD11} give two categories under this dimension.
 The first is the \textit{web-based scheme} in which application servers are hosted by Web sites providing required set of services as well as APIs and most of the services are usually free to users.
 The second category is \textit{cloud-based schemes} in which the application servers are hosted by a utility company that provides the necessary infrastructure, and the servers accessed by the SN provider are virtual.

 \subsubsection{The network model} Finally, in the network model~\cite{ChI08,PZD11}, the manner in which the users\textquotesingle~network is formed is taken into consideration.
 In this regard then, there are two further subgroups within this model: \textit{user-oriented} and \textit{content-oriented}.
 User-oriented SNs (also called profile-based SNs) emphasizes the social relationships that are in existence and content sharing occurs between users in the same community.
 Examples of SNs in this category include Facebook, MySpace and LinkedIn.
 The content-oriented SNs (also called content-based SNs) emphasize the common interests that may exist between users rather than the social relationships.
 Examples in this category include YouTube.

 \subsection{Desired functions in OSNs}
 \label{subsec:UsrExpMgmt}
 Kietzmann et al.~\cite{KHM+11} proposed a framework containing functional block that highlights the essential functional blocks for consideration by social media providers for maximum user experience.
 These functional blocks are identity, conversation, sharing, presence, relationships and groups.
 Further reputation is named, which we discuss separately.
 A brief summary of these concerns follows.

 \subsubsection{Identity} This considers the extent to which users are ready to reveal their identities within the social media environment by disclosing their personal details such as name, age, gender, profession, location or other information.
 Normally, this is made possible by users setting up profiles which have their personal details.
 A concern raised here is how privacy and anonymity can be enforced and how OSN providers use the information about their users.
 It has been seen however, that there needs to exist some tools that allow for a proper balance between sharing of identities hence anonymity and enforcing of privacy, and a good choice of such tools bolsters accountability among users, prevents cyber-bullying and encourages off-topic/-color comments.

 \subsubsection{Conversation} This aspect looks at how users communicate with each other within the social media setting.
 In this regard, the conversations can be set into a certain format, such as in Twitter or can be random, that is, a general text format, such as in Facebook.
 The OSN platform should facilitate for conversations among individuals as well groups.

 \subsubsection{Sharing} How the users exchange, distribute and receive content in the OSN is very important.
 In terms of user experience, the ease of sharing data as content, as well as guaranteed availability of the same content is a major concern.
 The OSN platform should ensure that the content is available in the face of high churn, even when the initial provider is offline.

 \subsubsection{Presence} The consideration in this case is the extent to which users are aware if other users are accessible, that is, knowing where they are (virtual and/or real world) and whether they are available.
 This can be made possible through provision of statuses, for example, \textquoteleft invisible\textquoteright, \textquoteleft available\textquoteright, \textquoteleft busy\textquoteright~and so on.

 \subsubsection{Relationships} Focus, in this case, is on the ability of users to relate to other users, that is, two or more users having some type of association that makes it possible to have a conversation, share content and even list each other as friends or fans.
 This aspect of the user experience will dictate the what-and-how of information exchange.
 An example of how these relationships are linked is how LinkedIn shows how users are connected to others and the number of degrees of separation the user is to another user.
 A general rule is that a social media that esteems the value of identity as low also considers relationships as of low concern.

 \subsubsection{Groups} This looks at how the users are able to form communities and sub-communities.
 Dunbar\textquotesingle s number is one of the proposed relationship-group metric, which considers that people have a cognitive limit that bounds the number of stable relationships they can have to 150 people.
 There are two types of groups in existence.
 In the first group, individuals go through their connections and place their friends, followers or fans into self-created groups.
 In the second group, the online groups are analogous to clubs in the real world and can be open to access, closed (approval is required) or secret (by invitation only).
 \\
 While~\cite{KHM+11} also discusses \textit{reputation} as an important element, we do not share this view.
 Reputation is used to emphasize on how users are able to identify the standing of other users, as well as themselves, within the social media setting.
 In reality, such explicit reputation rating do not take place.
 Neither in OSNs nor in other social applications do we find reputations.
 Only in online business reviews (such as restaurant ratings) are reputations found, but in contrast to social reputations, these are explicitly public, while a social reputation is mainly created for a personal use.
 Thus, a technical support for reputation is not considered relevant.

 \subsection{Functional requirements for OSNs}
 \label{subsec:SN_FuncReqs}
 The traditional goals of an OSN are achievable by ascertaining that the system delivers a minimal set of core functionalities.
 The following is a brief discussion of the core functionalities required for OSNs as highlighted in~\cite{ZSZ10} and discussed also as service requirements in~\cite{ARu12}.

 \subsubsection{Personal storage space management} This entails the ability of users to create/cancel a user account, create/edit a user profile, and upload/edit user generated content.
 In general, this considers the ability of the user to control some arbitrary space that has been assigned to that user upon creation of the account and profile, allowing the storage, deletion and manipulation (or editing) of user\textquotesingle s content.
 Most OSNs will usually allow the reporting of the user\textquotesingle s action in their personal space to other users that are their social contacts unless this feature is specifically disallowed by the user.

 \subsubsection{Social connection management} By providing this functionality the SN ensures users can establish/maintain/revoke a social connection.
 This functionality aims at enabling definition of the relationship between a user and others, using friend lists for example.
 It also enables locating and reestablishment of connections with lost friends as well as establishing new relationships due to common interests such as ideologies, media content and so on.

 \subsubsection{Social graph traversal} This functionality is also termed \textit{social traversal}.
 The online social graph provides another mechanism of retrieving a search list by performing traversals on the online social graph and examining the friend lists of other users.
 Traversals may be restricted to a subset by defining a traversal policy

 \subsubsection{Means of communication} An appropriate choice for the means of communication ensures that the required channels for users to interact with one another by sending appropriate messages to one another in the form of text, audio, video, photos or other format.
 The messages can be public or private.
 The SN should also support both synchronous communication (such as instant messaging) and asynchronous communication.

 \subsubsection{Shared storage space interaction} Besides having a personal storage space, SNs also offer shared spaces for interaction such as walls, forums or commonly shared folders.
 By having the option of an access-controlled shared storage space, further more sophisticated applications can be added that require data-based interaction, such as collaborative cooperation, gaming, digital workplaces and more.

 \subsubsection{Search facilities} Having this implemented in the system allow users to find and connect with new contacts by exploring the social network space.
 This allows users to find others not currently in their friend lists, thus establish new relationships.
 \newline

 In summary, SNs should offer users the ability to create a personal data space to present themselves and to search and connect with other users to build a digital social network.
 The interaction of the users is supported through various options for communication, such as sending messages to each other, to groups, or to interact in a notification based communication, with various publishers and subscribers.
 The interaction is further supported through various elements based on a shared, access-controlled storage, such as asynchronous communication threads (walls, threads), shared document spaces or shared state for commonly used applications.


 \subsection{Non-functional requirements for OSNs}
 \label{subsec:SN_NonFuncReqs}
 OSN applications must also provide a suitable environment to the public, that is, the general user, so as to inspire confidence in the system to the users as they utilize the services that are offered therein.
 The online social network (OSN) requirements needed to ensure such an environment is possible are divided into two main categories: privacy requirements, security requirements~\cite{ARu12} and metering.

 \subsubsection{Privacy requirements} The system must provide confidentiality, ownership privacy, social interaction privacy and activity privacy.
 \textit{Confidentiality} refers to the availability of appropriate access control policies, as well as encryption mechanisms to prevent information leakages.
 \textit{Ownership privacy} means that the content owner should be able make a choice of revealing ownership information to other users.
 \textit{Social interaction privacy} is guaranteed if a user is able to hide the interaction patterns between him/herself and other users.
 \textit{Activity privacy} means that the interaction between the user and his application suite are not exposed to the public.

 \subsubsection{Security requirements} To provide appropriate security, the system must also include cover channel availability, authentication, data integrity and authenticity, and some also name non-repudiation.
 \textit{Channel availability} requires that the service is available and can be used even under malicious attacks.
 \textit{Channel authentication} is ensured possible by providing some form of two-way authentication between an initiator and recipient of the message.
 \textit{Data authenticity and integrity} points to the necessity of the system preventing modification of content or messages by any unauthorized users.
 \textit{Non-repudiation} as a last point is optionally named in special cases.
 It means that the sender can be traced and the interaction is documented, which might be useful in case of any malicious content/messages.
 In many cases, where anonymity is higher prioritized, non-repudiation is to be avoided.

 \subsubsection{Metering} As an added bonus to the OSN, it would be beneficial to the users as well as the system developers to receive insights into how well the system operates, the general system performance as well as system limitations/overloads if they exists.
 This way, mitigating solutions may be designed to ensure optimum performance.
 Thus a reliable and accurate system monitoring interface can be integrated to collect system data, which can later be analyzed to gauge the health of the system.

 \subsection{Motivation for decentralization}
 \label{subsec:DOSNs}

 For the longest time, Client-Server (C/S) computing model has been the mainstay of OSNs due to the relative ease in developing applications and managing them on the centrally controlled systems.
 This made it possible to steadily grow the user base of any online application.
 However, as the use of OSNs grew rapidly due to the introduction of better technology and improved services, two key concerns stood out as regards the C/S model: accumulated costs for centralized operations, and security \&
 privacy concerns~\cite{GCR13}.

 \subsubsection{Accumulated costs for centralized operations} The concerns here are a projection of the use of centralized systems for the OSNs and manifest in the scalability concerns~\cite{MKI+16} which are discussed.
 \begin{enumerate}[label=\itshape\alph*\upshape)]
 \item
  \textit{Large number of highly connected users}: This present difficulties in managing the growing social graph in real time especially due to the tight coupling that exists between different communities, and at the same time, the data generated by the users is vast and over time becomes unmanageable.
 \item
  \textit{Infrastructure issues}: The core infrastructural issues that emanate from scaling of the SN are cost of equipment which must be regularly upgraded or added as the network grows; operational costs such as replacement of failed equipment as well hiring, training and maintaining skilled staff; and energy consumption as there is high energy consumption for powering the servers and by the Heating, Ventilation and Air Conditioning (HVAC) system for cooling the servers and networking equipment.
 \item
  \textit{Internal network traffic}: The large network interconnection presents large internal traffic, usually due to friend recommendations, real-time notifications, personalized marketing, replication, maintenance and indexes synchronization, which presents a scaling bottleneck.
 \item
  \textit{User-generated content management and dissemination}: In most of the SNs, the large portion of interactions is related to content creation and sharing.
  Therefore, handling and disseminating of the user-generated content (UGC) efficiently in a centralized social network creates scalability challenges.
 \item
  \textit{Database scalability}: Centralized SNs require some form of reliable database management system to handle huge amounts of data, ensure rapid deployment of data and UGC as well as maintain heterogeneity of content.
  For this purpose, traditional relational database management systems (RDBMSs) are unsuitable as it is difficult to horizontally partition due to relationships and dependencies among stored data~\cite{AED+11}, because they are designed to guarantee consistency they have limited scalability and availability especially in case of network partition~\cite{LMa10,LFK+14} and cannot provide required latency and scalability for SNs that have clusters replicating over a data centers that are geographically dispersed~\cite{LMa10}.
  This has necessitated the need for alternative data management solutions such as Cassandra~\cite{LMa10} and Haystack~\cite{BKL+10} by Facebook, Bigtable~\cite{CDG+08} and Megastore~\cite{BBC+11} by Google among others.
  This dependency on database solutions to store the data leads to the next concern.
 \end{enumerate}

 As a consequence of these issues, the operational costs are tremendously increased.
 These costs to the OSN provider have to be reimbursed and this is usually done mainly through selling of the accumulated data about the users to product advertisers and other ``analysts''.
 This leads to the second category of concerns.

 \subsubsection{Security \&
  privacy concerns} These concerns are summed up as privacy threats which are divided into two broad categories~\cite{KIa17,ORP19}:
 \begin{enumerate}[label=\itshape\alph*\upshape)]
 \item
  \textit{User-related threats}: These result from disclosure of private data to other OSN users or even unregistered users and this can be intentional such as by hacking or unintentional due to poor or misconfigured privacy settings.
  It has also been further shown that even though majority of the OSN platforms offer services to restrict access to personal data using predefined privacy settings, in most cases it were better for a user to have fewer of these privacy settings configured as more configurations tend to reveal much more~\cite{BAL13}.
 \item
  \textit{System provider-related threats}: These arise due to the implicit requirement by the system provider for self-disclosure by the users during the registration process under the assumption that the providers can be trusted to handle and protect their private information fairly and accurately.
  In addition to having access to the users' personal information, the system providers also have access to the data that the users upload, their browsing behavior, metadata, as well as logs.
  This data is in almost all cases owned by the providers who may utilize data mining techniques to extract implicit data that may violate a user's privacy.
  Using this data, the providers have introduced a new digital market called personal data markets~\cite{SAB+15} where the personal data may be used directly by the providers or sold to third parties to be used for personalized advertising, increased band awareness, emotional manipulation~\cite{KGH14,BMc18} as well as online activity surveillance~\cite{BEdeK13,Sto16,RGu18}.
  In addition, the system is also vulnerable to other attacks such as Sybil attacks~\cite{AAA+17}, social spamming~\cite{GHH+11} and other general cyberattacks on online services~\cite{NaA15}.
 \end{enumerate}

 These two groups of concerns must therefore be met with workable mitigating solutions.
 We look at possible solutions that have been discussed in literature.

 \subsubsection{Mitigating the concerns} In order to solve all the problems, it is important to note that OSNs are complex systems in which it may not be possible to completely stop all the privacy breaches.
 Possible solutions may include the use of anonymization, encryption, fine-grained privacy settings and matching access controls as well as encouraging of user awareness and change of behavior~\cite{ORP19} to solve the social concern and use of better systems to solve the technical concerns.
 However, an all encompassing solution involves taking a look at the computing platform itself, that is, moving from centralized to decentralized OSN systems.
 The decentralized model will not only solve the social concerns but also the technical concerns.
 Therefore, the existing centralized OSNs can be improved in two ways, that is, by extending the capabilities of the provided services and by decentralizing of the supporting infrastructures~\cite{DBV10}.
 Currently, it seems that the features and services provided by OSNs have been extended quite significantly.
 Therefore, the option of extending the current OSN services by decentralizing the server-side infrastructure merits further exploration.

 \subsubsection{Decentralized online social network (DOSN)} This is an online social network designed to run in a distributed manner with minimal or no central control.
 The distributed nature of the DOSN provides three main benefits for the user in comparison to centralized OSNs:
 \begin{enumerate}[label=\itshape\alph*\upshape)]
 \item
  \textit{The provider's operational costs are ideally reduced to zero, as all resources are provided by the users}.
  Thus, no monetary requirement is given to sell the users data.
  DOSNs can be developed as open source solutions and run through the users, thus eliminating the need for a provider at all.
 \item
  \textit{A better and user-oriented user privacy control can be applied}.
  No one needs to be trusted, as mathematically trusted access right management mechanisms can be applied and the openly available source code can be verified to correctly implement security mechanisms.
 \item
  \textit{Innovative development~\cite{BuD09} is encouraged as resources, in terms of communication and storage options, are widely available}.
  Thus depending on the contribution of the users, several gigabytes of storage space can be offered allowing to solve typical use cases such as file synchronization, workspace sharing and messaging large files.
 \end{enumerate}

 Two main classes of DOSNs are identifiable, web-based and P2P-based DOSNs~\cite{PBS10}.
 \textit{Web-based DOSNs} rely heavily on a distributed web server infrastructure and thus are only accessible to computer scientists and experts who can configure and set up such machines and in addition, there is the need for reliable web space which if unavailable would mean user profiles would be inaccessible.
 A \textit{P2P-based DOSN} is a major step in distributed computing where the participants (peers) simply install a program and cooperate with each other to realize a desired service.
 The P2P DOSN can be used by novice users and thus provides a broader acceptance.
 However, building a reliable, secure and appealing P2P OSN is only achievable if some challenges in providing essential services to OSN are addressed~\cite{DBV10}.

 \paragraph*{Summary}
 Many users of OSNs have been attracted to them because of the perceived benefits.
 However, of late, concerns have been raised, not only by the users (social concerns), but also by the providers themselves (technical concerns).
 To this end, one proposed solution to address the concerns raised is decentralization of the systems which can be realised by decentralization of the infrastructure.
 The available options include web-based and P2P-based DOSNs, with P2P-based solutions being a more viable choice.
 In the next section, we discuss how the essential services needed for a functional OSN are achieved by defining technical requirements for a P2P framework for an OSN. 

%

\section{Technical Requirements for a P2P Framework for Social Networks}
\label{sec:InfraSupport}

\begin{figure}
 [htbp!] \centering \includegraphics[scale=0.35]{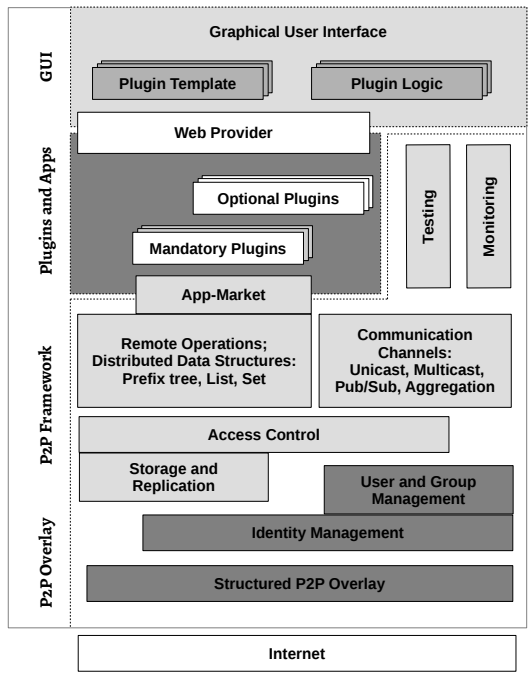}
 \caption{A P2P architectural model for OSNs}
 \label{fig:P2PFWmodel}
\end{figure}

To ensure that the user experience for a P2P-based online social network is matching the expectations, a set of P2P-functions is required which provide a reliable basis for the social networking operations.
Considering usability, the functionality of typical online social networks must be provided or even surpassed in terms of application richness and response time.
In order to achieve this goal, essential infrastructural elements need to be conceptualized and realized.
According to Graffi et al.
\cite{GPM+08}, who provided LifeSocial.KOM (now LibreSocial) the first P2P-based OSN, a four-fold architecture is to be applied: overlay, storage and communication, social networking elements and graphical user interface.
These are shown in Fig.
\ref{fig:P2PFWmodel} as the P2P architectural model for OSNs.
These functionality blocks are responsible for interconnecting the nodes reliably, providing rich P2P-based interaction functionality and building high quality social networking functions with an appealing look and feel.
In the following, we elaborate in detail the technical requirements defined by these functional elements.

\subsection{Overlay Network}
\label{subsec:OverlayNet}
An overlay is a logical network on top of the Internet built by peers that aim to provide for themselves essential networking operations.
This includes the provisioning of a new addressing scheme, joining and leaving protocols, routing and maintenance protocols as well as means to remain connected, operational and optimized even under strong network dynamics.
This section of the architecture focuses on ensuring the following areas are considered.

\subsubsection{Overlay topology} The overlay defines how nodes are to be addressed and connected, to be suited for the OSN. This depends primarily on the purpose of the social network.
An overlay network requires the nodes to have a node identifier, a routing table and a list of open connections to other nodes in the network.
Two main overlay approaches, namely structured and unstructured overlay networks, have emerged over time.
The first associates an ID in the network to a specific responsible node, which is in charge of it and handles messages and actions related to it, while the latter does not require IDs of data items to be associated with peers, thus data or nodes have to be searched throughout the network.
Many of the basic aspects of user management are heavily dependent on the topology selected and as such, a good selection of the topology means better provision of required services.
In Section~\ref{subsec:overlays}, a discussion on the overlays is presented.

\subsubsection{Addressing of Users, Nodes and Data} Typical P2P overlay networks have been designed for file-sharing networks where the topology has no concept of user identifiers as the nodes were expected to share files with any other nodes within the network.
Thus, only the data location is relevant in file sharing network.
OSNs, on the other hand, require reliable addressing of users, data and nodes in the network, for example, in order to send a message to a user, to retrieve a data item by its ID or to send a data item to a specific node based on its ID. Technically, the addresses of users, data and nodes have to be consolidated and a search and/or lookup for these IDs must be supported.
There is the possibility of the peers changing their physical address over several interactive sessions and with this also the location of the data they carry.
Therefore, an appropriate mechanism for addressing the nodes to support searching of friends and data, as well as discovering new friends, is needed.
This ensures that for existing connections, the trust links are maintained.

\subsubsection{Routing} A main functionality offered by the overlay network is the routing of messages to users, nodes and nodes carrying specific data by their ID. This functionality allows building of applications such as chatting apps in which messages are sent to specific user IDs, building of functionality such as replication where data is stored on a specific node determined by its nodes ID, as well as the sending of messages to a node by a given data ID, so that the responsible node in the structured P2P overlay can implement the task described in the message on the corresponding data, identified by its data ID. It has to be considered that the response time for the resolution of a lookup operation is crucial as users do not desire to wait to for long periods to get their results.
This is in contrast to file-sharing application, where a slightly longer response time was considered less relevant in perspective of a long download time for any file.

\subsubsection{Security} Authentication of the contact partners, as well as confidentiality and integrity in the communications, are essential in social networks.
Ideally, these security goals are integrated deep in the P2P-based OSN, thus in the overlay.
An overlay for P2P-based OSN should provide a user identification and a security concept to provide these aims.
Ideally nodes should be able to engage in a secure communication directly with their first message to each other without exchanging further security related messages upfront and without using services outside of the P2P-network.

Having peers, data and users in the P2P system with individual identifiers raises the question on how to map the users to peers and how to allow that users change the device they are using to join the P2P network.
In specific, it should support the use of multiple devices by users (although not simultaneously), and ideally also should not require any physical property (such as private key) to authenticate the users.
The authentication of the user should be purely based on his/her knowledge.

\subsubsection{Non-functional Requirements} The non-functional requirements at the overlay focus mainly on the aspect of system robustness.
Two aspects are therefore worth considering.

\begin{enumerate}[label=\itshape\alph*\upshape)]
\item
 \textit{Robustness to Churn}: Churn refers to the rate at which peers join and leave the network.
 This network dynamics leads to broken links and invalid entries in the routing tables.
 Thus, when a connection is needed, nodes might discover that their contacts are missing and need to be replaced, thus a delay is induced.
 Sometimes the delay and the impact on the routing table is so high, that a correct state cannot be reached anymore and the network gets partitioned.
 Thus, maintenance protocols need to be in place to counteract these dynamics and to keep the routing table consistent and operational.
\item
 \textit{Robustness to Attacks}: While the security measures guarantee that the communication is authenticated and secure, there might be nodes with adapted protocols that aim at compromising the network and causing harm.
 The attacks might address the connectivity of the nodes, aiming at partitioning the network, at hindering the correct routing and leading to failed lookups, or to attract routing and maintenance traffic to spy on the behavior of the nodes.
 The overlay must be able to withstand the presence of selfish or malicious nodes that willingly or unwillingly aim at sabotaging the functions of the network.
\end{enumerate}

\subsection{P2P Framework}
\label{subsec:P2PFramework}
Having an overlay that allows to communicate with specific users or nodes and to address data by its ID, there is a need for advanced functions to support the building of a P2P-based OSN. The aim of these functions, which make up the P2P framework, is to provide a toolbox for user and group management, data storage and replication, single and group communication, monitoring and quality control, as well as testing.
Having a set of mature and reliable mechanisms, independent of the later use cases, provides a basis to build rich applications on top that do not have to be aware of the underlying P2P functions.
This then will boost the usage of the P2P-based OSN. Following functions are considered relevant for the P2P framework as basic building blocks for a P2P-based OSN. 

\subsubsection{Single Data Storage}
In unstructured P2P overlays no specific place is defined for data to be stored, any node might be active to do so.
Thus, in order to retrieve data, the network typically is flooded in a more or less intelligent fashion.
This causes often high traffic lode and the risk of missing data although it is in the network.
Having a structured P2P overlay allows to use the routing functionality, to send a message to the responsible node for a given (data) ID. This node can be asked to store a data item or to send the stored data item to the requesting node.
Thus, data is always found and retrieved if it is in the network.
It has to be considered that the users might modify their data items, thus rendering previous copies outdated.
The consistency of the data in the network is essential, that is, that all nodes interested in a data item receive the most recently updated one.
The single data storage and retrieval functionality is one of the basic building blocks for P2P-based OSNs.
Storage aspects are highlighted in Section~\ref{subsec:Store}.

\subsubsection{Reliable Data Redundancy} The overlay provides the functionality to store and retrieve single data elements from the network.
This data must be replicated in a reliable and consistent manner, so that the failing of nodes does not lead to the loss of data.
Copies of each data item must be available anytime, thus mechanisms need to be in place that monitor the number of copies of a specific data item in the network and that initiate the creation of new replicas if the number of copies is too low.
The replication mechanisms have to identify a trade-off in the desired number of replicas, which on the one hand should be high enough so as to maintain the availability of the data, and on the other hand should be small enough to ensure quicker replica updates as well as lower load on the nodes.
Furthermore, the data replication needs to take into account the security requirements hence ensuring that the confidentiality, integrity and authenticity of the data and its replicas are maintained.
Optional requirements for the locality of the data as well as the consideration of trust to the replication nodes might be considered.
However, locality and trust-based approaches do not add further functional benefits and are thus considered optional.
Data redundancy mechanism such as replication and use of erasure codes are therefore necessary to ensure that data is available even in the face of churn.
Section~\ref{subsec:Store} gives further details on these mechanisms.

\subsubsection{Data Access Control for Users and Groups} While the overlay provides simple data storage and retrieval and the data replication module enforces data availability requirement in the network, this module provides access control to the data stored in the network.
When data is created, the author of the data stores it in the network.
Further modifications of the data, which might be a profile or photo album, should be allowed only for the initial author of the data.
A list of privileged users might be selected who should have read access only to this data while all other users, including the replica nodes, should not be able to read the data.
The data access rights must be valid both for the initial data as well as the replicas, and of course must be provided in any case.
This means that even if the user or his/her friends go offline, the data should remain available in the network and the access control should apply.
Please note, that the set of nodes with read access rights is not explicitly required to be the list of ``friends'', which is commonly available in an OSN. The reason is that the list of ``friends'' is typically not identical with the list of trusted nodes and it may also not be the intent of the users to provide access right for each data item to only this set of users.
Typically, for various data items, such as party pictures or formal business pictures, the set of users with read permission will be different.
Thus, the P2P framework must support definitions for each data item an individual set of read enabled users.

As the access control and authentication of users is up to now defined only for single users, in order to scale the security and access control protocols, it is reasonable to introduce groups or roles.
Single users can be added to these groups, while the members of a group share all the same access rights.
Thereafter, access rights for individual data items can be given to users or groups.
This makes it is easy to assign the read or write access to a specific group, such as thousands of users at once, instead of assigning them the access rights individually.

\subsubsection{Search for Data and Users} Structured P2P overlays support the lookup of data and users.
This function assumes that the ID of the user and the data is known, so that users and data whose ID is not know cannot be found.
The OSN, on the other side, interconnects users and allows them to create and access data items.
Users and data items, with corresponding access permissions, should be searchable based on keywords.
Thus, users could be searched based on their location and interests, and data could be searched based on its content and tags.
In order to support this main functionality and to allow users to also find other users and data they are not aware of, search mechanism are required.
A possible solution is to use indexing mechanisms relying on keys for structured overlays.
Another option is the usage of flooding techniques for keywords within the overlay.
A further searching method depends on decentralized tagging~\cite{GSS08}.
Generally, the overlay topology provides for either an indexing support or for better flooding approaches, but not both simultaneously.
Further there are special cases where additional indexing structures may need to be considered as the topology may not support advanced searching methods, such as multidimensional queries.
An overview on resource lookup and search approaches is discussed extensively in Section~\ref{subsec:ResLook}.

\subsubsection{Direct Communication, Multicast and Publish / Subscribe} One core function of an OSN is the 1-to-1 communication between nodes.
This can be done either asynchronously, when typically the storage is used, as well as synchronously, when messages are sent directly from the peer of one user to the peer of the other user.
Through appropriate addressing schemes, messages can be delivered based on the user's ID. This allows implementing application such as direct chatting or the notification of events, such as friend request or online status.

The next evolution step after 1-to-1 communication is 1-to-N communication, namely multicast.
For multicast communication the user can define a set of recipients to which a single message is sent.
This function is typically used to inform a set of users, such as the user's friends, about a new online status, or to send data update notifications to all replica nodes of a single data item.

The final evaluation in the communication options to be provided by the P2P framework is N-to-M communication.
In this, a set of communicating users want to send messages to a set of recipients.
A use cases is for example the group chat in which all members of the chat can send messages which should be received by all other members in the chat.
Another example is the implementation of a twitter-like function, in which users can subscribe for specific keywords as well as publish their own information relevant for another set of keywords.
The possible mechanisms to implement these functions are discussed in Section~\ref{subsec:PubSub}.

Of course, in all these communication mechanisms, authentication, confidentiality and integrity must be maintained to ensure that users are sure of whom they are communicating with and that the data sent is not modified or read by other parties.

\subsubsection{Optional: Secure Distributed Data Structures} Up to now we have discussed the storage, replication and access control for single data items that are stored in the DHT provided by the overlay.
A single data item, however, is very limited in its usage.
While it is reasonable to store small data items from the social network, such as a profile, as a single data item in the overlay, this finds its limitations if the size of the data structure, such as a list of albums each containing hundreds of photos, which would overload a single peer.
In such cases, where the data keeps increasing, it is advisable to distribute the data, store the various connected data items, for example, the various images and the various albums, on different nodes and thus distributed the load and allow for parallelized data retrieval.
Having distributed data structures (DDSs), such as proposed in \cite{JMG14,AGr16,AYA+16}, at hand provides means of creating and accessing the data items tailored for social networks.
Typical use cases, besides single data items, are unordered sets (such as for photos in an album), ordered lists (such as for list of comments on the user's wall) as well as tree structures (such as for folder structures for a collaboration space).
Distributed data structures should provide smaller grained access options such as adding entries to a set or list or rearranging the tree structure without replacing the whole data structure.
While OSNs could be implemented also using single data items, it is much more convenient to use data structures in the P2P network that are tailored to the needs of an OSN. In Section~\ref{subsec:DistDatStr} we discuss the options to implement distributed data structures.

\subsubsection{Optional: Distributed Quality Monitoring and Control Loop} While having all modules to build a decent, secure and reliable P2P-based OSN, such a network would operate blindly if there is no information on its performance, no option to identify situations of bad performance or options to control the performance.
Such a network, would be set up once, with a selection of configuration parameters, such as the routing table size, the periodicity of maintenance operations or timeouts, and from then on would not be able to be corrected.
In cases where the emerging behavior of the P2P network tends to a low performance side, this information would not be obtained.
A monitoring module integrated into the system is very reasonable so as to obtains performance statistics on the P2P network and OSN in form of aggregated statistics.
The monitoring module would provide information on the average, minimum and maximum retrieval times and give a standard deviation on this.
It would provide information on the average, maximum and expected traffic and storage load on the peers.
The sum of stored data items and sent messages can be counted and regularly provided to interested parties.
Through the availability of this complete, aggregated and timely information, the users, potential operators and the system itself can observe the performance of the P2P networks and OSN.

Having a monitoring information on the P2P network from the monitoring module allows the system, specifically, the implemented protocols and mechanisms, to optimize its behavior and thus to directly take influence on the performance of the P2P network.
By observing the average traffic and storage load on each peer, the peers can decide whether they should add more resources for fairness reasons.
By observing the average and maximum retrieval times for data objects, the mechanisms could re-configure their routing tables to store more contact and thus to reduce the hop count and retrieval time.
Thus, by using the monitoring information to implement a distributed control loop in the P2P network the network can be auto-tuned in a manner, that given performance goals are reached and kept.
The discussion on monitoring is done in Section \ref{subsec:Monitoring}


\subsection{Social Networking Apps and Graphical User Interface}
\label{subsec:SocNetApps}
The P2P overlay and P2P framework modules presented in subsections \ref{subsec:OverlayNet} and \ref{subsec:P2PFramework} respectively are quite general and not limited to the usage of P2P-based OSNs.
They do not provide dedicated functions such as profile handling, friend lists or photo albums.
They are general to use modules which can be used to build OSNs but also P2P-based games, chat programs or file sharing networks.
For the implementation of OSN specific functionality, ideally, a modular system is used which utilizes the functions offered by the P2P framework and P2P overlay and allows the programmer to add new social apps or plugins easily.

\subsubsection{Common OSN Applications} Online social networks at a minimum require a personal storage space, such as a profile, and social contacts, such as friends.
In combination with the opportunity for social graph traversal, nodes can present themselves, connect to friends and browse these friends' profiles.
In addition to this, users expect means of communication (messaging, a wall for conversation) as well as a basic set of shared storage space, where they can present photo albums and link each other.
The main interaction will take place in direct messaging communication, through comments and updates on the wall of the users as well as through photo uploads and comments.

\subsubsection{Support for New Applications} Most OSNs support the inclusion of third-party applications, which means that the applications offered can increase and decrease.
Many of these applications usually allow the peers to collaborate in some activity such as playing games.
The downside to these applications is that they risk opening services to untrusted third-parties, which exposes all the application providers to the privacy problem from the single service provider~\cite{BuD09}.
For the P2P environment, the peers choice of enabling a third party application should not affect other peers.
By using the functions of the P2P framework no new app should be able to circumvent the security features of the P2P framework.
All security and performance related information is handled by the P2P framework, while the OSN apps are only users of it and bound to its policies.
A range of possible peer-to-peer network applications are discussed in Section~\ref{subsec:Apps}.

\subsubsection {Graphical User Interface} Most of the current OSNs are operated through browsers and therefore online users are more agreeable to this mode of interaction.
Consequently, the graphical user interface should also be browser-based, supporting the newest features of HTML5.
The user interface should be capable of providing an overview on the OSN functions offered as well as be able to integrate new OSN apps that may be added later on.
The addition of further OSN apps is challenging as the developers should have the freedom to visualize their new apps, but on the other side should be restricted from tampering with the presentation of other apps.


Having elaborated the technical requirements for a P2P framework for social networks, leads us to the analysis of what solutions are already available in P2P literature.
In the next chapter we give an overview on the available P2P-based building blocks and discuss after that the P2P-based online social networks available.

\section{Peer-to-Peer Networks}
\label{sec:P2PNws}
A P2P network (or system) is \textit{a virtual, self-organized network formed over the existing physical network by introduction of specialized protocols that allow heterogeneous nodes to autonomously interact and share resources}.
From this definition, the most common characteristics of P2P networks can be deduced, which are, resource sharing, interconnection between peers, decentralization, self-organization, stability and fault resilience, scalability, anonymity and shared cost of ownership of the system~\cite{DFa17} 

In order to achieve the desired characteristics, the P2P network need to needs some key essential services to be designed and built within a component framework.
With a focus on computing systems, a \textit{framework} is a layered structure that shows which component programs can or should be built and how they would interrelate.
The P2P framework gives support for proper implementation and provides a means to evaluate how the entire system meets the non-functional and functional requirements.
A simple model of the framework is shown in Figure~\ref{fig:P2PFWmodel}.
A key aspect of the P2P framework is that it should operate within the structural confines of the TCP/IP model to enable global participation of peers.
A detailed discussion on the core components of the P2P framework supporting implementation of an OSN as highlighted in Section~\ref{sec:InfraSupport} follows.

\subsection{Overlay Structures}
\label{subsec:overlays}
The overlay network is a collection of logical links that connect nodes to the application layer~\cite{KKH+13}, and is in essence a set of protocols that run by mimicking the physical network's behavior~\cite{BYu10}.
The logical links may involve one or more actual physical links between participating nodes and the construction of the overlay itself is generally not dependent on the underlying layers although the information from these layers (such as network delays and physical proximity) is used in the operation of the P2P network~\cite{KKH+13}.

There are several ways of classifying P2P networks and this diverse classification methods exist because of the level of broadness adopted by an author~\cite{ASp04}, and this broadness is due to the evolutionary process of P2P architecture~\cite{HYA+04}.
The traditional classification considered only the manner in which it handles routing, hence structured and unstructured P2P overlays.
However, this method of classification does not cover all the possibilities in existence today.
The best proposal for classifying the P2P overlays is proposed by~\cite{KKH+13,Mal15}, which divides the into three distinct groups based on the number of overlays, hence \textit{single-overlay}, \textit{multi-overlay} and \textit{bio-inspired} P2P networks.
In addition, we consider other P2P overlays that take into account other factors.


\subsubsection{Single-overlay P2P networks}
\label{subsubsec:SingleOverlay}
This classification encompasses all the traditional or more common forms of P2P networks.
However, the P2P networks here are further divided based on two levels: type of index and type of structure.
Table~\ref{tab:SingleOverlay} gives examples of P2P networks in this group.

\begin{table*}[!htb]
 \begin{minipage}{.5\linewidth}
  \centering \scriptsize
  \caption{Single-layer Overlays}
  \label{tab:SingleOverlay}
  \begin{tabular}{|p{1.4cm}|p{2.7cm}|p{3.5cm}|}
   \hline \rule{0pt}{2ex}\textbf{Indexing Mechanism}&%
   \textbf{Network Structure}&%
   \textbf{Example(s)} \\
   \hline\hline \rule{0pt}{2ex}\textbf{Centralized}&%
   Structured&%
   None \\
   \cline{2-3} \rule{0pt}{2ex}&
   Unstructured&%
   Napster, BitTorrent~\cite{Coh08} \\
   \cline{2-3} \rule{0pt}{2ex}&
   Structured+Unstructured&%
   Trackerless BitTorrent \\
   \hline \rule{0pt}{2ex}\textbf{Distributed}&%
   Structured&%
   DHT-based P2P networks e.g.
   Pastry~\cite{RoD01}, Tapestry~\cite{ZKJ01}, Chord~\cite{SMK+01}, Kademlia~\cite{MMa02} \\
   \cline{2-3} \rule{0pt}{2ex}&
   Unstructured&%
   Gnutella v0.4, Freenet~\cite{Cla99,CSW+01} \\
   \cline{2-3} \rule{0pt}{2ex}&
   Structured+Unstructured&%
   None \\
   \hline \rule{0pt}{2ex}\textbf{Hybrid}&%
   Structured&%
   P2PSIP~\cite{rfc6940} \\
   \cline{2-3} \rule{0pt}{2ex}&
   Unstructured&%
   JXTA~\cite{TAA+03}, Gnutella v0.6, FastTrack/KaZaA~\cite{LRW03,LKR+06} \\
   \cline{2-3} \rule{0pt}{2ex}&
   Structured+Unstructured&%
   Skype\\
   \hline
  \end{tabular}
 \end{minipage}
 \begin{minipage}{.5\linewidth}
  \centering \scriptsize
  \caption{Multi-layer Overlays}
  \label{tab:MultiOverlays}
  \begin{tabular}{|p{2.0cm}|p{3.5cm}|}
   \hline \rule{0pt}{2ex}\textbf{Classification}&%
   \textbf{Example(s)} \\
   \hline\hline \rule{0pt}{2ex}Vertical&%
   NICE\cite{BBK02}, HIERAS\cite{XMH03}, \\
   \hline \rule{0pt}{2ex}Horizontal&%
   Structella\cite{CCR04}, Cyclone\cite{ALA+05} \\
   \hline
  \end{tabular}
  \vspace{5mm}
  \caption{Bio-Inspired Overlays}
  \label{tab:BioOverlays}
  \begin{tabular}{|p{3.0cm}|p{4.5cm}|}
   \hline \rule{0pt}{2ex}\textbf{Inspiration}&%
   \textbf{Example(s)} \\
   \hline\hline
   \rule{0pt}{2ex}Ant Colony Optimization&%
   AntCAN\cite{ABu05}, P2PSI\cite{HHw07}, BlatAnt\cite{BMH10}, Self-Chord\cite{FLM+10}, AntOM\cite{PMH+12}, Self-CAN\cite{GMM12} \\
   \hline \rule{0pt}{2ex}Bee foraging&%
   Antares\cite{FMa09}, P2PBA\cite{DMP+11}, \\
   \hline \rule{0pt}{2ex}Neurons&%
   SCAN\cite{GWS06} \\
   \hline \rule{0pt}{2ex}Fungus&%
   Myconet\cite{SGV09}\\
   \hline
  \end{tabular}
 \end{minipage}
\end{table*}

\paragraph*{\textbf{Type of index}}
The indexing mechanism used for locating of nodes, shared resources, or groups in this case may be \textit{centralized}, \textit{distributed} or \textit{hybrid}.
As a comprehensive discussion on the indexing mechanism is given in Section~\ref{subsec:ResLook}, we will only introduce them here.
A centralized index mechanism stores the indexes in a one or more centralized servers, referred to as trackers.
In distributed indexing, the indexes are distributed among participating peer nodes.
In hybrid indexing mechanisms, the indexing is distributed to a subset of the nodes, usually referred to as supernodes.

\paragraph*{\textbf{Type of structure}}
The P2P system may be unstructured, structured or a combination of unstructured and structured.
This method has been used as the conventional way of classifying P2P systems.
Each type of index mechanisms can further be broken down into these three groups.
\begin{enumerate}[label=\itshape\alph*\upshape)]
\item
 \textit{Unstructured overlays}: the topology constitutes flexible node relationships and lookup operations~\cite{Mal15} and each node relies only on adjacent nodes for delivery of messages to other nodes in the overlay~\cite{BYu10}.In these overlays, searches are conducted using a broadcasting search algorithm, also called flooding search, therefore it supports system churn with a degree of flexibility and node failure does not adversely affect searching~\cite{Kan11}.
 Implementing a Time-To-Live (TTL) value for each query message limits its lifetime in the network and reduces network load~\cite{Nhat09}.
 However, unstructured networks are unsuitable for rare data (exact match) queries, but are quite efficient for replicated data while also supporting keyword searches but at a high cost to bandwidth.
\item
 \textit{Structured overlays}: these have a tightly controlled topology maintained via a network graph, with resources placed in a deterministic fashion using distributed hash tables (DHTs)~\cite{Mal15}, and nodes cooperatively maintain routing information about how to reach all nodes in the overlay~\cite{BYu10}.
 Thus structured overlays support key-based routing protocols, and can only handle exact match queries with high precision but are not designed for keyword searches.
\item
 \textit{Combination of structured and unstructured}: such P2P systems rely on hybrid indexing schemes, hence have supernodes~\cite{KKH+13}.
 These supernodes are connected in a structured network formation and the communication existing between the regular nodes is unstructured.
\end{enumerate}

\subsubsection{Multi-overlay P2P networks}
\label{subsubsec:MultiOverlay}
These P2P networks are made up of several interconnected overlay networks to form a single functional entity~\cite{KKH+13}.
The use of multiple P2P overlays in a single network may provide a means to solve the totality of issues regarding pervasive networks~\cite{Coo06}.
Also most of the single-overlay solutions are domain- or application-specific, while the expectation is that the nodes may be active in more than one domain or activity, thus encouraging multiple overlays~\cite{HCL+10}.
The need for multiple P2P overlays is probably a consequence of perceived benefits realized due to its use in virtualization~\cite{MLI+12}, which enables the utilization of the same physical resources by many different applications~\cite{Mal15}.
Classification of the multi-overlay schemes is determined by considerations towards temporal synergies and dynamicity, as well as communication, state and service interactions~\cite{Mal15}.
Hence, the classifications are vertical and horizontal.
Examples of multi-layer overlays are in Table~\ref{tab:MultiOverlays}.

\begin{enumerate}[label=\itshape\alph*\upshape)]
\item
 \textit{Vertical multi-layer overlay}: in this, the higher-level overlay exploits functionalities of the lower-level overlay~\cite{Mal15}.
 Therefore, several overlays are clustered one on top of the other, with each layer being independent structured P2P overlay network.
 In most cases these layers are usually DHT-based.
 Communication is made possible through the use of \textit{gateway nodes} which are responsible for message routing between two vertically adjacent nodes.
\item
 \textit{Horizontal multi-layer overlays}: This classification considers the parallel operations of overlays~\cite{Mal15}.
 The multiple overlay networks, each referred to as a \textit{leaf}, are joined together to form a single DHT-based P2P network, with possible connections existing between the leaf overlays.
 The overall function of the resultant DHT-based P2P network is to ensure optimized routing and maintain the conceptual hierarchy of the leaf overlays~\cite{ALA+05}.
 Unlike vertical multi-overlay networks, there are no gateway nodes.
 Instead, leaf overlay connect by carefully selected links so as to ascertain a small number of total links per node.
\end{enumerate}

\subsubsection{Bio-inspired P2P Networks}
\label{subsubsec:BioOverlay}
These overlays are a result constructing P2P overlays networks using algorithms and techniques that are inspired by naturally occurring biological phenomena.
These bio-inspired solutions are characterized as being highly adaptive and reactive, having support of heterogeneity, distributed operations, resilience to component failure and can self-organize~\cite{BCD+06}.
Therefore, bio-inspired approaches have been taunted as a possible alternative for managing P2P overlay networks having been proven as an effective solution in the computer network domain~\cite{BLL+11}.
Existing solutions are mostly based on the collective behavior of ant colonies or bees called swarm intelligence, but other approaches have also been studied such as biological neurons and fungi-growth.
Examples of proposed overlays are listed in Table~\ref{tab:BioOverlays}.

\subsubsection{Other overlay considerations}
\label{subsubsec:OtherOverlays}
Several theories have been taken into consideration for the systematic development of overlays with desired properties such as locality awareness, anonymity, mobility and other features~\cite{Amf17}.
Examples include Geodemlia~\cite{GSR+12} and LobSter~\cite{AGr17} which are location-aware overlays, FRoDO~\cite{AGG+15} which supports anonymous communications.

\subsubsection{Security Discussion: Topology}
\label{subsubsec:Overlay_Security}
P2P networks are essentially embedded into the TCP/IP protocol suite and therefore some of the problems that networks face can also be found in P2P networks.
The security of the topology, specifically, the overlay structure is fundamental to the whole P2P network.
This is due to the fact that the overlay provides all the essential services that are utilized by the other components of the P2P network.
It is indeed the first point of entry into the network.
Threats that affect the topology have mainly to do with denial of essential services.
These include denial-of-service (DoS) and distributed denial-of-service (DDoS), man-in-the-middle attacks and routing attacks such as eclipse attack~\cite{SND06}, wrong routing forwards (attrition attacks~\cite{GMB+05}), identity theft~\cite{PZZ09} and churn attacks~\cite{SRe06}.

In order to mitigate these types of attacks, the solutions must therefore consider securing the communication.
These solutions are discussed in depth in Section~\ref{subsec:PubSub} where provision of communication channels is considered in depth.

\paragraph*{Overlays for P2P social networks}
With regard to SNs, in which each an individual data item is relevant and should be essentially retrievable, only single-layer overlay, structured networks seem to be suitable.
In this regard, we would say that it is more important to be able to retrieve rare data items in at most
$O(logN)$, while tolerating an expensive keyword-based search which may require development of other mechanisms to support it, than to have a cost-efficient searching, as availed by unstructured networks, but high costs in locating and retrieving the profile data of connected friends.
Another essential element of SNs is the ability to change data, such as profiles.
Thus, the network must support retrieving of all copies of this data for any changes to be effected fully.
In structured overlays only the one single responsible node for the data's identifier needs to be contacted, which is feasible, in contrast to having to search through the whole network for copies in an unstructured overlay network.
Although multi-layer overlay and bio-inspired overlay P2P networks may seem promising as solutions for social networks, it may require much more effort to implement needed mechanisms for social network services to match the centralized SNs.

\subsection{Overlay Function: Search and Lookup Mechanisms}
\label{subsec:ResLook}
A resource in the case of a network is either a node or data and the problem of resource discovery is synonymous to the search and lookup problem.
At the core of solving the P2P search/lookup challenge is the development of appropriate and agile indexing mechanisms and querying mechanisms for efficient information retrieval, which then make the search techniques dependable and adaptable to the changing network.
P2P indexes can be classified as follows:
\begin{enumerate}[label=\itshape\alph*\upshape)]
\item
 \textit{Local indexes}: the index is stored by a peer for its own data or objects only, such as in the first design of Gnutella.
 They support rich queries along with simple key lookups.
 Global data search is undertaken by query flooding.
 However, the use of local indexes in a large and growing network becomes inefficient.
\item
 \textit{Centralized indexes}: these depends on a single server to keep the references to the data on many peers such as Napster\footnote{Originally available on http://www.napster.com.}.
 However, a centralized index for a P2P network reintroduces the problems of centralized systems, and hence is discouraged for a fully decentralized application system.
\item
 \textit{Distributed indexes}: these maintain the information for a part of the identifier space as well as a systematic routing table to reach nodes responsible for the other parts of the identifier space.
 The distributed indexes can be semantic or semantic-free indexes.
 Most unstructured P2P networks utilize \textit{semantic} indexes, that is, they are human readable (Section~\ref{subsubsec:SemInd}).
 The problem introduced by semantic indexes is that they do not support persistent object references and prevent contention free references~\cite{WBS04} .
 The need for \textit{semantic-free} content-based referencing for P2P systems necessitated the development of a better way to handle discovery, hence the proposal of using DHTs which then enable object location using persistent keys in a high churn network~\cite{Hel03} (Section~\ref{subsubsec:SemfreeInd}).
\item
 \textit{Hybrid indexes}: they utilize the best of both worlds, that is, they combine two or more types of single-layer overlays to achieve effective indexing.
 In most cases, the structured network consists of nodes that perform the indexing called the supernodes, while other nodes are maintained in an unstructured format.
 A result of these hybrid indexes, is the development of \textit{multidimensional indexing mechanisms} to tackle the problem of rich text and multidimensional data searching.
\end{enumerate}

In the following we describe the semantic-free, semantic and multidimensional indexing mechanisms.

\subsubsection{Semantic-free mechanism for lookup}
\label{subsubsec:SemfreeInd}
Mechanisms that support some form of key-based routing (KBR) will generally offer semantic-free indexing.
These methods cover the use of DHTs, tree-based mechanism and skip lists/skip graphs.
The use of DHTs, that is,
the indexing of data based on a identifier derived from the hash of the data, provides semantic-free indexing.
This is encouraged by the fact that DHTs are designed for: greedy, reliable routing meaning nodes dynamically determine the shortest path to uniquely identified nodes or data; low node degree that reduces effects of high churn rates, low network diameter hence reduced hop count; and robustness meaning that a path can be found to a target even in cases of node failure~\cite{Hel03}.
Traditional DHTs such as Chord~\cite{SMK+01}, Pastry~\cite{RoD01} or Tapestry~\cite{ZHS+04} use a routing table of the size of
$\varTheta
(logN)$ neighbors to crete a topology that allows that the messages are routed to the responsible peers within
$\varTheta
(logN)$ hops.
However, hashing as used in the DHT based indexes, destroys the locality of data, particularly data item's closeness users (content locality) and prevents global routing if querying and answering nodes are within the same locality (path locality)~\cite{HJS+03}.


The Skip List~\cite{Pug89} based overlays, Skip graphs~\cite{ASh07}, SkipNet~\cite{HJS+03} and HSkip+~\cite{FSG14} can be used instead of DHT-based overlays to alleviate the problem of locality.
Skip Lists ensure balance through probabilistic balancing during insert and delete operations with every node having averagely
$\varTheta
(logN)$ neighbors similar to DHT based approaches.
Unlike the other DHTs, Skip graphs support prefix searches, proximity searches~\cite{GMS08} as well as location-sensitive name searches~\cite{RiM06}.

Tree-based overlays such as BATON~\cite{JOV05} also offer key-dependent searching that is also logarithmic.
In BATON, each of the peers in the network maintains a node of the tree.
Node links to other nodes may be parent links, children links, adjacent links or neighbor links.
Each node (leaf or internal) is assigned and manages a range of values that should be greater than those of the left adjacent node but smaller than the right adjacent node.
Searching is then performed based on the range in which the value falls either towards the right of the tree if value is greater or towards the left if the value is smaller.
Essentially, even searching in BATON is key dependent.

\subsubsection{Semantic mechanisms for searching}
\label{subsubsec:SemInd}
Because semantic-free mechanisms rely on the structure of the overlay network, they guarantee that a key can be found if it exists in the network.
However, they do not show the relationships existing among objects.
Semantic P2P indexes, on the other hand, highlight these relationships but do not guarantee finding scarce items in the network as they rely on heuristics.
Semantic indexing is mainly the domain of unstructured P2P networks.
For the unstructured networks, performing exact searches can only be realized through an exhaustive all node contacting flood, as a global index cannot be constructed~\cite{HaY10}.
However, keyword searches can be undertaken.
Keyword searches entail the issuing of a query for a single keyword or several keywords by a query issuer.
The keyword search mechanisms can be broken down into single- and multi-keyword searches.
\begin
 {enumerate}[label=\itshape\alph*\upshape)]
\item
 \textit{Single-keyword searches}: There are basically two ways that this can be realized, blind routing and routing indices~\cite{HaY10}.
 \begin{itemize}
 \item
  \textit{Blind routing} techniques do not take into account resource distribution thus are likely to get wider coverage.
  However, they generate high network traffic loads.
  These searches are achieved using flooding techniques.
  The flooding techniques can be classified as pure flooding, flooding across hops, TTL limit-based flooding such as expanding ring search, and probabilistic limit-based flooding such as random walks~\cite{BOI+12}.
 \item
  \textit{Routing indices approaches} build indices to help guide forwarding queries.
  Two ways of building indices are: peer-content based such as naive routing~\cite{CGM02}, or peer-queries based.
 \end{itemize}
\item
 \textit{Multi-keyword search}: This is a great motivation for making searching in the systems much easier and faster.
 However~\cite{HaY10} suggests three ways that can be looked into to solve the problem of multi-keyword searches by introducing modifications to the single-keyword search methods.
 First, perform single-keyword search and then merge the results.
 Secondly, assume that the multiple keywords are a single query.
 The last solution proposed is to view the problem from a query routing process perspective.
\end
{enumerate}

\subsubsection{Multidimensional Indexing Mechanism}
\label{subsubsec:MultiDim}
In Section~\ref{subsubsec:SemfreeInd} we reviewed the DHT mechanisms that rely on the use of keys and due to this they are primarily key-based search techniques that support exact-match queries only.
In Section~\ref{subsubsec:SemInd} we looked at non DHT mechanisms that rely on simple keyword lookup techniques and therefore they also have the disadvantage that they do not support complex queries.
So as to deal with complex queries, novel solutions have been proposed for P2P systems.
A highly desirable quality in P2P systems is the ability to support not only simple searching or key-based lookup queries but also complex rich text queries~\cite{HHH+02} and multidimensional data~\cite{ZXT+11,BPo15}.

\textit{Multidimensional indexing} (MI) allows users to perform querying efficiently in cases of multi-dimensional data such as Geo-spatial data.
This can be achieved using multidimensional indexing structures such as Skip Lists and Skip Graphs~\cite{BPo15}.
In~\cite{ZXT+11}, three classes of MI are discussed:
\begin{enumerate*}[label=\itshape\alph*\upshape)]
\item
 P2P-based MI methods that are extensions of centralized MIs that have been decentralized,
\item
 P2P-based systems that have been augmented to achieve MI, and 
\item
 combining of centralized MI and P2P-based systems.
\end{enumerate*}
This way, it is possible to run more advanced query types such as aggregation queries, multi-attribute queries, join queries, \textit{k}~Nearest Neighbor Query and Range Query.

\subsubsection{Security Discussion: Search and Lookup Corruption}
\label{subsubsec:Search_Security}
Consistent and reliable data availability is highly dependent on the data redundancy techniques, replication and/or erasure codes, that are implemented within the P2P network (Section~\ref{subsec:Store}).
However, even after the implementation of such schemes, the data availability may be highly affected due holes arising from routing-, storage- or resource lookup-based inconsistencies.
\textit{Content availability depletion}~\cite{FCM07} may arise due to attacks that target the content availability, making it hard for the legitimate users to find a needed resource and are usually accomplished by poisoning or pollution attacks.
Poisoning and pollution of the replicated resources lowers the relative availability of usable content in the P2P network.
In unstructured networks such as FastTrack and Gnutella, which use searching techniques, this is achieved by random decoy injection, replicated decoy injection or replicated transient decoy injection~\cite{CWC05}.
\textit{Random decoy injection} is the process of poisoning by the insertion/injection of large quantities of decoys in the network resulting in poor ranking of usable files from the search results.
\textit{Replicated decoy injection} occur when numerous replicas of the same decoy are injected into the network, resulting in higher rankings for the injected decoy in the search results.
This attack however can be easily detected.
An alternative to overcome detection, for example by a reputation system, is to frequently replace the replicated decoys injected in the network, a technique referred to as \textit{replication transient decoy injection}.
Structured networks on the other hand, because of the reliance on the routing tables to perform lookups, are affected more by routing table poisoning attacks.

\paragraph*{Resource discovery in P2P-based OSN}
For the purpose of designing an efficient application that supports finding other users quickly and efficiently, and in this case an OSN, the preferred option is to incorporate lookup (semantic-free indexing) mechanisms as they offer efficient search and retrieval.
However, in cases where the P2P network is to be designed for file sharing applications where there is need to find the closest neighbor that is available to share a file, then searching (semantic indexing) mechanisms would suffice.

\subsection{Storage Techniques and Redundancy}
\label{subsec:Store}
Designing a reliable storage mechanism is aimed at ensuring data availability and in P2P systems.
There are several proposal of data storage mechanisms that are in existence.
DHT based mechanisms include PAST~\cite{DRo01} on Pastry, Cooperative File System (CFS)~\cite{Dab01} on Chord and OceanStore~\cite{KBC+00} on Tapestry~\cite{ZKJ01}.
Most unstructured P2P networks are essentially storage networks as they we designed for file sharing purposes such as Freenet, FastTrack and BitTorrent.
Irrespective of the overlay upon which the storage technique has been designed on, the most important consideration is guarantee on data availability.
In most distributed systems, this is achieved through the inclusion of data redundancy mechanisms which utilize replication and/or erasure codes~\cite{CGD+14}.
We discuss how this is achieved.

\subsubsection{Data Availability through Replication}
\label{subsubsec:Rep}
Replication is the process whereby copies (full or partial) are sent to different peers in the system in order to ensure fault tolerance within a distributed system.
The term \textit{replica} is used in reference to copies of the replicated objects.
The advantages realized by having replication in the P2P system include high availability, reliability and fault tolerance, scalability, increased performance and presence of ``fail-safe'' infrastructures~\cite{SBX+15}.
\textit{Caching} is very similar to replication and it is aimed at releasing loads experienced at particular hot spots and decreasing file query and retrieval latency.
Caching is usually performed near the file owners or the file requestors or along a query path from a requestor to an owner~\cite{SSu11}.
However, caching is done opportunistically and is uncoordinated, leaving no information about where caches exist in case there arises a need to update the cached data items.

In replication the storage points of the data copies are and remain well known so that \emph{all} copies of the data item can be found and addressed, e.g.
in order to update them.
Desirable during replication is the file consistency being maintained, hence update management is essential and this is achieved through the use of suitable replica control mechanisms.
Three criteria are given for classifying replica control mechanisms: \textit{replication point}, \textit{update propagation method} or \textit{replica distribution}~\cite{GHO+96,MPV06,SBX+15}.
Based on the replication point, protocols are either \textit{single-master} or \textit{multi-master} with the option to perform push-based or pull-based updates to the slaves.
Based on the update propagation, the mechanisms are can be either \textit{synchronous} or \textit{asynchronous}.
Asynchronous replica control mechanisms further consider either \textit{pessimistic} or \textit{optimistic} updates.
\textit{``Pessimistic'' techniques} ensure single-copy consistency, that is, prevent access to a replica unless it is up to date.
This works well in a small network but fails in a globally distributed networks, such as the Internet, because the Internet remains slow and unreliable, pessimistic algorithms scale poorly in the wide area and some human activities require asynchronous data sharing \cite{SaS05}.
\textit{``Optimistic replication''}, on the other hand, allows for sharing of data efficiently in wide area as well as mobile environments and is therefore preferred for globally distributed networks.
However, since optimistic replication faces challenges due to divergent replicas and concurrent update conflicts, it is not suitable in systems that may rarely experience conflicts and have a high tolerance to data inconsistencies~\cite{SBX+15}, which is the case for most P2P applications.
Finally, replication distribution method can be performed as \textit{full replication}, where each site stores a copy of the shared objects or \textit{partial replication}, in which case sites only store a subset of the shared objects, thus sites store different replica objects which saves space overally.

The data replication techniques suitable for P2P systems can be classified into three groups: \textit{site selection techniques}, \textit{file granularity techniques} and \textit{replica distribution techniques}.
For site selection techniques we consider the structured and the unstructured P2P networks.
For unstructured networks, the proposed solutions include owner replication, path replication and random replication which are discussed in~\cite{LCC+02} as well as HighlyUpFirst replication and HighlyAvailableFirst replication which are discussed in~\cite{OSS03}.
Structured networks site selection techniques include successor replication, multiple hash functions, correlated hashing and symmetric replication as discussed in~\cite{KZH+07}.
File granularity techniques include full file replication, block level replication and erasure code replication~\cite{BMS+03,GBu07}.
Finally, replica distribution techniques that have been examined include uniform replication, proportional replication and square-root replication discussed in~\cite{LCC+02}, Pull-then-Push replication~\cite{LDP06} and optimal content replication~\cite{KRA+02}.
Most P2P systems utilize one or more of these replication strategies in combination so as to achieve some form of reliable replication.
The second aspect of achieving data availability is based on the use of erasure codes which we discuss next.

\subsubsection{Data Availability through Erasure Coding}
\label{subsubsec:EraCode}
Systems that rely solely on replication generally achieve high availability only with high space overhead~\cite{Mon10}.
Error Correcting Codes (ECCs)~\cite{Ham50} have been used to prevent information loss experienced during transmission of a data stream.
Erasure codes~\cite{Rab89} are a special class of ECCs which are used if a system can differentiate in advance the missing or corrupted encoded data segments.
Generally, a data block
$b$ of size
$S_b$ is first broken into
$m$ equal sized fragments of size
$S_{f}=S_{b}/m$ which are then coded into
$n$ blocks by adding
$r$ redundancy blocks in a way that it is possible to reconstruct
$b$ from any subset of
$m$ blocks among the
$m+r$ ($=n$) fragments.
The original blocks are referred to as \textit{data blocks} and the coded blocks as \textit{check blocks}.
The ratio
$n/m$ is called the \textit{stretch factor} and
$m/n$ is the \textit{useful space}.
The main idea behind the erasure codes is that given any choice for
$m$ blocks, it is possible to reconstruct the original data.
Replication can be seen as a special case of erasure codes where
$m=1$~\cite{Mon10}.
Common of erasure coding techniques used include \textit{Reed-Solomon codes}~\cite{ReS60}, \textit{Regenerating codes}~\cite{DGW+10}, and \textit{Hierarchical codes}~\cite{DBi08}.

In addition to selecting an appropriate coding technique, another challenge is maintaining a minimum number of data fragments in the network for durable long-term storage inspite of failures by ensuring proper fragment placement.
It has been shown that the choice of fragment placement has an impact on system performance~\cite{DWa01,LCZ05,CGD+14}.
Therefore, not only is the coding technique important, but also the replica placement policy.
Examples of placement policies include \textit{global \&
 random policy}, \textit{chain policy} and \textit{Buddy} (or \textit{RAID}) \textit{policy}~\cite{CGD+14}.


\subsubsection{Security Discussion: Storage and Resource Lookup}
\label{subsubsec:StoreLookup}
Security concerns that affect storage also affect the distributed data structures, as they are used to store objects, and the resource lookup mechanisms, which are used to locate the stored objects.
Information integrity in the P2P network may be compromised through the introduction of low quality (degraded) or by otherwise misrepresenting the content identity (false labeling)~\cite{PEJ+06}.
The main security threats that target the content stored therefore focus at corruption or erasure of stored data in the system.
Some of the threats include worm propagation, the rational attacks, storage and retrieval attacks, index poisoning attacks, pollution attacks and query flooding attacks~\cite{TrK12}.

Most of these challenges can be solved easily by incorporating a trust model within the system such as a reputation system.
It has also been shown that a trust model can mitigate worm propagation~\cite{RMo18}.
Reputation systems are useful in the detection of selfish peers, thus are good for mitigating against free-riders, but they fail in the detection of Byzantine peers and malicious peers.
Byzantine peers are peers who behave randomly, that is, they misbehave, but not necessarily following a pattern to maximize their benefit or to disrupt the system while malicious peers perform actions based on a target that is either detrimental or beneficial to the system (or both)~\cite{CiR11}.
To handle this, \textit{micropayment systems} (MPS) can be used.
MPSs are indirect incentive systems in which virtual or real currency, such as Bitcoin~\cite{Nak08}, is used to create a form of indirection between the contribution of a service and the request of similar contribution from another peer~\cite{ReC09,CiR11}.
The MPS architecture includes a Broker that issues the currency and certifies its value.
Additionally, the MPSs usually incorporate a significant amount of cryptographic verification and hence are mostly used in static content distribution systems.
However, solutions that require trusted third parties are to be avoided.

\paragraph*{Replication vs.
 erasure coding}
When retrieving a replicated data item, it is sufficient to contact one single peer which has the replicated data item.
This can be done through one lookup.
On the other hand, with erasure codes, at least
$m$ nodes are required so as to retrieve the complete data item which takes more time and generated more network traffic.
Much more severe is the case when data is to be updated.
With replicating, there is typically only a handful of nodes to be contacted, that is, those which are the replica holders, to update the data item.
In the case of erasure codes, all
$m$ nodes have to be contacted, which is typically a much higher number.
Thus, replication is faster in terms of the replication, update and retrieval process, but requires drastically more space.
Hence, as noted in~\cite{ShD12}, the use of replication is preferred for OSNs as the data is frequently updated, while the use of erasure codes is a better choice in systems handling large static data for archival/backup purposes despite the fact that they are very space efficient.

\subsection{Advanced Storage: Distributed Data Structures}
\label{subsec:DistDatStr}

A data structure, as used in computing, and specifically in system design is a collection of variables with relevant data types that are connected in various ways.
A distributed data structure (DDS) is a data structure that has been designed to work in a distributed environment as a self managing storage layer.
The DDS consists of
\begin
 {enumerate*} [label=\itshape\alph*\upshape)]
\item
 a \textit{data organization scheme} that specifies a collection of local data structures that act as stores of data items by copying them to various sites in the network, and \item a \textit{set of distributed access protocols} which that support the processors in issuing of modification and query instructions to the network and getting appropriate responses~\cite{DiN00}.
\end
{enumerate*} DDSs as a rule have strictly defined consistency models (operations on its elements are atomic), support a single, logical data item view for clients despite replication (one-copy equivalence), and they use two-phase commits for replication coherence~\cite{GBH+00} .

Generally, DDSs can be grouped into two classes, \textbf{\textit{hash-based}} and \textbf{\textit{order-preserving}}.
This classification relies on the fact that insertions and retrievals in the DDS is either based on hashing or keys.

\subsubsection{Hash-based Schemes} The basis of the DDSs in this class is the \textit{hash table}, a data structure used to map keys to values, store these
$\langle
key,value\rangle$ pairs, and retrieve the values using the provided keys.
A hash table consists of two parts, an array and a mapping function known as the hash function.
The array is actually a table in which data to be searched is stored.
The hash function maps the data item keys onto the integer space that defines the indices of the array, that is, it provides a means for assigning numbers to the input data so that the data can then be stored at the array index corresponding to the assigned number.
The most commonly used hash-based scheme in DDSs is the distributed hash table.

A \textit{Distributed Hash Table} (DHT) is a DDS which provides a decentralized mechanism for associating hashed key values to some stored data item, hence supporting hash table functions.
Another way of looking at a DHT is as a hash table which partitions the keyspace, then distributes the parts across a set of nodes such that each node within the set stores a portion of the hash table.
As DHTs are utilized in structured overlay networks, the interconnection that exists between the nodes supports the efficient delivery of the key lookup and insertion requests from the requestor to the node storing the key.
Generally, replication of stored items as well as maintenance of the
$\langle
key,value\rangle$ pairs within the overlay network bolsters robustness against node churn.

DHTs offer several advantages over the traditional Client/Server based services.
They support decentralization of operations, scalability, load balancing, system churning and fast and efficient routing as well as data retrieval~\cite{Tan05}.
With few exceptions, hash-based schemes require
$\Theta(logN)$ links per node and
$O(logN)$ hops to perform routing.
Viceroy and Koorde gives
$O(1)$ links per node while maintaining the
$O(logN)$ hops~\cite{ATG05}.
Also, most schemes implement a ring-based arrangement with the difference arising in the number of pointers stored per node for routing.

Hash-based schemes, however, face some shortfalls.
A major problem of using hash-based schemes is the tendency to destroy the key ordering, resulting is scattering of the data in the system due to the hash function.
This disadvantage hinders the ability to perform range queries and possibly leading to the storage of data items far from its frequent users~\cite{BSM07}.
Another issue of concern is the use of the high cost stabilization, maintenance and recovery protocols that operate in the background to mitigate against failures hence achieving system state consistency.
Also noteworthy is the fact that the hash-based structures are not able to self-organize.

Examples of DHT implementations include Chord~\cite{SMK+01}, Tapestry~\cite{ZKJ01}, Pastry~\cite{RoD01}, CAN~\cite{RFH+01} and Viceroy~\cite{MNR02}.
An alternative approach observed in literature is layering of range query schemes over the DHT systems.
In these systems, the DHT is the routing substrate and the upper layer handles the order-based queries.
Examples of such include P-Tree~\cite{CLG+04} which uses a
B$^{+}$-tree layer on top of Chord, and Squid~\cite{ScP03} which uses Hilbert space filling curves on top of Chord, which then support multidimensional indexing mechanism (Section \ref{subsec:ResLook}).

\subsubsection{Order-based Schemes} In order to directly solve the problems associated with DHT implementations, order-based schemes have been proposed.
In particular, this involves the construction of P2P systems based on tries or other types of search trees that provide distributed search tree (DST) capabilities, thus supporting order-based searches~\cite{BSM07}.
These DST schemes, rely on order-preserving structures, in specific, balanced search trees, which then directly support search queries that depend on the key order, for example, range searches.
The DST schemes can be distinguished into two groups depending on the rules of operation the trees apply for balancing: \textit{rotation based} and \textit{split-and-join based} schemes~\cite{BSM07}.
\begin{enumerate}[label=\itshape\alph*\upshape)]
\item
 \textit{Rotation order-based schemes} utilize the red-black tree and AVL tree structures to maintain order.
 In these structures, when a node joins or leaves the system, a restructuring operation called a rotation is performed.
 The restructuring will usually affect several nodes in the tree structure, hence, a non-local operation, causing a cascading of the rotations.
 A result of the whole restructuring process is that concurrent insertions and deletion of nodes is affected as inconsistencies may arise.
 This therefore means that the system may need to implement and utilize some form of mutual exclusion mechanism.
 A direct result of such an action is an impairment on the scalability of the entire system.
 Examples of P2P systems that have been implemented using these schemes include BATON~\cite{JOV05} which is based on the AVL tree.
\item
 \textit{Split-and-join order-based schemes} depend on the B-tree and its derivatives such as the
 $2,3$ tree, skip-tree and skip lists.
 They maintain order in the system by performing restructuring through a split and join operation.
 Such operations tend to be local as opposed to non-local as seen with rotation schemes.
 Therefore only a very minimal set of nodes in the tree are accessed and balancing is achieved via randomization.
 As a result of this, the split-and-join order-based schemes tend to be highly scalable as they do not need the mutual exclusion mechanism.
 Example implementations include Skip Graph~\cite{ASh07} and Skip Tree Graph~\cite{BSM07} based on the Skip List, Skip web~\cite{AEG05} based on a range-determined Link structure, and Hyperring~\cite{ASc04} based on a deterministic 2,3 tree.
\end{enumerate}

The order-based schemes have some itinerant advantages.
They show high resilience to node failures even in cases where adversaries target specific node sets.
An advantage they have over hash-based schemes is the provision of content and path locality.
In addition to this, the ordering of the data based on their keys allows the use of range queries as well as other types of multidimensional queries for searching data items.
Moreover, these schemes have in-built structural repair mechanisms that ensure the order is maintained, which in some cases may simply involve rearranging of the pointers.
However, on the downside, the schemes are strongly affected by some security faults.
One such security issue is the Byzantine fault problem, which may need some Byzantine fault models to be applied to these designs.
Also, they are vulnerable to security breaches such as DDoS attacks and can also be used as a platform for launching the DDoS attacks.

\subsection{Communication: Unicast, Multicast and Publish/Subscribe}
\label{subsec:PubSub}
In general, communication systems support three types of communication models, that is, \textit{unicast} (one-to-one), \textit{multicast} (many-to-many) and \textit{broadcast} (one-to-all).
In P2P systems, the use of broadcast algorithms is highly discouraged and when used there is a need to setup a limitation on the number of hops to prevent network flooding and consequently slow down the entire network.
Thus, it is preferable for P2P systems to utilize unicast and multicast algorithms so as to optimally and efficiently realize effective communications.
We therefore discuss the different communication options that are preferred for P2P systems.

\subsubsection{Unicast and Multicast Communications} Unicast communications allow users to make use of application features such as direct messaging, video/audio chatting, file sharing among others.
The inclusion of multicast communication allows sending packets to a group of recipients that may be scattered throughout the network.
Multicasting allows users to choose whether to participate in a multicast group or not.
Therefore, because the packets travel only to subscribed users, there is reduced network load and end-to-end-delay, in comparison to broadcasting systems~\cite{PCE+07}.
Multicasting can be done in several way, that is, one-to-many(1-to-M), many-to-many (N-to-M) and many-to-one (N-to-1), each being used to achieve a specific purpose in the network~\cite{rfc3170}.
1-to-M is useful in application that offer scheduled audio/video distributions, push media, file distribution and caching, announcements and monitoring of real-time information.
N-to-M is utilized in multimedia conferencing, synchronized resources such as databases, concurrent processing (in particular, distributed parallel processing), collaboration (such as shared document editing), distance learning, chat groups, distributed interactive simulations, multi-player games and jam sessions.
Finally, N-to-1 multicasting is seen in resource discovery, data collection, auction systems, polling, jukebox systems and accounting.

\subsubsection{Publish/Subscribe Systems} One particular use of all the features that multicasting provides is seen in the development of publish/subscribe systems.
A \textit{publish/subscribe (pub/sub) system} is an event-driven distributed system composed of three types of processes, inter alia, publishers, subscribers and brokers~\cite{NDE+14}.
The pub-sub system allows distribution of information/data from the publishers (data/event producers) to the subscribers (data/event consumers).
The publisher sends out a notification of an application event, and any user who subscribes to that application event becomes a target for the notification.
Brokers are essentially routing algorithms match the event notifications against the subscriber requirements and deliver the notifications to the target subscribers.
There are three key categories that have been used to classify pub/sub systems: subscription, event routing and overlay topology~\cite{She10,Uzu16}.

\begin{enumerate}[label=\itshape\alph*\upshape)]
\item
 \textit{Subscription models}: The pub/sub systems in this classification present the subscribers with the capability of precisely matching their interests.
 The subscription model of the pub/sub system determines the overall specification of events and also has an effect on how the events are routed within the event channel.
 The main subscription models are topic-based, content-based and type-based models.
 \begin{itemize}
 \item
  \textit{Topic-based}: Events have locally or globally unique IDs that are usually identifiable character strings.
  Topics also represent logical connection channels between publishers and interested subscribers with network multicasts and diffusion trees being utilized for event distribution.
  Because they take only coarse-grained subscriptions, they give limited expressiveness and choices for subscriptions.
  Examples of topic-based models for P2P networks include Scribe~\cite{RKD+01} and Bayeux~\cite{ZZJ+01}.
 \item
  \textit{Content-based}: Notifications in these pub/sub systems is composed of sets of value-attribute pairs and a subscription can be any randomly chosen number of attribute names with filtering based on their values.
  The advantage that content-based models have over topic-based models is that the subscription selectivity is increased because there is increase in the dimension of choices.
  Events that meet the subscription criteria are then delivered to the subscriber.
  However, the disadvantage with these systems is that it is challenging to develop matching algorithms that can easily scale up and remain efficient.
 \item
  \textit{Type-based}: Events are objects of a specific type group which can also encapsulate attributes and methods.
  In this subscription model, declaring of a desired type becomes the distinguishing attribute.
  They take a middle ground between the previous two subscription models, giving a coarse-grained structure on events (topic-based) on which fine-grained constraints can be expressed over attributes (content-based).
 \end{itemize}
\item
 \textit{Routing models}: These models take into consideration the problem of event dispatching, whereby they ensure that the matched events are properly routed to the relevant subscribers.
 The events are matched to the subscriptions using an appropriate filter, and a routing algorithm that is used in the system then forwards directly to the subscriber or finds an appropriate route via nearby elements to the subscriber.
 The routing algorithms can be~\cite{Uzu16}:
 \begin{itemize}
 \item
  \textit{Selective filtering}: subscriptions are filtered somewhere along the notification channel, thus presenting the need for a subscription or a routing table.
 \item
  \textit{Gossiping}: utilizes a probabilistic neighbour forwarding strategy.
 \item
  \textit{Flooding}: events are broadcasted through the notification channel.
 \item
  \textit{Rendezvous}: one node acts as a routing point for a given class of events.
 \end{itemize}


%
\item
 \textit{Overlay topology}: Pub/Sub systems may be classified based on the event channel's architectural realization or topology organization~\cite{MBC+12,BCR14}.
 Thus the main classifications are:
 \begin{itemize}
 \item
  \textit{Centralized fixed topology}: One broker acts as a centralized server, storing all subscriptions, performing event to subscription mappings and undertaking event delivery to matched subscribers.
  However, the centralized pub/sub systems suffer from single point of failure and also do not provide high scalability and reliability as is required for distributed applications.
  Examples of centralized pub/sub systems are Elvin~\cite{SAr97} and S-ToPPS~\cite{PBJ03}.
 \item
  \textit{Distributed static topology}: This is sometimes also referred to as hybrid P2P or partially decentralized.
  The static topology means that there exists a graph-like distribution of brokers that are predictable and not expected to significantly change over time.
  The main varieties of the topologies include, hierarchical (such as tree-based such as JEDI~\cite{CDF01}), acyclic in which event-flow through brokers is not permitted to form cycles (such as REBECA~\cite{PGS+10}), cyclic in which event-flows results in a general graph, or a combination of the above (such as SIENA~\cite{CRW00,CRW01}).
 \item
  \textit{Distributed dynamic topology}: This is also referred to as pure P2P or fully decentralized.
  The broker overlay in this case relies on a secondary real P2P overlay or is actually part of the overlay itself.
  Examples of include Scribe~\cite{RKD+01}, Bayeux~\cite{ZZJ+01}, NICE~\cite{BBK02}, Meghdoot~\cite{GSA+04} and LightPS~\cite{ALG08}.
 \end{itemize}
\end{enumerate}

\subsubsection{Security Discussions: Communication and Publish/Subscribe Systems}
\label{subsubsec:pub/sub}
The very first need of the P2P communication system is to ensure that both the communication data and the stored data are secured.
This is achieved through the use of appropriate cryptographic services (\textit{encipherment}).
Depending on the needs of the system, symmetric (secret key) and asymmetric (public key infrastructure) encipherment may be used.
The public key infrastructure (PKI) provides essential services such as node certification, node revocation, certificate storage and certificate retrieval, which ensure that there is secure assignment of NodeIDs, as well as provide for authentication.
It also provides necessary security controls such as availability, resiliency, unforgetability, proactive security and secure communications, and also supports efficient scalability, distribution of functionality and tolerance to churn~\cite{AKD+12}.

In order to further know to whom the various keys belong a suitable public key infrastructure (PKI) can be used, which links a user's identity to the cryptographic key.
Solutions proposed include use a set of trusted certification authorities (CA) to assign the NodeIDs to principals as well as to sign the NodeID certificates that bind a random NodeID to the public key of the principal and its IP address~\cite{CDG+02}, as well as using a distributed PKI~\cite{AKD+12}.
The CAs however are a single point of failure as they are vulnerable to both legal and technical attacks while the distributed PKIs are easily affected by Sybil attacks~\cite{CDG+02}, thus an alternative is to require nodes to solve crypto puzzles to obtain a NodeID~\cite{Mer78} which has been shown to mitigate against Sybil attacks~\cite{Bor06,REM+07}.
PKIs, however, face the limitation that either trust among the users or trust in a third party is assumed, which typically is not given in fully decentralized networks.

Further, an appropriate \textit{key exchange mechanisms} must be adopted in the system, such as Diffie-Hellman key exchange protocol, especially, when symmetric encryption is used.
Asymmetric keys can also be used for \textit{digital signatures} so as to sign data and verify the signed data, hence preventing data repudiation.
In combination with the cryptographic schemes, an appropriate \textit{access control method} is needed to authenticate the identity of a user or information about a user.
To complete the requirements for secure communications, \textit{secure routing} must also be supported, that is, the P2P systems must ensure that each peer can forward messages to other peers correctly based on the routing information that the peer has.
This means that the P2P network must have support unique and secure \textit{ID assignment} to prevent abuse of the illegal IDs by malicious peers.
Thus, when security functionality is well designed, considering appropriate encipherment techniques, key exchange mechanisms, access control mechanism, ID assignment and secure routing, any P2P network which is in normal cases designed to harbour semi-trusted and untrusted peers can remain robust and secure.

The pub/sub systems of the P2P networks are not free from security concerns.
Such systems heavily rely on the use of the communication services so as to send the events to the subscribers and receive requests for subscriptions.
The overall concern in pub/sub systems is ascertaining the confidentiality of exchanged information without limiting the decoupling of the paradigm~\cite{IRC10}.
Security threats that affect pub/sub systems include identity attacks, network communication attacks, network protocol attacks, passing illegal data, stored data attacks, remote information inference, loss of accountability and uncontrolled operations~\cite{Uzu16}.
In order to ensure that the pub/sub system maintains a secure environment, it should incorporate the following features: trust management, information flow control, ubiquitous security self-adaptation, decentralized security, plugins and dynamic security reconfiguration and combination of static and dynamic solution features~\cite{Uzu16}.
In general, the system should ensure that there is \textit{publication confidentiality} so that the content of events cannot be known by the broker or any unauthorized third party, and also \textit{subscription confidentiality} to ensure that filter details are hidden from brokers and unauthorized third parties.

\subsection{Services: Monitoring and Management}
\label{subsec:Monitoring}
Thus far, it is safe to assume that, with the help of the functionality blocks that have thus far been discussed, a decent and rich P2P-based social network application can be built.
However, another essential functional requirement emerges that is highly relevant for the operation of a quality-focused, fully decentralized P2P application, namely the monitoring and quality management.
Once the network is operational a specific quality and performance emerges based on the capacities of the participating network nodes, the current workload and mostly based on the initial configuration the P2P application was started with.
As we assume a fully distributed operation, there is no further chance to take impact on the performance of the network, which might, due to various reasons, divert to performance issues, node overload and an overall collapse of the performance.
Hence, it is essential to equip the network with monitoring capabilities to provide a timely and precise view on the performance of the P2P network, as well as management capabilities that allow to (automatically) take influence on the configuration of the P2P network nodes and thus on the emerging performance of the P2P network.

The goal of the monitoring is to retrieve an exhaustive statistical view on a wide set of metrics on all peers in the network and to disseminate it to all peers in the network.
The set of metrics is an extendable list, which is common to all peers in the network and contains metrics which are based on local measurements of all peers, such as the bandwidth consumption or observed lookup delays or a peer.
The statistical view on the metrics, i.e.
average, minimum, maximum and so on, is taken over the measurements of all peers in the network, thus leading to a global view on the system statistics.
The goal of the gathering of this global view is to disseminate the global view to all peers in the network and thus let them know about the status of the system.

The goal of the management component is to take the current monitoring status of the P2P network and to establish a mechanism for the distributed analysis of it that leads to an assessment of the situation and a plan on whether and how to change the configuration of all nodes, so that the result would improve the performance of the network.
To give an example, a P2P network might face long lookup delays in average, thus it takes long to pick up even small data items in the network.
Through the monitoring the nodes learn that the high hop-count for routing is causing the lookup delay and decide, through a distributed mechanism, to increase the size of their routing tables with the perspective to have the better contacts in the routing table resulting in a lower hop count.
The decision to increase the routing table size is communicated to all nodes and takes effect once all extended routing tables are sufficiently filled with more nodes.
After a while the nodes can again evaluate whether the lookup delay is satisfying or whether further adaptations are needed or not.

Monitoring information is used in~\cite{GSG+10} to foresee low availability of replicating nodes and nodes with relevant duties and thus to counteract by selecting further nodes for replication.
Through information on the priority of the messages, differentiated services can be provided through adaptive strategies to forward messages in P2P overlays, as presented in~\cite{GPK+07}.
Optimized routing in P2P-based social networks based on monitoring information including social interaction patterns is discussed in~\cite{GAD+15}.
Self-stabilization is a property, that allows converging from any given connected topology to a desired topology, such as Chord as described in~\cite{BDK+13}.
This approach can benefit from global monitoring statistics as the self-stabilization process can be accelerated.

Centralized monitoring approaches such as the simple network management protocol (SNMP)~\cite{HPW02} or network/transport layer-focused approaches~\cite{vBH97} are not suitable, thus a decentralized approach has to be considered.
One could integrate the monitoring functionality into the used overlay, such as it is done in DASIS~\cite{AAG+04} or Willow~\cite{vRB04}.
Here the prefix-based routing tables of the corresponding overlays are extended to maintain monitoring data as well.
A corresponding data exchange protocol is included in the routing table update communication.
P2P-Diet~\cite{IKT04} and HilbertChord~\cite{SSN+08} are further variants of integrated monitoring solutions in existing P2P overlays.
One main issue of combining the monitoring and routing functionality is that the two functionalities cannot be independently improved and optimized and thus remain, in both functions, typically basic.

The two main approaches for decentralized, P2P monitoring solutions are the structured approaches that build new topologies on top of the used P2P overlay for dedicated monitoring data flows, as well as the unstructured approaches, that simply use the contacts that are available in the overlays routing table.
The latter, unstructured approaches typically apply a gossip based information exchange, where each peer exchanges periodically its knowledge with neighbors.
Examples for this category is gossiping~\cite{KDG03}, T-MAN~\cite{JMB09} and push-sum~\cite{BCF+11}.
While all nodes can directly start monitoring and the monitoring topology is robust against churn, information spreads slowly and redundancy occurs, leading to outdated monitoring results.

Structured monitoring approaches typically build a tree structure, in which the monitoring information is gathered, cyclic free towards the root, aggregated on its way towards the root, and then spread to all participating nodes in the tree again.
Examples are SkyEye \cite{GAn17,GKX08}, CONE~\cite{BVV03}, \cite{LLi04}, or SOMO~\cite{ZSZ03}.
SkyEye, as an advanced example, uses a tree-based approach which allows for efficient aggregation and dissemination of information, up and down the tree respectively.
The tree height defines the freshness of the aggregated statistics, which can be obtained without any redundant information transfer, thus highly optimized.
Also the costs are bound by the fixed node degree of the nodes.
Regardless of the position in the tree, each node encounters the same load.
Lastly, one may note that trees in the first place are vulnerable to churn.
Through the creation of multiple trees, such as in \cite{DSI+18}, a dynamic set of parallel monitoring topologies are created that is used to highly reduce the failure of an individual node on the monitoring tree.
Through the expected similarity of the monitoring results in the parallel topologies, errors and outliers in the monitoring data can be identified and corrected.

Finally, monitoring information giving insights on the quality and weaknesses of the network, can be used to implement a distributed control-loop for P2P systems, as suggested in~\cite{GSR+09,Gra10}.
Here, the monitoring information is obtained through a distributed approach, analyzed and parameter changes are decided which are then communicated and executed throughout the network.
By this, the network is capable of identifying and resolving its own weaknesses and then reach, and hold, a specific goal with regard to the lookup hop count or lookup delay.
Thus, several improvements become possible through a monitoring approach.
We further tend to believe that such a monitoring and management mechanism with an integrated quality monitoring and control loop is essential for the operation of a fully distributed P2P-based social network.

\subsection{Applications}
\label{subsec:Apps}
In parallel to the development of P2P networks, was the development of suitable applications, and some of the applications, such as KaZaA~\cite{LRW03} and Napster, have resulted in further research on P2P networks so as to improve their overall performance and quality characteristics.
In this section, we would like to briefly review some of the P2P applications in general sense.
We discuss these taking into consideration three broad categories: Content delivery networks, communication and collaboration, cryptocurrency, and mobile P2P applications.

\subsubsection{Content delivery networks (CDNs)} Content delivery networks are also sometimes referred to as \textit{content distribution networks}.
CDNs are designed to enable content owners and creators in widely disseminate their content in the Web~\cite{Pas12}.
The greatest concern in CDNs is how effectively ensure that content is readily available to end users.
This can be addressed by ensuring load balancing of content by use of efficient replication strategies such as~\cite{MTQ+12} or caching algorithms such as distributed caching algorithms~\cite{BGW10}, and web re-direction for requests based on some replication rule.
User requests in a CDN are serviced via a \textit{request-routing} system that includes
\begin{enumerate*}
\item
 a \textit{request-routing algorithm} that selects the most appropriate edge server for a given request, and
\item
 a \textit{request-routing mechanism} that directs the user request to an appropriate edge server~\cite{AKS+17}.
\end{enumerate*}
One recent web technology that has been presented as a solution for P2P content distribution is WebRTC\footnote{https://webrtc.org/}.
\cite{VWS13} present a model that leverages WebRTC, a technology that enables development of rich, high-quality Real-Time Communication (RTC) applications for the browser, mobile platforms, and IoT devices, to design P2P content distribution applications.
However, WebRTC presents other challenges when used for P2P content distribution that must be addressed~\cite{VSB16}.
Applications in this category that we focus on are file sharing and video streaming applications.

\begin{enumerate}[label=\itshape\alph*\upshape)]
\item
 \textit{File sharing}: These applications are the mainstay of P2P technologies and the early P2P applications such as Napster and KaZaA are typical examples.
 For this reason, P2P networks have become synonymous to file sharing and were previously considered as file sharing communities~\cite{DaD05}.
 These P2P file sharing networks rely on special soft wares that provide relevant protocols for file sharing.
 Currently active client software include BitTorrent\footnote{http://www.bittorrent.com/}, uTorrent\footnote{http://www.utorrent.com/}, BitLord \footnote{http://www.bitlord.com/} and eMule\footnote{http://www.emule.com/}.
\item
 \textit{Video streaming applications}: They can be classified into two categories: live and on-demand~\cite{LGL08}.
 In a live streaming session, live video is disseminated to peers in realtime and video playback on all peers is synchronized.
 For video on-demand (VoD) systems, users can with flexibility choose the videos that they want to watch at the time they want.
 Recent proposals for VoD applications include SocialTube~\cite{LSW+12} and Liquidstream II~\cite{DGE+15} and for P2P live streaming applications include SopCast\footnote{http://www.sopcast.com} and Chunked-Swarm~\cite{PDG15}.
 \cite{LGL08,RPI12,GRP14} provide comprehensive studies on P2P video streaming applications.
\end{enumerate}

\subsubsection{Communication and Collaboration} In addition to having the ability to share content, another area of research in P2P technology has has been in increasing communication and collaboration in the system to allow the users to interact directly.
The development of the ARPANET, the predecessor to today's Internet, was in itself a collaboration of several institutions with the goal of creating a communication platform.
Thus, research on P2P networks has also been focused at making communication and collaborations a possibility.
In this study we looks at three areas: P2P voice-over-IP (P2P-VoIP) applications and P2P online gaming, and P2P online social networks

\begin{enumerate}[label=\itshape\alph*\upshape)]
\item
 \textit{P2P Voice-over-IP (P2P-VoIP)}: Traditional VoIP required standard infrastructure between the users, and a telco operator was contacted to set up a VoIP connection between the users who were then charged for the use of this service.
 In P2P-VoIP, users only need a reliable internet connection to implement the same services.
 VoIP systems have three basic functions resource location, session establishment and presence monitoring~\cite{BLo04}.
 Two P2P-VoIP application are Skype\footnote{https://www.skype.com/} and SOSIMPLE~\cite{BLo04,BLJ05}.
 Initially, Skype was built on the FastTrack protocol used by KaZaA which was a major factor for Skype's success as FastTrack operated smoothly even behind firewalls as compared to SIP. Skype was purchased by Microsoft in May 2011 and although the structure of the Skype network is not clearly known, several researchers have reverse engineered it to show that it is based on a P2P infrastructure~\cite{BSc04,GDJ06}.
 SOSIMPLE combined the standard IP-telephony protocol SIP/SIMPLE, which is a family of IETF standards for IM and VoIP with a Chord overlay as providing the DHT function.
\item
 \textit{Online Gaming}: This is a rapidly growing phenomenon in the area of collaborative virtual environments, where thousands of users distributed all over the globe collaborate while playing a game over the Internet.
 These gaming environments are referred to as \textit{massive multiplayer games} (MMGs)~\cite{DHS06} or massively multiplayer online games (MMOGs)~\cite{Yak13}.
 Two directions have been followed to provide architectural support for MMGs: the traditional C/S paradigm, multiserver architecture and P2P architectures~\cite{DHS06,Yak13}.
 The C/S paradigm is the most prevalent game architecture but it has been observed that P2P architectures give several advantages that are not realized with C/S architectures.
 They support distribution of computations as well as network loads among peers, are highly scalable, not cost intensive and have good performance.
 A comprehensive listing of available P2P architectures for MMGs can be found in~\cite{Yak13}.
\item
 \textit{P2P online social networks (P2P-OSNs)}: As has been previously highlighted in Section~\ref{sec:SNs}, OSNs are a rapidly growing means of communication on the Internet and P2P-based solutions have also emerged to tackle the challenges that exist in centralized OSNs (see Section~\ref{subsec:DOSNs}).
 OSNs provide the capability to merge all the services for collaboration and communications such as file sharing, document editing, instant messaging, audio and video calls, interactive online gaming among other.
 P2P OSNs is the main focus of this survey and is discussed in depth in Section~\ref{sec:p2pOSNs}.
\end{enumerate}

\subsubsection{Cryptocurrencies} These are P2P digital exchange systems in which cryptographic algorithms are utilized in the generation and distribution of the currency units~\cite{Far15}.
The use of cryptocurrencies emerged as a means of bypassing the existing world Fiat currency, currency declared as legal tender by a government, that are considered outdated, having a limited money supply and are historically mismanaged by governments~\cite{ANV13}.
Cryptocurrencies are not legal tender and are therefore not backed by governments.
Several online currency exchange systems have been proposed or developed, such as PayPal, but these have been based on Fiat currency.
In 2008, Nakamoto~\cite{Nak08}, proposed Bitcoin, which was the first Cryptocurrency.
Since then, others have been created such as Ethereum~\cite{Woo14}, Blackcoin~\cite{Vas14}, Permacoin~\cite{MJS+14} among others.

\subsubsection{Mobile P2P Applications}
\label{subsubsec:MobP2P}
Mobile computing has become very widespread as more and more people are able to access sophisticated mobile devices such as smart phones and iPads.
Further, mobile telephony infrastructure has also been rapidly developed which has easened mobility and Internet access.
Mobile hosts constantly change their IP address as they are constantly changing the point of network attachment but are still connected to the network at the link layer of the TCP/IP model.
This means that the TCP connections would ideally fail and packet routing may not occur.
However, this problem is solved by the introduction of \textit{Mobile IP} addresses.
The idea behind Mobile IP is that the host has a ``home'' network.
Ordinary forwarding of packets takes place while at home.
When the mobile host migrates to another network, it keeps its home IP address but special routing forwarding algorithms are used to make the host appear like it is still at home.
Additionally, unlike other ordinary computing devices, many mobile devices have limited computing resources and therefore do not handle the same type or size of the content~\cite{BaF06}.
Thus Internet content for mobile devices must be customized and repackaged so as to support the resources and communication protocols for mobile devices.

Essentially, applications that can run on normal computing devices can also run on mobile devices, albeit with the mobile device requirements taken into consideration when developing them.
Bakos and Farkas~\cite{BaF06}, show this to be the case by implementing distributed computing, file sharing, content sharing and keyword search applications for smart phones.
LightPeers~\cite{Chr07} was developed as a platform for lightweight mobile pure P2P networking for student groups while outside a school building, thus supporting nomadic learning.
Kellerer et al.~\cite{KDM+07} proposed a P2P service platform that provides support for lookup and information distribution while ensuring reliability, controlability, bootstrapping and reputation management.
Lehtinen~\cite{Leh06} developed a mobile P2P file sharing application using the Session Initiation Protocol as the underlying signal protocol on the Nokia Series 60 platform.
P2PBluetooth~\cite{PBC09} was a prototype application that was developed for mobile P2P file sharing using Bluetooth.
These examples are merely a sample of the implementations that show viability of mobile P2P applications.

\subsubsection{Security Discussion: P2P Applications}
\label{subsubsec:P2PApps}
The applications that are designed for use within the P2P framework depend on the services that are offered by the underlying layers.
A failure in one or more of these services has a direct impact on the expected performance of these applications.
Securing the underlying layers is therefore the key to ensuring that the applications perform with minimal interference.
However, there are notable cases where the applications are themselves a point of security concern.

Peer-to-peer file-sharing applications as well as streaming applications rely on a form of resource reciprocation among users.
In case the peers are reluctant to share resources, a situation known as \textit{free riding} occurs in the network, where a peer that refuses to contribute to the network at an acceptable level is called a free rider~\cite{KKU09}.
The result is only a small population of the peers serve a large population, leading to problems in scaling and also may introduce a single point of failure.
Also shared files are limited or increase very slowly resulting in degraded process search quality.
The effects of free riding are very keenly felt in video and audio streaming applications in comparison to the simple file sharing applications.

The authors in \cite{GLM11} identify five main sources of vulnerabilities in P2P streaming: malicious/malfunctioning peer nodes, integrity of the distributed data, presence of supernodes, presence or absence of secured overlay routing, and vulnerabilities in the application codes.
As a result of this, the streaming applications are open to attacks such as collusion attacks, forgery attacks, membership and Eclipse attacks, neighbor selection attacks, Sybil attacks , DoS attacks, omission attacks and pollution attacks.
Also, decentralized authentication of nodes in a P2P application is generally difficult.
This is true for media streaming applications as well as P2P-VoIP~\cite{See09}.
They do not really face challenges with confidentiality but rather in authentication and key exchange.
Due to this lack of authentication, \textit{unsolicited communications} such as Spam over IP Telephony (SPIT) are possible.

By far the largest concern in MMOG applications is \textit{cheating}.
Cheating is basically a security breach in which the game rules are broken and is an behavior used by players to gain an unfair advantage over other peer players~\cite{YRa05}.
This is made possible in MMOGs because of conventional security breaches that occur as a result of focus on confidentiality, integrity and availability in the applications as opposed to considering the social aspects of security breaches as well~\cite{HuZ08}.
Cheating within the MMOGs can be achieved by exploiting misplaced trust, collusion, abusing game procedure, misusing/abusing virtual assets, exploiting machine intelligence, modifying client infrastructure, denying peer players service, timing cheats, compromising passwords, exploiting lack of secrecy and authentication, exploiting a bug or loophole, compromising game server, internet misuse, as well as by social engineering~\cite{HuZ08}.

Mobile security has certain characteristics that distinguish them from conventional computer security.
These characteristics are, \textit{inter alia},
\begin
 {enumerate*}[label=(\roman*)]
\item
 devices can be moved around by the users hence easy to steal or easily tampered with (\textit{mobility}),
\item
 the device\textquotestraightbase s owner is the unique user (\textit{strong personalization}),
\item
 allows users ease of access to various Internet service which open up the device to malware infections (\textit{strong connectivity}),
\item
 various technologies are combined within a single device giving an attacker options of access routes (\textit{technology convergence}), and
\item
 lack some features that are available in the computers (\textit{reduced capabilities})~\cite{LMS13}.
\end
{enumerate*} Therefore, in addition to general vulnerabilities associated with P2P networks, mobile P2P applications will also experience the effects of due to mobile security problems of the devices.
The main threat in mobile devices are malware such as viruses, worms, Trojans, rootkits and botnets.
A malware is a software or program code that are designed to use a device without the owner\textquotestraightbase s permission, and usually are hostile, intrusive or annoying in nature.
Also, because the devices are also unique to the users, it is not easy to guarantee privacy and anonymity.

For mobile P2P networks,~\cite{HZL+05} propose the use of a Secret-sharing-based Mutual Anonymity Protocol (SAM) for enforcing anonymity in mobile P2P networks.
Mobile devices, need some form of security as well to prevent them from being affected by basic inherent security threats that are common in many P2P networks.~\cite{LMS13} discuss various security solutions to handle malware in mobile devices that definitely prove useful for mobile devices in the P2P network, especially in the content sharing applications for mobile P2P, where this is a big threat.

Having reviewed the essential building blocks for advanced P2P applications, next we present and discuss the proposed solutions of P2P-based social networks.

\section{P2P-based Social Networks}
\label{sec:p2pOSNs}
Peer-to-peer based SNs have been put forward as a possible alternative for addressing the security and scalability challenges associated with centralized approached for SNs.
To this end, various proposals for possible solutions have been made, as is evident from literature.
Each proposal tries to achieve a fully functional P2P-based OSN using various combinations for P2P components, and may sometimes propose new methods to achieve a critical aspect of OSN within the P2P environment.
In discussing the proposals studied, we take the direction of classifying them based on the structure of the overlay.
In addition, we also consider whether the SN is a microblog or a full SN (which we simply refer to as SN).
In the analysis, the trend is to give a general overview of the design goals, the architecture, and a brief discussion of possible flaws noted with the proposal.

\subsection{Single-overlay distributed social networks}
\label{subsec:SingleOverlayDistributedSNs}
As previously in Section~\ref{subsubsec:SingleOverlay}, these social networks are designed on a single overlay (structured or unstructured) and therefore all routing and storage procedures are handled by the overlay itself.
We discuss the proposals in this category.

\subsubsection{LifeSocial.KOM / LibreSocial}
\label{subsubsec:LifeSocial}
LifeSocial.KOM / LibreSocial~\cite{GPM+08,GGM+10,GGS11} was first proposed in 2008 and is a plugin-based and extendible P2P based OSN build on a composition of various essential P2P functionalities within a P2P framework.
Due to name conflicts, the initial name of LifeSocial.KOM has been changed to LibreSocial (\url{https://libresocial.com}).
The OSGi-based framework is highly modular and utilizes FreePastry as the P2P overlay network and PAST to provide reliable storage with data replication mechanisms.
FreePastry and PAST have been heavily extended to meet the requirements with regard to robustness, fault-tolerance and security.

Security is enforced through registrations and login mechanisms as well as use of access control on data stored as described in~\cite{GMM09}.
An asymmetric key mechanism without the use of a server, certificate authorities or even the trust in other nodes is used for for the creation of cryptographic keys, that are used throughout the system.
The public key is also the node ID, which allows for direct encrypted communication and authentication.
In joining the network, weak and strong nodes are treated differently so as to support weak nodes in the network, mainly as clients, without involving them in the burden of carrying load.
As FreePastry is a DHT it offers reliable key based routing.
To quicken the process the routing protocol has been adapted to become iterative such as in Kademlia instead of the previous recursive approach.
Furthermore, the data storage logic PAST has been extended to support data updates, such as an update of the profile data.
The main modification in the data storage is the introduction of security and access control.

Data is replicated with support for WRITE, READ, UPDATE, DELETE and APPEND operations.
Access control is managed by encrypting the data for the various access enabled users in the network.
One outstanding feature of LibreSocial is that the friends in the network do not play any privileged role in terms of security and trust in them is not required.
The replication mechanisms maintain availability of the stored data and ensure load balancing, overload avoidance and the support for weak nodes in the network.
Through the support of secure, access controlled and replicated distributed data structures, such as distributed sets, linked lists and prefix hash trees, a variety of data forms are supported, such as (comment) lists, (photo and friend) sets or (forum thread) trees.
As the underlying P2P framework and secure data structures are very general, it is easy to add new application functionalities in LibreSocial.
One unique function in LibreSocial is the availability of a message inbox which can be read by the corresponding user only but filled with entries from various users.
This feature is cryptographically enforced to ensure that crucial messages, intended only for the eyes of the recipient user must not be read by his friends.
This is guaranteed in LibreSocial.

With regard to communication, end-to-end secured unicast, multicast and publish/subscribe functions are provided.
A mechanism for keyword-based and range-queries is integrated to support the searching for searchable items.
Currently it is used to suppport the search for users based on the profile information that explicitly has been marked as public by its owner.
The goal of providing a secure communication, user-centric access control management and monitoring services for the OSN while providing all the common OSN functionalities such as user profiles, friends lists, user groups, photo albums, chatting and status updates is reached.
It also includes the forums and data spaces for collaboration, messaging and calendars for coordination, as well as text-, audio-based and video-based real time communication.
In addition to the rich set of applications it comes with an integrated decentralized app repository, that any user can host, and which allows to created and share application plugins with other users.
Also it comes with a fully decentralized monitoring that allows to observe and evaluate the performance of the P2P network.

\subsubsection{Porkut/My3}
\label{subsubsec:My3}
Porkut~\cite{NPA10} and My3~\cite{NPA11} are similar proposals by the same authors to provide privacy-preserving data access.
The design seeks to achieve three goals: the elimination of a single administrative control; privacy preservation of individual's privacy content, giving users complete control of their profile and it's content; and the exploitation of trust relationships among network users for improvement of content availability and storage performance.

The architecture of the application involves three key features: a \textit{DHT} such as OpenDHT~\cite{RGK+05} that is used to store meta information of the user based on a \textit{user-to-TPS} mapping to form the
${\langle key,
 value\rangle}$ pair; an \textit{online time graph} that contains all the user's friends as vertices and edges are only existent if there is an overlap in online times between two trusted pairs; and a \textit{storage layer} which is a construction of the \textit{trusted proxy set}
($TPS$) for a user
$u$.
The
$TPS$ is a set of self-defined nodes in which a user's profile is hosted.
Using an appropriate algorithm the set
$TPS$ for a particular user is constructed from his social graph in which users are characterized as having two parameters: \textit{a geographical location} that determines the time zone of the user, and an \textit{online time period} which is the time the user is online in the social network.
The proposed criteria possible for selecting the proper set of members into
$TPS$ from all possible trusted friends are: \textit{low access and consistency costs} and \textit{high data availability}.

A privacy-preserving index of the social network contents is constructed such that it is possible to perform privacy-aware searching which enables content discovery among friends in the OSN and allowing new discovery of new friends and the establishment of new social connections.
The index mechanism uses \textit{k}-anonymization techniques so that a list of keys are mapped to a list of values.
This helps achieve content and owner privacy, so that, with the indexing scheme, strangers can contact each other based on interested content.
The main aspects of the system that are emphasized, in addition to the basic requirements for an OSN, are: \textit{storage layer} formed by the
$TPS$ construction, \textit{profile accessibility} through an available mount point of a given user, \textit{update propagation} as a user's profile is replicated to other mount points, and \textit{eventual consistency} since concurrent updates ensure mount points are up to date.

This system assumes that users have friends they trust and that online times of these trusted friends overlap.
However, it is important to note that the existence of (online) friends as well as the requirement to trust these friends is an assumption that is not always existent in reality.
Without friends or trust in them, then data availability, confidentiality and integrity is not guaranteed.

\subsubsection{Megaphone}
\label{subsubsec:Megaphone}
Megaphone~\cite{PEn10} is a P2P microblogging application designed with the aim of overcoming the problem of single point of failure due to reliance on web-based services in centralized systems.
It utilizes Pastry and Scribe to perform message and group routing by organizing the social graph of users into a multicast trees on top of Pastry using Scribe in which the ``poster'' node is the root of the tree with ``followers'' being a child nodes.
Therefore the poster creates the tree and performs the task of managing the joins, lists of followers, storage of the public keys of child nodes and sending of messages.
Followers can send response messages to the posters.
The messages can be signed and encrypted using public key infrastructure based on RSA~\cite{RSA78} algorithm.
A poster will also be responsible for the generation of session keys which is encrypted with the poster's private key and further encrypted using the followers public keys.
All nodes in the multicast tree cache this session key but it is readable only by authorised nodes.
Authentication is achieved using certificates which are generated by the nodes themselves and self signed or generated by a certification authority.
Node IDs are based on the hash of the username and the hash of the public key which are concatenated to guarantee uniqueness.
All members of a multicast tree, hence followers, have knowledge of the originator of any post and all the followers which means that the privacy of the users is not fully guaranteed.

\subsubsection{\textit{eXO}}
\label{subsubsec:eXO}
\textit{eXO}~\cite{LNT+11} is a completely decentralized, scalable system that is designed to offer key social networking services, while relying on a P2P platform.
The system has two main goal, namely, foster the idea of highly distributed SN functionality, that is, autonomy, and support full user control even when sharing content.
Content is primarily images, audio and video content and secondarily text.
The underlay consists of a large number of nodes, with each node running a routing protocol for a structured overlay DHT such as Pastry~\cite{RoD01} and Chord~\cite{SMK+01}, such as the previously presented approaches.
To support autonomy and privacy, content shared by a user is kept only at the user's node while replication is done only on nodes adjacent to the user node in the
$ID$ space at the owner's request or in case the content's availability is needed.
Content as well as user profiles can be either \textit{public} which is indexed and available to all, or \textit{private} which is not indexed.
Nodes can take any of three roles depending on the tasks that the nodes perform.
A node can act as a front end, taking on the role of a \textit{request solver}, serving user requests and dispatching the requests to relevant peer nodes.
It can also act as a \textit{network storage interface} for contents and profile replicas.
Lastly, it can act as a \textit{catalogue node} when it stores indexing data structures.

To support indexing, every content item is represented by a set of terms, or keywords, describing it, called the \textit{content profile}.
This is useful in performing ``top-\textit{k} similar content items'' queries.
Also, an individual user is described by a set of terms that the user defines which is then called the \textit{user profile}.
These two profiles allow \textit{eXO} to support the use of tags which are terms contributed by a user describing a specific content item or user.
Tags help achieve better quality query result by leveraging community wisdom.
\textit{eXO} supports public networks and personal SNs.
The public network is composed of the DHT, and the content and user profiles stored at the DHT nodes can be indexed and accessed via the DHT. The DHT structure makes it possible to perform queries on the user profiles as desired.
Through the process of searching for interesting user profiles, a user can identify and add these user profiles to their personal social network.

The main limitation in \textit{eXO} is that data can be either fully public or hidden.
Through the lack of a foundation for security mechanisms, confidentiality, access control and data integrity is not given.
Thus, the platform can be used for only very simple applications.

\subsubsection{PAC'nPOST}
\label{subsubsec:PacnPost}
PAC'nPOST~\cite{AsC12} is a framework for a microblogging social network.
It is implemented on an unstructured P2P network.
The two goals of the systems are: first, to enable the users the retrieve blogs of other users that are being followed and; secondly, to allow a user to perform keyword-based searches.
The search and retrieval mechanism is based on a \textit{probably approximately correct} (PAC) search architecture, where a query is sent to a fixed number of random nodes in the network and the probability of attaining a certain accuracy being a function of nodes queried (fixed) and the documents' replication rate.
As the nodes do not push their blog entries, but rather they must be searched, chances are high that blog entries are missed or that over time, in case that the number of participants rises, the network will be overloaded with search messages.
In unstructured overlay networks a high search precision necessarily requires an adequately high traffic load, thus the network functions do not scale.

\subsubsection{DECENT}
\label{subsubsec:DECENT}
DECENT~\cite{JNM+12} is a proposed decentralized OSN architecture, whose goal is to: utilize object-oriented design (OOD) to enhance flexible data management; ensure efficient access revocation and fine-grained data policies using suitable cryptographic methods; and combine confidentiality, integrity, and availability through use of DHT functionalities.
The architecture of DECENT is modular, that is, the data objects, cryptographic mechanisms and the DHT are three separate components that interact through interfaces.
This allows the OSN to use any type of DHT or cryptographic mechanism.
Every object in the OSN has three access policies (\textit{read}, \textit{write} and \textit{append}) associated with it which can be attribute-based (AB), identity-based (IB) or a combination of both.
These policies are defined by the user at object creation and stored in the object's metadata.

DECENT implements modifications to the attribute-based encryption (ABE) as well as modifications to ABE's support for immediate revocation by the use of the EASiER scheme.
With ABE the message is encrypted using a randomly chosen symmetric encryption key, then encrypted with ABE. DECENT's considers the ABE encrypted symmetric key as actually part of the object reference and not included in the object itself, hence keeping the policy hidden from untrusted storage nodes.
Object can have different read policies associated if there exists several references for an object.
The EASiER scheme uses a proxy in every decryption to ensure the revoked contact is no longer able access data that requires the revoked attribute(s).
Two extensions were done to the EASiER scheme.
First, the proxy functionality is divided among several randomly selected nodes using threshold secret sharing.
With the assumption that majority of nodes are not actively malicious, this will ensure the security of the proxy.
Second, is an extension to support attribute delegation.
To enforce authenticity, the write policy public key must be part of the object reference, rather than the object itself.
Also, as the append policy is authenticated by the write-policy signature, it is included as part of the object metadata.

The participants in the OSN are organized into a DHT with preliminary tests on a FreePastry simulator and a Kademlia implementation on PlanetLab.
The DHT provides for a scalable
$\langle
key,value\rangle$ pair store having an efficient lookup mechanism for the location of objects stored in the nodes.
Using the
$objID$ as the key, the objects are stored in the DHT. To enforce availability there are several replicas of an object, with each object having a version number, authenticated by a write-policy signature, as part of its metadata for freshness guarantees.
The write policy prevents the modifications by a malicious node.
The storage node does not know the write-policy signature public key
($SPK$) as it is part of the object reference.
Therefore the storage node cannot differentiate between a legitimate update and a malicious update.
DECENT addresses this by adding an unencrypted metadata field to an object containing a public key.
This is then used for authenticating write requests (write authentication public key, or
$WAPK$).
Thus, write/delete request must be signed by the corresponding secret key.
If not, the storage node refuses the request.

DECENT is similar to Cachet in providing security, sharing the same shortcomings.
First, the origin or the user credentials are unclear and second, through the focus on secure storage, the communication options are neglected.

\subsubsection{HorNet}
\label{subsubsec:HorNet}
HorNet~\cite{IMC+12} is a proposed microblogging service for contributory social networks that is built on a structured P2P overlay network.
A contributory social network is a SN whereby the SN's resources, such as CPU, network and storage, are voluntarily contributed by the participating network members.
The service is developed with the focus on availability, decentralization and performance.
The architecture of HorNet is split into three layers: the communication layer, a middle layer and the upper layer.
The communication layer serves the function of connecting all the nodes that join and leave the system.
This layer is implemented as a structured overlay network, specifically FreePastry.
FreePastry offers key-based routing (KBR) that also allows for scalability and churn-tolerance.
The middle layer guarantees availability of application logic\textquoteright s data and components.
This is done through the provision of a file-based storage service that ensures there is message persistence and other required data, and a freely available middleware called CoDeS~\cite{Igl11} which is used for service deployment is contributory communities.
The upper layer of HorNet is a set of small services.
Each user has an instance of the services in this layer deployed using CoDeS, and through CoDeS, the instances are kept available.

CoDeS forms a platform for service deployment by aggregation of a set of non-dedicated, global and heterogeneous computers, which guarantees required services are always available, self-managed and decentralized.
CoDeS also provides redundancy for failure-tolerance and uses weak consistency replication of its internal information so as increase performance while ensuring almost accurate information provision.
CoDeS also allows users to manage which resources to share with the network.
The overall design of the HorNet system is thus divided into two main parts.
The first is client providing the HorNet features to the users which is deployed as a web container.
The second aspect of the design is a component set that is deployed as CoDeS services which enforces HorNets core functionalities in a decentralized manner.
Authentication is done using a PKI and it assumes that users hold a public-key certificate from a trusted CA upon registration for the services.
All messages sent in the system are signed to protect authenticity and integrity.

HorNet uses the CA and the CoDeS servers as centralized components, as long as they are available HorNet networks might operate.
As soon as one element is switched off, the network fails.
For a fully decentralized OSN, servers should be avoided.

\subsubsection{PESCA}
\label{subsubsec:PESCA}
This proposal by~\cite{RJM15} was designed with the aim of achieving privacy in the social communication and as well as social data availability.
PESCA assumes structured P2P overlay based on DHTs so as to offer an efficient the lookup service.
It monitors the users's online patterns as well as the devices they use, taking into consideration the time of the day and the days of the week which are stored into the user online table (UOT) as small non-overlapping time slots which also indicate the device used.
The replica placement strategy that is utilized take into account the online direct/indirect friends and the data audiences.
The strategy determines the best replica matrix corresponding to a user's data in a greedy fashion.
The replica candidate lists includes the user's friends who are online and have enough storage space.
Each candidate is scored based on the number of overlaps between its uptime and the audience's uptime.
Candidates with scores of zero are eliminated.
Candidates with the highest scores are chosen, also taking into consideration the storage space available in case of similar scores.
Data confidentiality and access control are achieved using broadcast encryption (BE) scheme~\cite{MMi12}.
Users have a uniquely selected global identity (GID) based on a hash value of the user's email address.
A user also generates a virtual identity (VID), an ambiguous index, and a BE secret key for every social contact added as a friend.
The user also allocates space, called space budget (SB) that can be spared for the friend.
The user then sends the VID and the BE secret key via a secure channel along with the GID, the current UOT and assigned SB via a public channel.

However, there is no information provided by the authors as to what OSN services the proposal offers outside of guaranteeing secure social communication and social data availability.
This opens the proposals to questions on the practicality of its use in the real world.

\subsubsection{WebP2P}
\label{subsubsec:WebP2p}
This work is a break from the norm in the sense that it aims at providing a DOSN without the need to install any new software.
Disterh{\"o}ft et al~\cite{DiG15} propose a solution and prototype for a Web-based P2P framework that supports social networking.
Specifically, it supports secure buddy lists, data storage and personal text, audio and video communication.

It is implemented on a browser-based implementation of the Chord Protocol that relies on WebRTC (\url{https://webrtc.org/}).
WebRTC allows browser instances to establish connections to each other.
Based on this, a DHT is implemented based on OpenDHT and has been extended to support secure user identities, secure communication and a simple secure storage.
Authentication is offered using public keys generated using an asymmetric Elliptic Curve Cryptography mechanism.
The public keys are used as node IDs and also as user IDs.
Each user initially has to use these cumbersome lengthy 160bit IDs as user identifiers until, through a dialogue the user can identify each other and assign in the GUI and buddy list an alternative name for this user ID. Thus, fully decentralized authentication is provided.
As the user IDs are public keys, the communication can be encrypted with the public key of the receiver and signed by the sender, providing confidentiality and integrity.
Data is stored and replicated in the DHT created among the browsers and encrypted by the data owner.
A distributed identity-based access control mechanism is further used that uses self-signed certificates.

The Web-based DHT overlay provides the basic functions that support buddy list management, communications via text-based chat, and device-independent, distributed storage of contacts and chat history.
Thus, the functions are focused on chatting and maintaining a buddy list.
Larger data items cannot be stored as WebRTC is limited to use only 5MB of storage space to hinder potential attacks.
Thus only tiny data can be stored.
Besides the overlay-based storage and communication, WebRTC channels are also used to establish video and audio calls between the participants.
The WebP2P-framework is self contained, secure and provides for the purpose of secure chatting in a web-based P2P-DOSN with focus on chatting, video and audio.

\subsection{Single-overlay hybrid social networks} These SNs implement a hybrid structure in which a single overlay (structured or unstructured) is utilized and some degree of centralization is incorporated.
In most of these cases, centralization is used to provide a solution for indexing while the P2P overlay is used to handle routing.
We discuss some of the SNs proposed in this category in the following.

\subsubsection{P2P Social Networking (PeerSoN)}
\label{subsubsec:PeerSoN}
PeerSoN~\cite{BSV09} was an advancement of the ideas put forward in~\cite{BuD09}.
It is built to address privacy concerns raised over OSNs as well as look at how to ensure availability.
In the proposed prototype, the privacy problem is addressed by integrating encryption and access control to implement a user login procedure.
Availability is made possible by including novel file sharing procedures.
The architecture of PeerSoN is two-tiered and is designed in such a manner as to ensure the users' contents are decoupled from the control mechanisms.
The first tier, the lower level, consists of the users and the content, which allows the users to exchange content directly with each other.
The second level is the DHT which provides lookup services, so that users can find each other as well as find desired content.
The DHT also stores the user's meta-data and stores updates for a user in case a user goes offline.
PeerSoN implemented the DHT using OpenDHT. Security and privacy concerns in the system are addressed by using identity management.
Specifically, PeerSoN assumes the availability of a public-key infrastructure (PKI) with the possibility of revocation of keys and encryption using public keys of intended audience.

OpenDHT is a centrally managed deployment of the Bamboo DHT on PlanetLab~\footnote{https://www.planet-lab.org}.
The use of a centralized component in the OSN is not desired when the overall goal is to realize full decentralization.

\subsubsection{Safebook}
\label{subsubsec:Safebook}
Safebook~\cite{CMS09,CMS09b,CMO10} is a decentralized SN designed to achieve two goals.
First, it utilizes a P2P architecture to avoid user data and user behavior control by a single entity such as a service provider.
Second, it aims at providing privacy ans trust management for user data and communication in the system through trust relations existing in the social network.
Safebook enforces security by providing mechanisms for end-to-end confidentiality; proper authentication for access control; privacy mechanisms to ensure not only anonymity, unlinkability and untraceability of user communications, but also confidentiality of private information; and data integrity to prevent profile data tampering.
It also guarantees data availability.

Safebook's implements a three-tier architecture having a direct mapping of layers to the OSN level as follows: a user-centered SN layer which forms the SN level of the OSN, a P2P substrate which implements the application services (AS), and the Internet which represents the communication and transport (CT) level.
Participants in the OSN are viewed as a host node on the Internet, a peer node in the P2P overlay and a member of the SN layer.
The participating nodes in the network form two types of overlays: a set of \textit{Matryoshkas} which are concentric structures in the SN level that provide distributed data storage with privacy and end-to-end confidentiality by assuming that users trust their friends in handling their data carefully, and a P2P substrate such as a DHT that supports data lookup services.
In addition, there is a \textit{trusted identifier service} (TIS) which ensures each node is given a unique pseudonym and identifier for the SN level along with related certificates.
The TIS is not involved in data management and hence does not violate the goal of privacy preservation.
This guarantees protection against attacks such as Sybil and impersonation attacks.
After obtaining an identity, a new user begins the process of creating their own Matryoshka.
This is done by first sending requests to friends or trusted peers.
Secure communication between users is made possible by encryption and decryption of the pseudonyms using private and public keys, hence message integrity and confidentiality.

Disadvantageous in this approach is the need to have friends to be able to store data reliable, to trust these friends to not tamper with the data as well as the need for the TIS as centralized component.
In Safebook all data is to be assumed to be shared with the ``friends'', thus confidentiality is not given.
In addition, users without friends or with friends that are often online cannot maintain the availability of their data.

\subsubsection{Cuckoo}
\label{subsubsec:Cuckoo}
Cuckoo~\cite{XCZ+10,XCF+10} is a socio-aware online microblogging system that is proposed and built to be compatible to the Twitter architecture.
It is designed to utilize the Twitter servers to conserve bandwidth and storage resources while also taking advantage of the P2P technologies for scaling and reliable microblogging services.
Cuckoo uses Pastry~\cite{RoD01} as its underlying overlay.
Typical microblogging services allow for the following social relations to exist between users: \textit{friend}, \textit{neighbor}, \textit{follower}, and \textit{following}.
Along with the Pastry routing table, users maintain these four user lists.
So as to perform searching to perform status updates, Cuckoo uses a hybrid search method, that is, flooding to fetch statuses from influentials and the DHT to fetch statuses from normal users.
Newly published micro-content is disseminated to users using push method rather than pull as it is more efficient.
Micro-content is also replicated among followers thus followers can provide the content to each other in a cache-and-relay style in case the original publisher is unavailable.

It has to be noted that without a cryptographic foundation any communication and data stored can be tampered and read.
Cuckoo is insecure and highly requires the existence friends, neighbors, followers and followed users.

\subsubsection{Litter}
\label{subsubsec:Litter}
Litter~\cite{JWB+11,JEL+13} is a lightweight microblogging service that leverages P2P virtual private network (P2PVPN) technologies, such as Hamachi\footnote{https://www.vpn.net/} and SocialVPN\footnote{http://ipop-project.org}.
P2PVPNs utilize P2P technologies for direct IP traffic tunneling among peers.
They also provide a trusted platform in which peers can communicate and collaborate at the IP layer with each other.
Thus the trusted connections between the social peers form a social overlay P2P network.
The P2PVPN also supports IP multicasting by tunneling the multicast packets to each friend with whom a node has an encrypted P2P connection.
Litter's microblogging service is composed of two basic IP layer mechanisms.
The first IP layer is IP multicasting that is used to propagate messages to two-hops neighbors within the social graph.
The second IP layer is UDP datagrams that are necessary for traversing the social graph for update dissemination to social distant peers.
The use of these services is based on the assumption that users are running a P2PVPN. 

To achieve peer discovery and cryptographic key distribution, there are two proposed solutions.
In the first solution there is exchange of endpoint information and public keys via some trusted, out-of-band communication path by the users.
This model is referred to as the \textit{Freenet darknet model}.
The second solution achieves peer discovery and key exchange through a reliance on the \textit{XMPP federation} as the trusted medium.
The messages in the system are distributed in a push/pull format in followers in four ways namely, multicast push to followers, multicast pull by followers, random-walk push to distant followers and random-walk pull by followers.
Message privacy is done through the use of a permission flag that controls who can share the posts as well as a time-to-live (TTL) to control the scope of the updates.
Message verification and integrity is done through the use of signatures.
Followers are responsible for acquiring the publisher's public key via a trusted out-of-band system.

A disadvantage in such VPN networks is that the entry into the network is unclear.
Nodes without friends, just joining the network, do not find contact points to connect to.
Also, unfortunately, both security and trust initialization approaches, either through out-of-band channels or though centralized external XMPP servers, are not fully suitable for a fully decentralized OSN. Additionally, as only messaging is in the focus and routing is based on random-walks, the essential reliable data storage and retrieval is not provided.

\subsubsection{SuperNova}
\label{subsubsec:SuperNova}
SuperNova~\cite{ShD12} is a distributed OSN whose architecture relies on a super-peer-based network of volunteer agents.
The system is designed to provide flexibility in terms of storage, that is, users have the choice of where to store their content and whose content they want to store.
Additionally, users can choose the three level of access to other users: public hence accessible to all, private hence not accessible except to owner, and protected thus visible only to a chosen subset of friends.
This allows the system to provide full content ownership.
To deal with the problem of data unavailability in case a user is offline, SuperNova allows users to replicate their contents to a list of users called Storekeepers, which then make the data available when the user is unavailable.
Super-peers are nodes that provide services in the system and particularly to new nodes.
They essentially take part in the formation of the network's control infrastructure, and are the basic building block of the entire architecture as they provide lookup services, storage services, book-keeping services, recommendation services (such as for new friends or for storekeepers) among other services.

While the design addresses several requirements stated for a P2P-based OSN, the root of trust is undecided.
Users must trust the Storekeepers to treat their data properly.
Also, incentives for this cumbersome task are not given.

\subsubsection{LotusNet}
\label{subsubsec:LotusNet}
LotusNet~\cite{ARu12} is a proposed framework that allows for the development of P2P-based social network services.
Its goal is to provide support for strong user authentication and offers a solution for the trade-off between security, privacy and services within a DOSN. It achieves this through provision of users options for tuning privacy settings via a very flexible and fine-grained access control system.
The architecture also includes a suite of high-level services which support custom application development and mash ups.
LotusNet proposes to provide a reliable and secure P2P-based OSN by making some trade-offs on the OSN functional and non-functional requirements (Sections~\ref{subsec:SN_FuncReqs} and~\ref{subsec:SN_NonFuncReqs}).

The LotusNet architecture is based on a DHT which offers distributed storage thus allowing the implementation of social widgets to share and collect data.
The DHT used is Likir~\cite{AMR+08}, a customized version of Kademlia.
Likir is similar to other DHTs except that users need to fulfill a preliminary user registration procedure so as to receive a certified identifier for their DHT node.
The architecture of Likir makes this possible by provision of a centralized Certification Service (\textit{CS}) for this purpose.
The use of Likir offers two properties that are an advantage to the OSN. Firstly, the overlay communications are two-way authenticated.
The use of authentication along with binding of the user's identity to a fixed and random Kademlia ID effectively counteracts this threats such as Sybil attacks, resulting in a more robust P2P layer.
Secondly, Likir offers verifiable ownership of content by attaching certificates signed by the owner to every content published on the DHT. This allows for secure identity-based resource retrievals, resulting in a filtering facility that performs very sharp resource retrievals.
The use of authenticated interaction protocol and certificates makes it possible to meet security properties at the overlay level, similarly as in LibreSocial.

Directly on top of LotusNet's P2P layer is a custom suite of widgets which interact with each other by exchanging objects through the DHT that provides the essential social network services.
The provision of these services is directly based on the API of the overlay node which includes the identity management and authentication features.
The widgets are not restricted to communicate only via the DHT, but they can establish direct connections if needed.
In such situations, the distributed storage can be used for preliminary Diffie-Hellman exchange for the setting up of a secure out-of-band connection.
The resulting new secure channel is also authenticated and encrypted as the key agreement protocol is done on a fully authenticated layer.
Additionally, since identity management is at the overlay level, all data published by the same node are marked with the same user identity with no consideration on the nature of the widget that generated the content.
Therefore integration becomes easy since every widget is able to collect and aggregate content from the different widgets owned by the known social contacts by a simple method invocation.
This tackles the walled garden problem, since every application is potentially able to cooperate with other modules, that is, it guarantees maximum interoperability.

LotusNet preserves the privacy of the shared information by using signed grants.
As the grants are linked to social contacts rather than to shared resources, their numbers do not grow with respect to the resources owned or with the number of privacy policy rules.
LotusNet uses a \textit{Discretionary Access Control Module} (DACM), which is layered directly in the Likir node, to manage the individual social connections and to set privacy policies in assigning grants.
To further reduce risk of privacy violation for sensitive information, LotusNet allows tuning of the widgets' privacy level by storing the sensitive data at a set of trusted contacts specified by the user.

While LotusNet gains essential elements of its security through Likir, the use of a centralized Certification Service in Likir is disadvantageous.
A further limitation is that the access control is enforced through policies that are selected rather than individual settings for each data object.
Thus it is impossible to eventually fine tune the access rights on each document, or for example every picture, that is uploaded.

\subsubsection{Vegas} Vegas~\cite{DMD12} is an hybrid OSN utilizing an unstructured P2P network.
It was designed as a secure OSN that limits the access to a user's social graph to the ego network only.
It ensures that users users have full control over who can access their personal profile and published content by enforcing strong trust relationships that are mapped to the real world.
It also aims at offering mobility support while guaranteeing profile availability in cases when the user is offline.
The communication between P2P devices is done through \textit{exchangers} which are secure asynchronous communication channels that support delay-tolerant information exchange.
The implementation provides support for emailing, short messaging (SMS), instant messaging (XMPP) and microblogging (Twitter).
To ensure profile availability despite P2P anomalies like churn, Vegas utilizes \textit{datastores}, which enforces a write-one read-all storage policy, where only the owner can write.
Users can operate several datastores through which the friends can access the profile and shared content.
The messages are secured using public key pairs
$K^{-}/K^{+}$ called the \textit{link-specific} key pair.
Thus for a given friend X of user Y, Y applies to the message to encrypt it,
$K^{-}_{X}$ to sign it, and adds the fingerprint of
$K^{+}_{X}$ then sends it to the exchanger.
The identity of A is can only be verified by X since only X knows the mapping of the fingerprint.
To ensure access control for profile data during profile synchronization, a user applies a symmetric key to any given profile attribute, applies
$K^{+}$ for each of his friends to encrypt it, and then updates all the corresponding datastores.
In case of compromise on the link-specific key, 

The downside with Vegas is that it takes into account the desired security and privacy aspects as proposed by the authors at the expense of functionality.
In addition, it utilizes XMPP for instant messaging and Twitter for microblogging services which reintroduces the problems of centralized OSN to the entire setup.

\subsubsection{Decentralized OSN using P2P technology}
\label{subsubsec:DOSN_p2pTech}
Tran et al.~\cite{TNH15} propose a social network based on P2P architecture that supports social computing services in a distributed environment.
The goals of the proposed P2P based SN are to achieve scalability in architecture, reliability in content distribution and autonomy in administration.
It is also aimed at solving the problem of heterogeneity by using certain peers (called super peers) with more resources, such as storage, processing power and bandwidth, to support other peers with complex operations.

The proposed SN architecture is based on a super peer P2P network based on the Gnutella protocol to implement two authentication and posting services that bring out the ability of reducing reliance on centralized servers and also increase and encourage group communication in the social network.
The \textit{authentication service} allows users to access and use the network services.
The network includes and publishes several registration servers for user registration.
After registration, a list of super peers in the network is forwarded to the user and the super peers also receive an update on the registered user.
This allows the users to simply authenticate themselves on the super peers the next time they join.
In case the super peers are offline, users can authenticate with the registration servers and obtain an updated list of super peers.
Registration servers maintain a database storing details of the super peers and registered users, while the super peers maintain a database which stores registered users synchronized by the registration server.
Social activities of the users also helps in finding more super peers.
The users can then utilize the \textit{posting services} after authentication.
They can post messages to and request messages from other individual users or user groups.
These messages contain user statuses, user profiles and discussion updates.
Group communication is supported for updating discussions.
Group communication is implemented by defining the user group information based on user profiles.
The users then select to send messages to either the whole group, a set of users or only one user.
The posting service improves data privacy and search capability by allowing the users to keep personal data on peers and super peers.
Each peer has a MySQL database to store user data, peer data and messages.

Disadvantageous in this approach is the strong reliance on the registration server which has to maintain the overview on all super peers and all nodes.
Another stark limitation of the approach is that only messaging is supported, i.e.
nodes can push and pull data to/from each other, but there is not reliable data storage available.
Consequently, the option in building an appealing OSN on top are limited.

\subsubsection{HPOSN}
\label{subsubsec:HPOSN}
HPOSN~\cite{WLL+15} is an optimized hybrid OSN model based in P2P rather than a fully distributed OSN. It is designed to solve Local Service Fault Partition (LSFP), a situation in which the network is inaccessible because of fault partitioning caused by equipment/link failure or network attacks in centralized OSNs, which completely prevents users from logging into the network.
The designers propose leveraging P2P technologies along with the centralized servers to offer a solution to the LSFP problem, so that the application operates in centralized mode in normal conditions and has a supplemental P2P mode in cases where the LSFP problem occurs.
Data is stored in the servers as well as the local terminals.
However, only data that is considered important is stored in the local terminal, and an index pointing to the unimportant data stored at the server is maintained by the nodes.
Unlike centralized OSNs, HPOSN supports direct communication between nodes by leveraging SocialVPN to establish direct communications.
The system adopts Onion Routing~\cite{GRS99} to guarantee anonymous communications.
Data stored at the terminals and the servers is encrypted using asymmetric encryption, with the private key of the user being used to encrypt the data.

This proposal, although because it relies on the use of centralized servers, does not fully guarantee all aspects of privacy.
This is evident from the fact that the system providers still have some access to the private data and may employ data mining algorithms to find out more information about the users.
Also, because of the use of the servers as the main storage of the network, the scalability of the entire system is in question.

\subsection{Multi-overlay social networks} This class of P2P-based social networks utilize an architecture that implements two or more overlays, such as structured overlay based on a DHT and a social overlay, so as to realize efficient indexing and storage.
They are discussed hereafter.

\subsubsection{Cachet}
\label{subsubsec:Cachet}
Cachet~\cite{NJM+12} was designed to realize an OSN on a hybrid structured-unstructured overlay format, by augmenting a DHT with social links between users.
It main focus is on protection of the confidentiality, integrity and availability of the user content, while also ensuring privacy of the user relationships.
The architecture utilizes a DHT to enable decentralization, cryptographic techniques to enforce data confidentiality and objects to represent data.
It supports user profiles and wall features such as status updates, wall posts from the other social contacts, post comments as well as newsfeeds.
Policies are given through user identities or attributes.
The identity-based policies are set to define user-specific access while attribute-based policies define group-access of social contacts that share common features.
There are basically three types of policies defined on objects, namely, read, write and append policies, defined by the owner of the object at creation time and stored in the object metadata.

The access policies are enforced cryptographically through a hybrid scheme utilizing traditional public keys and attribute-based encryption (ABE).
With ABE, an object is encrypted using an AB policy.
The ABE scheme used for Cachet is an extended version of EASiER~\cite{JMB11} that provides supports for efficient revocation for Ciphertext Policy Attribute-based Encryption~\cite{BSW07} with the help of a minimally trusted proxy.
For the hybrid mode, message encryption is done using a randomly chosen symmetric encryption key, which is in turn encrypted with ABE. The Read policy is placed in the object reference instead of the object itself, thus enforcing policy privacy from storage nodes.
Therefore, the social graph is hidden from the storage nodes by ensuring that authorization does not reveal identities of users hence storage nodes are unaware of the identities of the users that store and retrieve data from it.
The Write and Append policies are enforced through access control of the corresponding signature keys.
Encryption for the Write policy key is by the object owner and Append policy with an AB policy.

Data is stored as an object in a DHT, such as Pastry~\cite{RoD01} or Kademlia~\cite{MMa02}, using a random object identifier
($objID$) as the DHT key.
Also, the storage nodes verify the Write Policy on objects.
The use of the DHTs covers certain desired features such as lookup and prevention of lookup attacks, availability replication, and prevention of malicious data overwrites using the write-policy verification.
Due to the basic construction of the base architecture, Cachet proposed a gossip-based social caching algorithm used in combination with an underlying DHT. This is used to leverage on the social trust relationships to improve the performance and reliability when downloading and reconstructing a social contact's wall or an aggregated newsfeed, a process which would otherwise be a lengthy process because of the decryption process.

While it is to highlight that Cachet fulfills the requirements we state for secure storage in decentralized OSNs, it does not specify the root of trust for the security mechanisms.
Through the use of social caching, data is spread but cannot be located reliably for an update or deletion.
Further, the functionality is limited as the focus is on secure data handling but does not cover communication options such as unicast, multicast or publish/subscribe.

\subsubsection{Twister}
\label{subsubsec:Twister}
Twister~\cite{Fre13,Fre16} is a microblogging architecture that leverages P2P technologies.
The goal of the design is to foster scalability, resiliency to failures and attacks, independence from central authority for user registration and provision of easily usable encrypted private communications and public posts.
The proposed system is made up of three mostly independent overlay networks.
The first overlay is based on the Bitcoin protocol~\cite{Nak08}.
This provides decentralized and secure user registration through the use of the Blockchain mechanism thus avoiding the need for a central authority.
The second P2P network is a structured DHT overlay network based on Kademlia~\cite{MMa02}.
It provides
$\langle
key,value\rangle$ storage for user resources and tracker location for the third network.
This DHT overlay allows for arbitrary resource storage and user retrievals, which includes profiles, avatars and posts.
The resource-to-peer mapping is based on a one-way hash function that ensures deterministic resource location while ensuring even content distribution across the network.
Also in order to enforce privacy and prevent compromise, the one-way hash function is performed on the user's IP address and port number instead of on the user's username only.
The third network is a collection of possibly disjoint \textquotedblleft swarms\textquotedblright~of followers.
This swarm mechanism is used for distributing new posts and it solves the problem of efficient notification delivery of new posts to users thus sparing the followers of the need to poll on a certain address of the DHT network to check for updates.
The swarm is a modified BitTorrent P2P unstructured overlay network.

One drawback of BitTorrent and variants is that while it is optimized for a fast delivery, it does not support data availability.
If no node is available in the swarm, the file is not available.
Also, the update or deletion of content is not considered, as it is not needed in a file sharing scenario, but essential in a social networking scenario.

\subsubsection{DiDuSoNet}
\label{subsubsec:DiDuSoNet}
DiDuSoNet~\cite{GAD+15} is a DOSN that is developed on a P2P overlay network with the aim of taking advantage of trust relationships to enforce certain services such as trustness, information diffusion and data availability.
It is a two-tier level system.
The first level is a Dunbar-based P2P social overlay.
The second level is the DHT. In the, Dunbar-based social overlay, connections between the nodes are akin to the social relations of the ego networks of the users which were first identified by the psychologist Robin Dunbar.
In social overlays (SOs), nodes of a P2P system only connect to one another if their owners are friends.
DiDuSoNet leverages a social aspect called Dunbar approach~\cite{Dun09} in the SOs, which considers the fact that a user stably maintains approximately 150 friend connections at any given time.
This number is referred to as the \textit{Dunbar number}.
An ego network~\cite{EvB05} is a network consisting of an actor (ego) and other actors that he is connected to (alters), and an ego network can be quite large.
By reducing an ego network using the Dunbar number, the result is a \textit{Dunbar-based ego network}.
The DHT makes lookup of other nodes easier and makes the system robust to churn.
The system used Pastry~\cite{RoD01} as the underlying overlay.
Atop the DHT, a data availability service is implemented, which autonomously selects two nodes in an ego network to store each published profile.
To search for profiles, a point of storage list called the PoS table stores the Overlay IDs of all PoSs for a given ID
($SocialID$).
Data stored inside the Dunbar-based Social Overlays is private and only visible to friends.
Private data is stored at the owner's node.
To hide the overlay IDs of their PoSs to prevent spying by undesired nodes, the authors suggest having the nodes use an attribute-based encryption (ABE) scheme or a ciphertext-policy attribute-based encryption (CP-ABE)~\cite{BSW07} scheme which mitigates the access of the PoSs list to selected friends only.
Also to prevent unauthorized access of data, the authors suggest the use of asymmetric keys.

However, in this work only the secure data storage is considered, further elements of a DOSN are left out, such as communication, applications and a real implementation.
Also, as previous examples, this solution requires that users share their data with their friends, which requires abandoning confidentiality for the sake of availability.

\subsubsection{SEDOSN}
\label{subsubsec:SEDOSN}
SEDOSN~\cite{FWS+15} was designed to provide a secure decentralized OSN framework based on P2P technology.
The application consists of three layers: an overlay network layer, function layer and the user interface layer.
The network layer is designed using TomP2P\footnote{https://tomp2p.net}, an open source DHT library, and a P2P network that connects the peers is built upon the physical network, making the peers independent of the physical network.
The function layer consists of four modules: a User Relationship Module made using SQLite (a lightweight database) for managing the metadata of the relationships of the users; the Attribute Encryption Module that utilizes RW's~\cite{RWa13} attribute-based encryption algorithms modified to include discretionary authorization so as to realize fine-grained access control on users' data; the BitTorrent Module for efficient transfer of shared file; and the Storage Service Module to store and get objects in the P2P network.
The User Interface layer is designed using the JavaFX technique which supports creating and delivery of rich internet applications.

The system focuses on ensuring secure file sharing without compromising the privacy setting of the users.
However, the system seems to lack advanced functionalities such as chatting and messaging which are standard in most OSNs, while offering only file sharing services.

\subsubsection{Blogracy}
\label{subsubsec:Blogracy}
Blogracy~\cite{FPT16} is a microblogging social networking system that is focused on achieving anonymity and resilience of censorship, content authentication and activity stream semantic interoperability.
It has a modular architecture built on two components: an underlying BitTorrent module for basic file sharing and an OpenSocial application programming interface (API).
The \textit{BitTorrent module} provides four key services to the OpenSocial container: StoreService for new key-value pairs storage request handling in the DHT; LookupService for searching values associated with a requested key in the DHT; SeedService for seeding newly shared file; and DownloadService for alerting users of the availability of a requested file.
The P2P file sharing mechanisms utilizes two logically separated DHTs.
The first DHT maps the user's identifier, including a reference, to his activity stream which is represented in a standard format that is encoded in a JSON file.
This JSON file contains a reference to the user's profile and references to user generated content, which are in the form of Magnet-URIs.
The references are the keys to the second DHT, which are resolved as actual files.
The \textit{OpenSocial module} implements the social aspects of Blogracy via a web application and as indicated, relies on the services that the BitTorrent module provides.
The OpenSocial containers design is based on the Model-View-Controller architecture.
The controller's function is to distribute responsibilities for various operations in key classes.
The system is also built to support core functionality extensions by use of autonomous agents thus providing recommendations on users and content, personalized results and trust negotiation mechanisms.

Blogracy strives to offer anonymity and pseudonymity while ensuring content is verifiable authentic and has integrity using two methods.
The first method is using the user's public key as the user's identifier.
Then the user signs his messages and indexes so that verification of authenticity and integrity are easily done by receivers.
The second method is the use of a cryptographic hash of the public key, and for Blogracy, the hash function corresponds to the one used by the DHTs.
The proposed system also utilizes attribute-based encryption to support protection against unauthorized access of data (posts, contacts, communications and activities).
The encryption scheme used is based on the Cyphertext-Policy Attribute-Based Encryption (CP-ABE) protocol~\cite{BSW07}.
Blogracy allows different levels of confidentiality for each individual social activity.
The content creator releases the content with parameterized attribute credentials directly to acknowledged followers by encrypting the content using the public key of the followers.
To support anonymity, Blogracy is implemented on I2P~\cite{Sch09}, which is an anonymizing P2P overlay network that implements a protocol resembling Tor~\cite{DMS04}.
Tor (the Onion Router) is a networking technology that is developed with the aim of guaranteeing some level of anonymity for the users by hiding their real network location.
Semantic interoperability is also possible as it uses activity streams and weak semantic data formats for contacts and profiles, hence can be integrated into existing social platforms such as Twitter and RSS-based content streams, as either data source or a data sink.

Blogracy thus pushes the content to the followers, which is typical to microblogging, but different to social networks, where data is pulled and browsed.
This requires a reliable data storage, which is not guaranteed in this case.

\section{Comparative analysis and Summary}
\label{sec:Anal&%
 Sum}

In this section, we consider the various aspects of the systems, taking into consideration the functional requirements and non-functional requirements.
We systematically assess each of the proposals and compare the contributions made and milestones met by each in realizing a fully distributed, secure and scalable OSN. This comparison will take into consideration two aspects that will be evident, that is, the \textit{overlay} (single-overlay distributed/hybrid and multi-overlay) and the \textit{services offered} (mixed services and microblogging).
The term \textit{mixed services} is used to denote an OSN that offers more than one type of service to the consumers such as chatting, messaging, audio-visual communications, (micro)blogging and so on.
\textit{Microblogging} systems in this case are only limited to offering a platform for microblogs to the consumers.
We also briefly consider the developmental timelines of the proposed solutions, inherent trends that may not be directly visible from, as well as the status of the proposal.

\subsection{OSN requirements \&
 system status} The functional and non-functional requirements for OSNs have been defined in Sections \ref{subsec:SN_FuncReqs} and \ref{subsec:SN_NonFuncReqs} respectively.
Accordingly, each P2P-based OSN has been objectively analyzed to show what aspects are met and the analysis is presented in Table~\ref{tab:SocCharp2pOSNs}.
Two important aspects of this analysis must be mentioned for clarity when looking at the were taken into account during the analysis.
The first aspect is on the requirements presented.
Based on the literature available for any proposal, in cases where a suggestion is made to use a particular solution in order to realize any desired functionality, the assumption was made that said solution was not implemented and consequently affected requirements were not met.
The second aspect is the system status.
Although it may immediately be assumed that the OSN may have actually been implemented, our analysis took into account the presence of irrefutable evidence of the existence of a prototype or system deployment.
The discussion that follows takes into consideration the type of overlay.

\begin{table*}
 \centering \tiny
 \begin{threeparttable}
  \caption{System status, functional and non-functional requirements}
  \label{tab:SocCharp2pOSNs}
  \begin{tabular}{|l|l|l|l|c||P{0.1cm}P{0.1cm}P{0.1cm}P{0.1cm}P{0.1cm}P{0.1cm}||P{0.1cm}P{0.1cm}P{0.1cm}P{0.1cm}|P{0.1cm}P{0.1cm}P{0.1cm}|P{0.4cm}|}
   \hline \rule{0pt}{2ex}\multirow{4}{*}{\textbf{Overlay Structure}}&%
   \multirow{4}{*}{\textbf{Type}}&%
   \multirow{4}{*}{\textbf{Proposal}}&%
   \multirow{4}{*}{\textbf{Services}}&%
   \multirow{4}{*}{\textbf{System status}}&%
   \multicolumn{14}{c|}{\textbf{Requirements}}\\
   \cline{6-19} \rule{0pt}{2ex}&%
   &%
   &%
   &%
   &%
   \multicolumn{6}{c||}{\textit{Functional}\footnotemark[1]}&%
   \multicolumn{8}{c|}{\textit{Non-functional}}\\
   \cline{6-19} \rule{0pt}{2ex}&%
   &%
   &%
   &%
   &%
   \multirow{2}{*}{\textit{PSM}}&%
   \multirow{2}{*}{\textit{SCM}}&%
   \multirow{2}{*}{\textit{SGT}}&%
   \multirow{2}{*}{\textit{Com}}&%
   \multirow{2}{*}{\textit{SSI}}&%
   \multirow{2}{*}{\textit{SF}}&
   \multicolumn{4}{c|}{\textit{Privacy}\footnotemark[2]}&%
   \multicolumn{3}{c|}{\textit{Security}\footnotemark[3]}&%
   \multicolumn{1}{c|}{\multirow{2}{*}{\textit{Metering}}}\\
   \cline{12-18} \rule{0pt}{2ex}&%
   &%
   &%
   &%
   &%
   &%
   &%
   &%
   &%
   &%
   &%
   \textit{Cf}&%
   \textit{OP}&%
   \textit{SIP}&%
   \textit{AP}&%
   \textit{CCA}&%
   \textit{DIA}&%
   \textit{NR}&%
   \\
   \hline\hline \rule{0pt}{2ex}Single-overlay distributed&%
   Structured&%
   LifeSocial.KOM/LibreSocial&%
   Mixed services&%
   Prototype&%
   \cmark&%
   \cmark&%
   \cmark&%
   \cmark&%
   \cmark&%
   \cmark&%
   \cmark&%
   \cmark&%
   \cmark&%
   \cmark&%
   \cmark&%
   \cmark&%
   \cmark&%
   \cmark\\
   \rule{0pt}{2ex}&%
   &%
   Porkut/My3&%
   Mixed services&%
   -&%
   \cmark&%
   \cmark&%
   \cmark&%
   \cmark&%
   \cmark&%
   \cmark&%
   &%
   \cmark&%
   \cmark&%
   \cmark&%
   &%
   \cmark&%
   \cmark&%
   \\
   \rule{0pt}{2ex}&%
   &%
   Megaphone&%
   Microblogging&%
   -&%
   \cmark&%
   \cmark&%
   \cmark&%
   \cmark&%
   &%
   \cmark&%
   \cmark&%
   &%
   &%
   &%
   \cmark&%
   \cmark&%
   \cmark&%
   \\
   \rule{0pt}{2ex}&%
   &%
   eXO&%
   Mixed services&%
   -&%
   \cmark&%
   \cmark&%
   \cmark&%
   \cmark&%
   \cmark&%
   \cmark&%
   &%
   \cmark&%
   \cmark&%
   \cmark&%
   \cmark&%
   &%
   \cmark&%
   \\
   \rule{0pt}{2ex}&%
   &%
   DECENT&%
   Mixed services&%
   -&%
   \cmark&%
   \cmark&%
   \cmark&%
   \cmark&%
   \cmark&%
   \cmark&%
   \cmark&%
   \cmark&%
   \cmark&%
   \cmark&%
   \cmark&%
   \cmark&%
   \cmark&%
   \\
   \rule{0pt}{2ex}&%
   &%
   HorNet&%
   Microblogging&%
   -&%
   \cmark&%
   \cmark&%
   \cmark&%
   \cmark&%
   \cmark&%
   \cmark&%
   \cmark&%
   &%
   &%
   &%
   \cmark&%
   \cmark&%
   \cmark&%
   \\
   \rule{0pt}{2ex}&%
   &%
   PESCA&%
   Mixed services&%
   -&%
   \cmark&%
   \cmark&%
   \cmark&%
   \cmark&%
   \cmark&%
   \cmark&%
   \cmark&%
   \cmark&%
   \cmark&%
   \cmark&%
   \cmark&%
   \cmark&%
   \cmark&%
   \\
   \rule{0pt}{2ex}&%
   &%
   WebP2P&%
   Mixed services&%
   Prototype&%
   \cmark&%
   \cmark&%
   &%
   \cmark&%
   &%
   \cmark&%
   \cmark&%
   \cmark&%
   \cmark&%
   \cmark&%
   \cmark&%
   \cmark&%
   \cmark&%
   \\
   \cline{2-19} \rule{0pt}{2ex}&%
   Unstructured&%
   PAC'nPOST&%
   Microblogging&%
   -&%
   \cmark&%
   \cmark&%
   \cmark&%
   \cmark&%
   &%
   \cmark&%
   &%
   &%
   &%
   &%
   &%
   &%
   &%
   \\
   \hline\hline \rule{0pt}{2ex}Single-overlay hybrid&%
   Structured&%
   PeerSoN&%
   Mixed services&%
   -&%
   \cmark&%
   \cmark&%
   &%
   \cmark&%
   \cmark&%
   \cmark&%
   \cmark&%
   &%
   &%
   &%
   \cmark&%
   &%
   &%
   \\
   \rule{0pt}{2ex}&%
   &%
   Safebook&%
   Mixed services&%
   -&%
   \cmark&%
   \cmark&%
   &%
   \cmark&%
   &%
   \cmark&%
   \cmark&%
   \cmark&%
   \cmark&%
   \cmark&%
   \cmark&%
   \cmark&%
   \cmark&%
   \\
   \rule{0pt}{2ex}&%
   &%
   Cuckoo&%
   Microblogging&%
   Prototype&%
   \cmark&%
   \cmark&%
   \cmark&%
   \cmark&%
   \cmark&%
   \cmark&%
   \cmark&%
   &%
   &%
   &%
   &%
   \cmark&%
   \cmark&%
   \\
   \rule{0pt}{2ex}&%
   &%
   LotusNet&%
   Mixed services&%
   -&%
   \cmark&%
   \cmark&%
   \cmark&%
   \cmark&%
   \cmark&%
   \cmark&%
   \cmark&%
   \cmark&%
   \cmark&%
   \cmark&%
   \cmark&%
   \cmark&%
   \cmark&%
   \\
   \cline{2-19} \rule{0pt}{2ex}&%
   Unstructured&%
   Litter&%
   Microblogging&%
   Prototype&%
   \cmark&%
   \cmark&%
   \cmark&%
   \cmark&%
   \cmark&%
   \cmark&%
   \cmark&%
   &%
   &%
   &%
   \cmark&%
   \cmark&%
   \cmark&%
   \\
   \rule{0pt}{2ex}&%
   &%
   SuperNova&%
   Mixed services&%
   -&%
   \cmark&%
   \cmark&%
   \cmark&%
   \cmark&%
   \cmark&%
   \cmark&%
   &%
   &%
   &%
   &%
   &%
   &%
   &%
   \\
   \rule{0pt}{2ex}&%
   &%
   Vegas&%
   Microblogging&%
   Prototype&%
   \cmark&%
   \cmark&%
   &%
   \cmark&%
   &%
   \cmark&%
   \cmark&%
   \cmark&%
   \cmark&%
   \cmark&%
   \cmark&%
   \cmark&%
   \cmark&%
   \\
   \rule{0pt}{2ex}&%
   &%
   Tran et al.\cite{TNH15}&%
   Mixed services&%
   -&%
   \cmark&%
   \cmark&%
   \cmark&%
   \cmark&%
   \cmark&%
   \cmark&%
   &%
   &%
   &%
   &%
   &%
   &%
   &%
   \\
   \rule{0pt}{2ex}&%
   &%
   HPOSN&%
   Mixed services&%
   -&%
   \cmark&%
   \cmark&%
   &%
   \cmark&%
   &%
   \cmark&%
   \cmark&%
   \cmark&%
   \cmark&%
   \cmark&%
   \cmark&%
   \cmark&%
   \cmark&%
   \\
   \hline\hline \rule{0pt}{2ex}Multi-overlay&%
   &%
   Cachet&%
   Mixed services&%
   -&%
   \cmark&%
   \cmark&%
   &%
   \cmark&%
   &%
   \cmark&%
   \cmark&%
   \cmark&%
   \cmark&%
   \cmark&%
   \cmark&%
   \cmark&%
   \cmark&%
   \\
   \rule{0pt}{2ex}&%
   &%
   Twister&%
   Microblogging&%
   Deployed&%
   \cmark&%
   \cmark&%
   &%
   \cmark&%
   &%
   \cmark&%
   \cmark&%
   \cmark&%
   &%
   &%
   \cmark&%
   \cmark&%
   \cmark&%
   \\
   \rule{0pt}{2ex}&%
   &%
   DiDuSoNet&%
   Mixed services&%
   -&%
   \cmark&%
   \cmark&%
   \cmark&%
   \cmark&%
   \cmark&%
   \cmark&%
   &%
   &%
   &%
   &%
   &%
   &%
   &%
   \\
   \rule{0pt}{2ex}&%
   &%
   SEDOSN&%
   Mixed services&%
   Prototype&%
   \cmark&%
   \cmark&%
   &%
   \cmark&%
   &%
   \cmark&%
   \cmark&%
   \cmark&%
   \cmark&%
   \cmark&%
   \cmark&%
   \cmark&%
   \cmark&%
   \\
   \rule{0pt}{2ex}&%
   &%
   Blogracy&%
   Microblogging&%
   Prototype&%
   \cmark&%
   \cmark&%
   &%
   \cmark&%
   &%
   \cmark&%
   \cmark&%
   \cmark&%
   \cmark&%
   \cmark&%
   \cmark&%
   \cmark&%
   \cmark&%
   \\
   \hline
  \end{tabular}
  \begin{tablenotes}
  \item[1]
   Functional requirements: (\textbf{PSM}-Personal storage management, \textbf{SCM}-Social connection management, \textbf{SGT}-Social graph traversal, \textbf{Com}-Means of communication, \textbf{SSI}-Shared Storage Space Interaction, \textbf{SF}-Search facilities)
  \item[2]
   Non-functional requirement: Privacy (\textbf{Cf} -Confidentiality, \textbf{OP}-Ownership Privacy, \textbf{SIP}-Social Interaction Privacy, \textbf{AP}-Activity Privacy)
  \item[3]
   Non-functional requirement: Security (\textbf{CCA}-Cover Channel Authentication, \textbf{DIA}-Data Integrity and Authenticity, \textbf{NR}-Non-Repudiation)
  \end{tablenotes}
 \end{threeparttable}
\end{table*}

\subsubsection{Single-overlay distributed OSNs} These OSNs are designed on a single overlay (structured or unstructured) and rely on a distributed indexing mechanism for resource location.
We consider the solutions based on the type of structure individually.
\begin{enumerate}[label=\itshape\alph*\upshape)]
\item
 \textit{Structured overlays}: The indexing mechanism are based on keys and hence most of the solutions proposed are DHT-based.
 In this group, there were nine (8) OSNs that were identified, of which two, Megaphone and HorNet were microblogs, while the rest, LibreSocial, Porkut/My3, eXO, DECENT, PESCA and WebP2P all offer mixed services.
 In terms of achieving the functional requirements, with the exception of Megaphone and WebP2P, all the remaining six proposals met all the requirements.
 However, the numbers of the proposals that meet all the non-functional requirements is drastically different, with only LibreSocial standing out in this category.
 Complete privacy is not achieved by Porkut/My3, Megaphone, eXO and HorNet, and complete security is not met by Porkut/My3 and eXO. Metering is only implemented in LibreSocial.
 Of these OSNs, only LibreSocial and WebP2P have prototypes.
\item
 \textit{Unstructured overlays}: There was only one proposal in this category, that is PAC'nPOST which is a microblog.
 It did not meet all the six functional requirements, lacking shared storage space interaction.
 It also fails to meet all the non-functional requirements, in addition to the fact that the system status is not known.
\end{enumerate}

\subsubsection{Single-overlay hybrid OSNs} The OSNs in this category are also designed on a single overlay (structured or unstructured), while the indexing mechanisms relies on a hybrid of distributed and centralized mechanisms.
Because of the incorporation of centralized solutions in these OSNs, there is a general tendency not to meet all the requirements as the centralized mechanisms re-introduce some of the challenges faced in centralized OSNs.
The solutions are discussed based on the base overlay.
\begin{enumerate}[label=\itshape\alph*\upshape)]
\item
 \textit{Structured overlays}: Four proposals are discussed here, PeerSoN, Safebook and LotusNet being OSNs with mixed services and Cuckoo being a microblog.
 LotusNet meets all the functional requirements as well as complying with privacy and security requirements fully.
 Cuckoo meets all functional requirement but not the security and privacy requirements, although it is the only solution with a prototype.
 Safebook guarantees privacy and security although it does not meet some functional requirements.
 However, none of the OSNs offers metering as a non-functional requirement.
\item
 \textit{Unstructured overlays}: The proposals in this classification are SuperNova, HPOSN and the proposal by Tran et al.~\cite{TNH15} offering mixed services, and Litter and Vegas as the microblogs, which incidentally are the only proposals that have prototypes.
 Three of these proposals, Litter, SuperNova and Tran et al.~\cite{TNH15} meet all functional requirements.
 Two proposals Vegas and HPOSN achieve all privacy requirements as well as security requirements.
 Litter achieves all security requirements although it does not meet the privacy requirements.
 Metering requirement is not met by any of the proposals.
\end{enumerate}

\subsubsection{Multi-overlay OSNs} This group of OSNs is interesting because the solutions are designed to utilize more than one overlay so achieve functionality.
The solutions seek to combine the advantages offered by different overlays in combination to overcome the disadvantages seen in each individual overlay.
The overlays utilized may be structured only, unstructured only, or a combination of structured and unstructured overlays.
Five proposals are analysed that fall in this category, that is, Cachet, DiDuSoNet and SEDOSN that offer mixed services, and Twister and Blogracy which are microblogs.
SEDOSN and Blogracy have prototype implementations while Twister\footnote{http://twister.net.co} is the only P2P OSN that is in active deployment.
DiDuSoNet is the only proposal that meets all the functional requirements, but is the only proposal in this category that fails in meeting all the non-functional requirements.
From the remaining four proposals, only Twister fails to meet all privacy requirements.
No proposal includes the ability for metering for measurements of system health.
\\
One interesting aspect that is visible from a cursory look at the functional requirements is that at a minimum, all proposed solutions met four functional requirements, that is, personal storage management, social connection management, communication and search facilities.

\subsection{Developmental progression}
\label{subsec:DevProgress}
In \cite{PZD11}, the authors analyzed selected OSNs from 1997 to 2011, indicating a boom between 2003 and 2006.
Table \ref{tab:p2pSNs_State} is a summary of the P2P OSNs highlighting the timelines and current status of the proposed implementation over the period of the analysis (2008 to 2016).
The boom of the P2P OSN platforms is seen to be clustered between 2011 and 2013, with the majority being in the year 2012.
Although the P2P OSNs may seem to be a decade too late, it is most probable that their development may be a direct result of concerns observed in centralized OSNs during the previous decade, in particular, privacy, security and scalability.
As has been shown, the different P2P OSNs aim at meeting one or more of these concerns using various techniques.
It is also evident that most of these systems were developed in academic environments.

\begin{table}[t]
 \tiny \centering
 \caption{Developmental timeline of the P2P OSN proposals}
 \label{tab:p2pSNs_State}
 \begin{tabular}{|P{0.4cm}|p{1.2cm}|l|p{4.1cm}|}
  \hline \rule{0pt}{2ex}\multirow{1}{*}{\textbf{Year}} &
  \multirow{1}{*}{\textbf{Proposal}} &
  \multirow{1}{*}{\textbf{Services}} &
  \multirow{1}{*}{\textbf{Institution}} \\
  \hline\hline \rule{0pt}{2ex}2008&%
  LifeSocial.KOM / LibreSocial &%
  Mixed services&%
  TU Darmstadt \\
  \hline \rule{0pt}{2ex}2009&%
  PeerSoN&%
  Mixed services&%
  TU Berlin/EPFL/NTU Singapore \\
  \rule{0pt}{2ex}&%
  Safebook&%
  Mixed services&%
  TU Darmstadt \\
  \hline \rule{0pt}{2ex}2010&%
  Cuckoo&%
  Microblogging&%
  Univ.
  of G{\"o}ttingen/Nanjing Univ./Fudan Univ.
  \\
  \rule{0pt}{2ex}&%
  Megaphone&%
  Microblogging&%
  California State University Long Beach \\
  \rule{0pt}{2ex}&%
  Porkut&%
  Mixed services&%
  EPFL, Switzerland \\
  \hline \rule{0pt}{2ex}2011&%
  eXO&%
  Mixed services&%
  Univ.
  of Patras/Univ.
  of Ioannina \\
  \rule{0pt}{2ex}&%
  My3&%
  Mixed services&%
  EPFL, Switzerland \\
  \rule{0pt}{2ex}&%
  Litter&%
  Microblogging&%
  Univ.
  of Florida \\
  \hline
  \rule{0pt}{2ex}2012&%
  Vegas&%
  Microblogging&%
  Ludwig-Maximilians-University Munich\\
  \rule{0pt}{2ex}&%
  Cachet&%
  Mixed services&%
  Univ.
  of Illinois \\
  \rule{0pt}{2ex}&%
  DECENT&%
  Mixed services&%
  Univ.
  of Illinois \\
  \rule{0pt}{2ex}&%
  HorNet&%
  Microblogging&%
  Univ.
  Oberta de Catalunya \\
  \rule{0pt}{2ex}&%
  LotusNet&%
  Mixed services&%
  Univ.
  degli Studi di Torino \\
  \rule{0pt}{2ex}&%
  PAC\textquoteright nPOST&%
  Microblogging&%
  Univ.
  College London \\
  \rule{0pt}{2ex}&%
  SuperNova&%
  Mixed services&%
  NTU Singapore \\
  \hline
  \rule{0pt}{2ex}2013&%
  twister&%
  Microblogging&%
  PUC-Rio \\
  \hline \rule{0pt}{2ex}2015&%
  PESCA&%
  Mixed services&%
  Isfahan Uni.
  of Tech./Foulad Inst.
  of Tech./Ryerson Univ.\\
  \rule{0pt}{2ex}&%
  Tran et al.\cite{TNH15}&%
  Mixed services&%
  Intl.
  Univ., Ho Chi Minh \\
  \rule{0pt}{2ex}&%
  DiDuSoNet&%
  Mixed services&%
  Univ.
  of Pisa/IIT-CNR Pisa/Univ.
  of D{\"u}sseldorf \\
  \rule{0pt}{2ex}&%
  WebP2P&%
  Mixed Services&%
  Univ.
  of D{\"u}sseldorf \\
  \rule{0pt}{2ex}&%
  SEDOSN&%
  Mixed Services&%
  Peking University \\
  \rule{0pt}{2ex}&%
  HPOSN&%
  Mixed Services&%
  Shandong Normal University \\
  \hline \rule{0pt}{2ex}2016&%
  Blogracy&%
  Microblogging&%
  Univ.
  of Parma \\
  \hline
 \end{tabular}
\end{table}

\subsection{Essential P2P components} In Section~\ref{sec:P2PNws}, a general overview was given in consideration of key components that must be considered during the design of any P2P-based OSN. In line with that, in Table \ref{tab:OSNInfrastructure}, we present a summary of the components realised in the surveyed P2P OSNs.
The key components that were considered were the overlay, storage mechanisms implemented, lookup/search mechanisms, data redundancy mechanisms utilized and inclusion of a publish/subscribe mechanism.

\begin{table*}[]
 \tiny \centering
 \caption{Components realised in the P2P-based OSNs}
 \label{tab:OSNInfrastructure}
 \begin{tabular}{|p{1.3cm}|p{1.0cm}|p{1.2cm}||P{1.7cm}|P{2.5cm}|P{2.5cm}|P{2.5cm}|P{1.0cm}|}
  \hline \rule{0pt}{2ex}\multirow{2}{*}{\textbf{Structure}}&
  \multirow{2}{*}{\textbf{Type}}&
  \multirow{2}{*}{\textbf{Proposal}}&
  \multicolumn{5}{c|}{\textbf{Infrustructural Aspects}} \\
  \cline{4-8} \rule{0pt}{2ex}&%
  &%
  &%
  \textit{Overlay}&
  \textit{Storage Mechanism}&
  \textit{Lookup/Search Mechanism}& \textit{Data Redundancy Mechanism}&
  \textit{Pub/Sub Mechanism} \\
  \hline\hline \rule{0pt}{2ex}\textbf{Single-overlay distributed}&%
  \textit{Structured}&%
  LifeSocial.KOM/ LibreSocial&%
  Pastry&%
  Standard - PAST; Advanced - Sets, linked lists, prefix hash trees&%
  Semantic-free&%
  Replication \&
  caching&%
  \cmark \\
  \rule{0pt}{2ex}&%
  &%
  Porkut/My3&%
  OpenDHT&%
  Trusted Proxy Set (TPS)&%
  Semantic-free&%
  Replication to TPS&%
  \cmark \\
  \rule{0pt}{2ex}&%
  &%
  Megaphone&%
  Pastry&%
  Local storage&%
  Semantic-free&%
  Replication&%
  \cmark \\
  \rule{0pt}{2ex}&%
  &%
  eXO&%
  Pastry/Chord&%
  Local storage&%
  Semantic free&%
  Replication to adjacent nodes&
  \\
  \rule{0pt}{2ex}&%
  &%
  DECENT&%
  Pastry/Kademlia&%
  DHT for object storage&%
  Semantic-free&%
  Replication \&
  versioning&%
  \cmark \\
  \rule{0pt}{2ex}&%
  &%
  HorNet&%
  Pastry&%
  Storage service API based on DHT&%
  Semantic-free&%
  Replication&%
  \cmark \\
  \rule{0pt}{2ex}&%
  &%
  PESCA&%
  DHT-based&%
  Local storage&%
  Semantic-free&%
  Replication based on friends' user online times&%
  \xmark \\
  \rule{0pt}{2ex}&%
  &%
  WebP2P&%
  Chord&%
  Local storage&%
  Semantic-free&%
  Replication&%
  \xmark \\
  \cline{2-8} \rule{0pt}{2ex}&%
  \textit{Unstructured}&%
  PAC'nPOST&%
  Unstructured&%
  Local storage&%
  Semantic (probabilistic search)&%
  Replication&%
  \cmark \\
  \hline\hline \rule{0pt}{2ex}\textbf{Single-overlay hybrid}&%
  \textit{Structured}&%
  PeerSoN&%
  OpenDHT&%
  Local storage&%
  Semantic-free&%
  Replication&%
  \xmark \\
  \rule{0pt}{2ex}&%
  &%
  Safebook&%
  KAD&%
  Nodes in concentric ``Matryoshka''-like circles&%
  Semantic-free&%
  Replication&%
  \xmark \\
  \rule{0pt}{2ex}&%
  &%
  Cuckoo&%
  Pastry&%
  Local and server cloud&%
  Semantic (flooding) for influentials else semantic-free&%
  Replication to servers&%
  \cmark\\
  \rule{0pt}{2ex}&%
  &%
  LotusNet&%
  Likir&%
  Local and set of trusted contacts&%
  Semantic-free&%
  Replication on trusted contacts&%
  \cmark \\
  \cline{2-8} \rule{0pt}{2ex}&%
  \textit{Unstructured}&%
  Litter&%
  Social overlay&%
  Local storage&%
  Semantic (pseudo-random walk)&%
  Replication to one-hop peers&%
  \cmark \\
  \rule{0pt}{2ex}&%
  &%
  SuperNova&%
  Super-peers&%
  List of users&%
  Search queries sent to super peers&%
  Replication&%
  \xmark \\
  \rule{0pt}{2ex}&%
  &%
  Vegas&%
  Unstructured&%
  Web-based datastores&%
  Search queries to subset of friends&%
  Replication of datastores&%
  \cmark \\
  \rule{0pt}{2ex}&%
  &%
  Tran et al.\cite{TNH15}&%
  Gnutella&%
  MySQL database at local nodes&%
  Semantic (flooding with TTL)&%
  Replication&%
  \cmark \\
  \rule{0pt}{2ex}&%
  &%
  HPOSN&%
  Social overlay&%
  Local storage and cloud servers&%
  Queries to server based on stored index&%
  Replication on servers&%
  \xmark \\
  \hline\hline \rule{0pt}{2ex}\textbf{Multi-overlay}&%
  &%
  Cachet&%
  Pastry/Kademlia \&
  Social overlay&%
  Local storage&%
  Semantic-free&%
  Replication&%
  \cmark \\
  \rule{0pt}{2ex}&%
  &%
  Twister&%
  Bitcoin, Kademlia, BitTorrent swarm&%
  Local storage, BitTorrent network&%
  Semantic-free&%
  Replicaion&%
  \cmark \\
  \rule{0pt}{2ex}&%
  &%
  DiDuSoNet&%
  Pastry \&
  social overlay&%
  Local storage&%
  Semantic-free&%
  Replication to ego network&%
  \cmark \\
  \rule{0pt}{2ex}&%
  &%
  SEDOSN&%
  TomP2P \& BitTorrent&%
  Local storage, SQLite database for metadata&%
  BitTorrent protocol (Client/Tracker)&%
  -&%
  \xmark \\
  \rule{0pt}{2ex}&%
  &%
  Blogracy& Two BitTorrent DHTs&%
  Local storage&%
  Semantic-free&%
  Replication&%
  \cmark\\
  \hline
 \end{tabular}
\end{table*}

\subsubsection{Single-overlay distributed OSNs} We consider the structured and unstructured overlays separately and discuss them herein.
\begin
 {enumerate} [label=\itshape\alph*\upshape)]
\item
 \textit{Structured}: Most of the proposals were based on common DHT-based overlays, such as Pastry, Chord and Kademlia, to form the network.
 Therefore, the lookup mechanics were based on key-based routing, hence semantic free.
 The proposals all implemented replication in differing ways so as to guarantee data availability, as the local nodes replicated the local data to other nodes in the network using appropriate algorithms.
 Only LibreSocial included advanced storage features, that is, distributed sets, linked lists and prefix hash trees.
 Only WebP2P and eXO did not include the publish/subscribe mechanisms as part of the features.
\item
 \textit{Unstructured}: The only solution here, PAC'nPOST is based on a purely distributed overlay, hence pure P2P (unstructured).
 Therefore, semantic searching using probabilistic search methods is performed.
 Each node handles its own data but replicates it to other nodes to for data availability guarantees, in addition to implementing a pub/sub mechanism.
\end
{enumerate}

\subsubsection{Single-overlay hybrid OSNs} The component discussion here also considers the base overlay and follows.
\begin
 {enumerate} [label=\itshape\alph*\upshape)]
\item
 \textit{Structured}: These OSNs had differing methods in how the storage is handled, but all use replication to support data availability.
 PeerSoN uses local storage with replication to other nodes based on OpenDHT's algorithm.
 Safebook utilizes nodes directly connected to the local node and arranged in a concentric circle as the replicating nodes, while Cuckoo uses the server cloud to support data replication.
 LotusNet relies on the local DHT and replicates to trusted contacts.
 PeerSoN, Safebook and LotusNet rely on semantic-free lookup mechanisms for locating data/objects, with Cuckoo relying on a combination of semantic-free lookup and semantic-based search.
 Finally, both Cuckoo and LotusNet incorporate a pub/sub mechanism.
\item
 \textit{Unstructured}: The OSNs in this category implement different techniques to handle storage, and no two OSNs have the same method.
 This is expected in unstructured overlays as the network does not offer any distributed data structures such as DHTs, but allows the designers to develop novel techniques to handle data management.
 Data availability is guaranteed by replication in all cases, and with the exception of SuperNova and HPOSN, the remaining OSNs integrate a pub/sub mechanism.
\end
{enumerate}

\subsubsection{Multi-overlay OSNs} The proposed OSNs in this group all incorporate an additional data storage mechanisms in addition to the local storage that relies on the DHT mechanisms.
Twister incorporates the BitTorrent network and SEDOSN incorporates an SQLite database to handle metadata, while the rest rely solely on the local storage.
All solutions with the exception of SEDOSN, for which no information was provided, ensure data availability via replication, and similarly with the exception of SEDOSN, all include a pub/sub mechanism.

\subsection{Security considerations} Any discussion about OSNs without paying special attention to the aspect of security management is incomplete.
In Table \ref{tab:p2pOSN_Security}, a summary of key security features identified in the analyzed P2P OSNs is shown.
At the minimum, it is desired that they provide some form of identity creation and verification, include an access control mechanism, guarantee confidentiality and ensure data integrity, while ensuring user anonymity.
In the survey of the P2P OSNs, it is seen that only LibreSocial, DECENT, LotusNet, Safebook, Cachet, SEDOSN and Blogracy incorporate all the required security mechanisms to ensure secure communications and guarantee user privacy.
However, it is important to note that the microblogs generally have a tendency to not implement all the security requirement because, as an unwritten rule, microblogs do not guarantee privacy, as users are able to access all the messages in the network, and in many cases, view the profiles of other users whether known directly/indirectly or not known and as well as follow/unfollow other users.


\begin{table*}[]
 \tiny \centering
 \caption{Security features of the P2P-based OSNs}
 \label{tab:p2pOSN_Security}
 \begin{tabular}{|p{1.3cm}|p{1.0cm}|p{1.2cm}|P{1.8cm}|P{1.8cm}|P{2.0cm}|P{2.0cm}|P{1.3cm}|P{0.6cm}|}
  \hline \rule{0pt}{2ex}\multirow{2}{*}{\textbf{Overlay Structure}}&%
  \multirow{2}{*}{\textbf{Structure type}}&%
  \multirow{2}{*}{\textbf{Proposal}}&%
  \multicolumn{6}{c|}{\textbf{Security Aspects}} \\
  \cline{4-9} \rule{0pt}{2ex}&%
  &%
  &%
  \textit{Identity creation}&%
  \textit{Identity verification}&%
  \textit{Access Control}&%
  \textit{Confidentiality}&%
  \textit{Integrity}&%
  \textit{Anonymity} \\
  \hline\hline \rule{0pt}{2ex}\textbf{Single-overlay distributed}&%
  \textit{Structured}&%
  LifeSocial.KOM/ LibreSocial&%
  Public key as Unique ID&%
  Public key&%
  User-centric settings + Access Control Lists&%
  Symmetric encryption&%
  Digital signatures&%
  \cmark \\
  \rule{0pt}{2ex}&%
  &%
  Porkut/My3&%
  No information&%
  No information&%
  \xmark&%
  \xmark&%
  \xmark&%
  \cmark \\
  \rule{0pt}{2ex}&%
  &%
  Megaphone&%
  Concatenated hash of username and public key&%
  Public key&%
  Session keys + Asymmetric encryption&%
  Asymmetric encryption&%
  Digital signatures&%
  \xmark \\
  \rule{0pt}{2ex}&%
  &%
  eXO&%
  Hash on user-specific detail&%
  User ID&%
  User-centric configurations&%
  \xmark&%
  \xmark&%
  \cmark \\
  \rule{0pt}{2ex}&%
  &%
  DECENT&%
  Random ID as User ID&%
  User ID&%
  Attribute-based policies&%
  Asymmetric attribute-based encryption&%
  Digital signatures&%
  \cmark \\
  \rule{0pt}{2ex}&%
  &%
  HorNet&%
  Public key&%
  Public key certificate&%
  Access control list&%
  \xmark&%
  Digital signatures&%
  \xmark \\
  \rule{0pt}{2ex}&%
  &%
  PESCA&%
  Hash of user's email address&%
  Global ID&%
  Virtual ID + symmetric key&%
  Broadcast encryption&%
  Digital signatures&%
  \xmark \\
  \rule{0pt}{2ex}&%
  &%
  WebP2P&%
  Asymmetric key from username and passwordA&%
  Public key as Chord ID&%
  Identity-based access control&%
  Asymmetric encryption&%
  Digital signatures&%
  \xmark \\
  \cline{2-9} \rule{0pt}{2ex}&%
  \textit{Unstructured}&%
  PAC'nPOST&%
  No information&%
  No information&%
  \xmark&%
  \xmark&%
  \xmark&%
  \xmark \\
  \hline \rule{0pt}{2ex}\textbf{Single-overlay hybrid}&%
  \textit{Structured}&%
  PeerSoN&%
  Hash of user's email&%
  Globally unique ID (GUID)&%
  \xmark&%
  Asymmetric encryption&%
  \xmark&%
  \xmark \\
  \rule{0pt}{2ex}&%
  &%
  Safebook&%
  TIS-generated ID and pseudonym&%
  Node ID and pseudonymn&%
  Attribute-based policies&%
  Asymmetric encryption&%
  Digital signatures&%
  \cmark \\
  \rule{0pt}{2ex}&%
  &%
  Cuckoo&%
  Server generated ID&%
  Not required&%
  \xmark&%
  \xmark&%
  \xmark&%
  \xmark \\
  \rule{0pt}{2ex}&%
  &%
  LotusNet&%
  Public key and OpenID to obtain certificate&%
  Likir ID&%
  Access control grants&%
  Nonce-based two-way authentication&%
  Digital signatures&%
  \cmark \\
  \cline{2-9} \rule{0pt}{2ex}&%
  \textit{Unstructured}&%
  Litter&%
  No information&%
  User ID (UID)&%
  \xmark&%
  Encrypted IP tunnels&%
  Digital signatures&%
  \xmark \\
  \rule{0pt}{2ex}&%
  &%
  SuperNova&%
  Username, location and interests&%
  Userlist at superpeers&%
  \xmark&%
  Threshold-based secret sharing\cite{VAB09}&%
  \xmark&%
  \xmark \\
  \rule{0pt}{2ex}&%
  &%
  Vegas&%
  No information&%
  No information&%
  Asymmetric encryption&%
  Symmetric + asymmetric encryption&%
  Digital signatures&%
  \cmark \\
  \rule{0pt}{2ex}&%
  &%
  Tran et al.~\cite{TNH15}&%
  Registration servers&%
  Super peers&%
  super peer controlled&%
  \xmark&%
  \xmark&%
  \xmark \\
  \rule{0pt}{2ex}&%
  &%
  HPOSN&%
  Servers&%
  Globally unique ID&%
  Asymmetric encryption&%
  Asymmetric encryption (Onion routing)&%
  \xmark&%
  \xmark\\
  \hline \rule{0pt}{2ex}\textbf{Multi-overlay}&%
  &%
  Cachet&%
  User ID&%
  User ID&%
  Attribute-based policies&%
  Asymmetric attribute-based encryption&%
  Digital signatures&%
  \cmark \\
  \rule{0pt}{2ex}&%
  &%
  Twister&%
  Unique user ID&%
  Username and password&%
  \xmark&%
  Ellliptic Curve Integrated Encryption Scheme&%
  Digital signatures&%
  \xmark \\
  \rule{0pt}{2ex}&%
  &%
  DiDuSoNet&%
  Unique Social ID&%
  Social ID&%
  \xmark&%
  \xmark&%
  \xmark&%
  \xmark \\
  \rule{0pt}{2ex}&%
  &%
  SEDOSN&%
  User's email address&%
  Global ID&%
  Attribute-based encryption&%
  Symmetric encryption&%
  Digital signatures&%
  \cmark \\
  \rule{0pt}{2ex}&%
  &%
  Blogracy&%
  Public key and username&%
  Public key&%
  Ciphertext-policy attribute-based encryption&%
  Asymmetric encryption&%
  Digital signatures&%
  \cmark\\
  \hline
 \end{tabular}
\end{table*}

\section{Lessons Learned}
\label{sec:LessonsLearnt}
Building social networks that are designed to operate in a fully distributed environment is not a new idea and has been studied quite extensively.
In particular, using P2P networks as a platform for building decentralized online social networks (DOSNs) has been taunted as a solution to the problems due to the accumulated costs for centralized operations~\cite{GCR13,MKI+16} and security and privacy concerns~\cite{KIa17,ORP19}.
However, as has been show, any functional P2P-based OSN must at least achieve desired functional requirements for OSNs and at best the non-functional requirements so as to effectively address the concerns raised (see sections~\ref{subsec:SN_FuncReqs} and~\ref{subsec:SN_NonFuncReqs}) while ensuring users enjoy the best services .
During the course of undertaking this study and compiling this survey, some lessons were learned that are worth considering.

\paragraph*{Need for agreeable P2P protocol standards}
Different proposed solutions achieve the functional and non-functional requirements using different combinations of P2P mechanisms.
This is an indication of the need for adoption of a standard for P2P technologies.
An attempt has been made to create a standard by the Internet Architecture Board (IAB) that was published as RFC 5694~\cite{rfc5694} in November 2009.
However, this standard was based on the early P2P technologies which have since experienced a considerable metamorphosis.
Therefore, a newer standard must be tabled within the research community and eventually adopted.
That said, the fact that there is much that has been done is an indication of the diversity of solutions for any given problem in P2P technology, and the ease in which it can be adapted to achieve a desired functionality.
In general, most applications, and in particular OSNs, designed on the P2P platform, seem to follow a general format for the architecture: \textit{an overlay}, required services on top of the overlay (here referred to as \textit{the framework}) and \textit{the application}.
Similar layouts have been suggested in~\cite{PFS14} This may well be a precursor to an adoptable standard for P2P applications.
In line with this need to achieve a standard, we have undertaken a general survey of individual P2P component solutions that have been proposed in literature that we see meet the basic technical requirements for the P2P architecture (defined in section~\ref{sec:InfraSupport}).

\paragraph*{From research systems to active systems}
Research on P2P-based OSN has matured and many of the proposals have to a large degree aimed at solving one or more challenge experienced in the centralized OSNs in a unique manner.
As yet, with few exceptions, the viability and behavior of these proposed systems in the real world is yet to be seen.
In order to compete with the centralized OSNs, the designers of the P2P-based OSNs must strive to ensure that the product they offer gets to the users and find ways to inform users of their presence.
Although in many cases the focus has been on a secure and private solution, there is need for finding a balance between security/privacy and system usability.
This may be achievable by having the proposal deployed and monitoring the behavior of the users against the behavior of the system and then making appropriate adjustments.

\paragraph*{Motivating user communities}
Even though the proposed systems are brought online, it is another thing to get users to start using them.
DOSNs have been around for some time yet the user communities have not grown at the same rate as centralized OSNs.
DOSNs founded on federated solutions such as Diaspora\footnote{https://diasporafoundation.org/}, Friendica\footnote{https://friendi.ca/} and Mastodon\footnote{https://joinmastodon.org/} have been able to get user communities but the numbers are far much smaller than those of the centralized OSNs.
Generally, DOSNs promise to offer many features such as more security and privacy control, but the greatest uphill task faced is convincing the users of centralized OSN users to migrate and use them, as the centralized OSNs have large, established user bases, are easily accessible worldwide, and boast of a mature infrastructure~\cite{KIa17}.
The ability to monetize the user's data by the providers of the OSNs stands out as a major reason for the continued growth and establishment of centralized OSNs.
Thus far, it appears to the users that the accrued benefits of using centralized OSNs are way better than what DOSNs offer.

\paragraph*{Pure vs Hybrid}
In the early days of P2P technology, most proposed P2P-based OSNs tended towards two directions: either pure P2P (structured, unstructured or combinations) or P2P augmented with centralized technology (hybrid).
A major drawback experienced in pure P2P applications is the need for an always online bootstrap node, which may not always be achievable, to ensure that the network formed is kept alive which in turn affects profile and content availability as well as content distribution~\cite{MKI+16}.
On the other hand, hybrid P2P systems, despite overcoming these challenges faced in pure P2P systems, reintroduce the shortfalls of centralized systems that affect the OSNs.
Therefore, solutions proposed in either line must weigh the pros and cons associated, and what system designers are willing to make compromises vis-{\`a}-vis what the users are willing to tolerate.

\section{Conclusion}
\label{sec:concl}
This survey was divided into three major sections.
The first section laid a foundation for the P2P-based OSNs by introducing social networks in general, providing a clear road map for the entire study.
The second section of the study was a comprehensive breakdown and discussion of the key enabling features of P2P networks that support implementation of applications, and in particular, implementation of online social networks.
However, the technology is fast evolving and there are many changes that have since been observed.
This section highlighted the various developments and proposals that have been presented in literature, considering each key feature.
The final section was an analysis of selected proposed P2P-based OSNs, highlighting key P2P features implement.

From the analysis done on the P2P-based OSNs, there are some positives that can be highlighted.
It is important to note that the P2P-based OSNs, have been as a result of seeking to bring to face a solution that meets the shortfalls seen in the centralized OSNs, by providing a decentralized platform that was not only privacy-preserving, secure and scalable but also achievable.
The P2P platform has shown capabilities of meeting all these goals, but only after the implementation of novel solutions to meet application specific challenges that guarantee robustness, storage, data availability, reliable communication as well as security.
A brief summary of some of the solutions to achieve this have been discussed.

From the analysis of the P2P-based solutions, it is seen that some of these proposals fail in meeting all the minimum requirements to guarantee maximum user experience.
The analysis considered if the defined requirements (functional and non functional) are achieved.
Further, the P2P components implemented in the proposals were compared.
Additionally, the security features were analyzed as a key ingredient in the OSNs, and in particular because the P2P-based OSNs are designed with a goal of being able to provide security and privacy.
A major concern observable with all the P2P-based OSNs, excluding twister, is that all of them are not online, or indeed have never been online.
This may probably be due to several factors:
\begin
 {enumerate*} [label=\itshape\alph*\upshape)]
\item
 most P2P-based OSNs have been research projects so that beyond the initial work no one undertakes further development,
\item
 the number of critical network users in the P2P-based OSNs is not easily achieved (\textit{chicken-egg problem}), and
\item
 lack of monetization of the P2P-based OSNs hence no motivation for further system development due to the absence of any meaningful financial gain.
\end
{enumerate*} This means that although in theory, these systems are better than the current centralized implementations, they are not seen to make an impact in terms of user communities that utilize them.

\FloatBarrier

\bibliographystyle{IEEEtran.bst}
\balance
\bibliography{newrefs}

\end{document}